\documentclass[%
reprint,
superscriptaddress,
longbibliography,
 amsmath,amssymb,
 aps,prx,
]{revtex4-2}

\usepackage[T1]{fontenc}

\usepackage{graphicx}%
\usepackage{dcolumn}%
\usepackage{hyperref}%
\hypersetup{
	colorlinks=true,
	linkcolor=blue,
	citecolor=blue,
	urlcolor=blue,
}

\usepackage{tikzscale}
\usepackage[normalem]{ulem}
\usepackage{algorithm}
\usepackage{algpseudocode}
\usepackage{comment}
\usepackage{enumerate}
\usepackage{xcolor}

\usepackage{mathtools}
\usepackage{amsmath}
\usepackage{amsthm}
\usepackage{amssymb}
\usepackage{bm}%
\usepackage{bbm}
\usepackage{physics}
\usepackage{orcidlink}

\usepackage{tikz}
\usetikzlibrary{arrows,calc,decorations.markings,math,arrows.meta,shapes.geometric}
\usetikzlibrary{tikzmark,scopes,decorations.pathreplacing,shadings}

\usepackage{subfigure}
\usetikzlibrary{calc}

\newcommand{\mixedMPS}{
\begin{tikzpicture}[]

\node at (-1.6,3.6) {(a)};
   \foreach \i in {1,...,\Nq} {
  	\node[draw=\tensorborder, fill=\tensorcolor, line width=\linew, regular polygon, regular polygon sides=3, shape border rotate=90, minimum size=\rtensor,inner sep=0pt] (A1\i) at (\i, 0) {};
  }
  \foreach \i in {1,...,\Nq} {
  	\node[draw=\tensorborder, fill=\tensorcolor, line width=\linew, regular polygon, regular polygon sides=3, shape border rotate=90, minimum size=\rtensor,inner sep=0pt] (A2\i) at (\i, 1) {};
  }
  \foreach \i in {1,...,\Nq} {
  	\node[draw=\tensorborder, fill=\tensorcolor, line width=\linew, regular polygon, regular polygon sides=3, shape border rotate=90, minimum size=\rtensor,inner sep=0pt] (A3\i) at (\i, 2) {};
  }
  \foreach \i in {1,...,\Nq} {
  	\node[draw=\tensorborder, fill=\tensorcolor, line width=\linew, regular polygon, regular polygon sides=3, shape border rotate=90, minimum size=\rtensor,inner sep=0pt] (A4\i) at (\i, 3) {};
  }
  \foreach \i in {1,...,\the\numexpr\Nq-1} {
    \draw[thick] (A1\i) -- (A1\the\numexpr\i+1\relax);
    \draw[thick] (A2\i) -- (A2\the\numexpr\i+1\relax);
    \draw[thick] (A3\i) -- (A3\the\numexpr\i+1\relax);
    \draw[thick] (A4\i) -- (A4\the\numexpr\i+1\relax);
  }

  \foreach \i in {1,...,\Nq} {
    \draw[thick] (A2\i) -- (A3\i);
  }
  \draw[thick] (A1\Nq.east) .. controls(\Nq+0.4,0.25) and (\Nq+0.4,0.75) .. (A2\Nq.east);
  \draw[thick] (A3\Nq.east) .. controls(\Nq+0.4,2.25) and (\Nq+0.4,2.75) .. (A4\Nq.east);

\node[draw=\opborder, rectangle,  fill=\opcolor,  line width=\linew, minimum width=\rtensor, minimum height=\rtensor] (Z1) at (1, 1.5) {\tiny $Z$};
\node[draw=\opborder, rectangle,  fill=\opcolor,  line width=\linew, minimum width=\rtensor, minimum height=\rtensor] (X1) at (2, 1.5) {\tiny $X$};
\node[draw=\opborder, rectangle,  fill=\opcolor,  line width=\linew, minimum width=\rtensor, minimum height=\rtensor] (Z2) at (1, 3.5) {\tiny $Z$};
\node[draw=\opborder, rectangle,  fill=\opcolor,  line width=\linew, minimum width=\rtensor, minimum height=\rtensor] (Z3) at (2, 3.5) {\tiny $Z$};
\node[draw=\opborder, rectangle,  fill=\opcolor,  line width=\linew, minimum width=\rtensor, minimum height=\rtensor] (Y1) at (3, 3.5) {\tiny $Y$};
\node at (-1.,1.5) { $\mathrm{Tr}(\rho \gamma_j \rho \gamma_k)=$};
\foreach \i in {4,...,\Nq} {
    \draw[thick] (A4\i.north) to[out=105, in=-105, looseness=1.4] (A1\i.south);
 }
 \draw[thick, overlay] (A41.north) -- (Z2);
 \draw[thick, overlay] (A42.north) -- (Z3);
 \draw[thick, overlay] (A43.north) -- (Y1);	
  \draw[thick, overlay] (Z2) to[out=105, in=-105, looseness=1.4] (A11.south);
  \draw[thick, overlay] (Z3) to[out=105, in=-105, looseness=1.4] (A12.south);
  \draw[thick, overlay] (Y1) to[out=105, in=-105, looseness=1.4] (A13.south);

\pgfresetboundingbox
\useasboundingbox (-1.8,-1.2) rectangle (5.2,4.2);

\end{tikzpicture}

}

\newcommand{\Tj}{
\begin{tikzpicture}[]

\node at (-0.8,2.6) {(b)};
 
  \foreach \i in {1,...,\Nq} {
  	\node[draw=\tensorborder, fill=\tensorcolor, line width=\linew, regular polygon, regular polygon sides=3, shape border rotate=90, minimum size=\rtensor,inner sep=0pt] (A2\i) at (\i, 1) [label=below:\tiny $A_{\i}$]{};
  }
  \foreach \i in {1,...,\Nq} {
  	\node[draw=\tensorborder, fill=\tensorcolor, line width=\linew, regular polygon, regular polygon sides=3, shape border rotate=90, minimum size=\rtensor,inner sep=0pt] (A3\i) at (\i, 2) [label=above:\tiny $A_{\i}^*$]{};
  }
  \foreach \i in {1,...,\the\numexpr\Nq-1} {
    \draw[thick] (A2\i) -- (A2\the\numexpr\i+1\relax);
    \draw[thick] (A3\i) -- (A3\the\numexpr\i+1\relax);
  }

  \foreach \i in {1,...,\Nq} {
    \draw[thick] (A2\i) -- (A3\i);
  }

\node[draw=\opborder, rectangle,  fill=\opcolor,  line width=\linew, minimum width=\rtensor, minimum height=\rtensor] (Z1) at (1, 1.5) {\tiny $Z$};
\node[draw, rectangle,  fill=\opcolor,  line width=\linew, minimum width=\rtensor, minimum height=\rtensor] (X1) at (2, 1.5) {\tiny $X$};

\node at (0.4,1.5) { $=$};
\node[draw=\tensorborder, rectangle, rounded corners=\corner,  line width=\linew, fill=\tensorcolor, minimum height=1.25cm, minimum width=\rtensor] (T) at (-0.4, 1.5) {\tiny $T_{j}$};

\draw[thick] (\Nq+0.1,2) -- (\Nq+0.4,2);
\draw[thick] (\Nq+0.1,1) -- (\Nq+0.4,1);
\draw[thick] (-0.15,2) -- (0.15,2);
\draw[thick] (-0.15,1) -- (0.15,1);

\end{tikzpicture}
}

\newcommand{\initLSigma}{
\begin{tikzpicture}
    \node at (-0.8,1.6) {(a)};
  \node[draw=\tensorborder, fill=\tensorcolor, line width=\linew, regular polygon, regular polygon sides=3, shape border rotate=90, minimum size=\rtensor,inner sep=0pt] (Ar) at (2, 0) [label=below:\tiny $A_{j}$]{};
\node[draw=\tensorborder, fill=\tensorcolor, line width=\linew, regular polygon, regular polygon sides=3, shape border rotate=90, minimum size=\rtensor, inner sep=0pt] (Ar_dag) at (2, 1) [label=above:\tiny $A_{j}^*$]{};
\node[draw=\opborder, rectangle,  fill=\opcolor,  line width=\linew, minimum width=\rtensor, minimum height=\rtensor] (Op) at (2, 0.5) {\tiny $X$};
\node[draw=\tensorborder, rectangle, rounded corners=\corner,  line width=\linew, fill=\tensorcolor, minimum height=1.25cm, minimum width=\rtensor] (L) at (-0.19, 0.5) {\tiny $L_{X}$};
\node[draw=\tensorborder, rectangle, rounded corners=\corner, line width=\linew, fill=\tensorcolor, minimum height=1.25cm, minimum width=\rtensor] (L2) at (1.35, 0.5) {\tiny $L_{Z}$};
\node at (0.7,0.5) { $=$};
	
\draw[thick] (1.625,0) -- (Ar);
\draw[thick]   (2.375,0) -- (Ar) ;
\draw[thick] (1.625,1) -- (Ar_dag);
\draw[thick]   (2.375,1) -- (Ar_dag) ;
\draw[thick] (Op) -- (Ar);
\draw[thick] (Op) -- (Ar_dag);

\draw[thick] (0.125,1) -- (0.375,1);
\draw[thick] (0.125,0) -- (0.375,0);

  \node[draw=\tensorborder, fill=\tensorcolor, line width=\linew, regular polygon, regular polygon sides=3, shape border rotate=90, minimum size=\rtensor,inner sep=0pt] (Ar) at (6., 0) [label=below:\tiny $A_{j}$]{};
\node[draw=\tensorborder, fill=\tensorcolor, line width=\linew, regular polygon, regular polygon sides=3, shape border rotate=90, minimum size=\rtensor, inner sep=0pt] (Ar_dag) at (6., 1) [label=above:\tiny $A_{j}^*$]{};
\node[draw=\opborder, rectangle,  fill=\opcolor,  line width=\linew, minimum width=\rtensor, minimum height=\rtensor] (Op) at (6., 0.5) {\tiny $Y$};
\node[draw=\tensorborder, rectangle, rounded corners=\corner,  line width=\linew, fill=\tensorcolor, minimum height=1.25cm, minimum width=\rtensor] (L) at (3.81, 0.5) {\tiny $L_{Y}$};
\node[draw=\tensorborder, rectangle, rounded corners=\corner, line width=\linew, fill=\tensorcolor, minimum height=1.25cm, minimum width=\rtensor] (L2) at (5.35, 0.5) {\tiny $L_{Z}$};
\node at (4.7,0.5) { $=$};
	
\draw[thick] (5.625,0) -- (Ar);
\draw[thick]   (6.375,0) -- (Ar) ;
\draw[thick] (5.625,1) -- (Ar_dag);
\draw[thick]   (6.375,1) -- (Ar_dag) ;
\draw[thick] (Op) -- (Ar);
\draw[thick] (Op) -- (Ar_dag);

\draw[thick] (4.125,1) -- (4.375,1);
\draw[thick] (4.125,0) -- (4.375,0);
\end{tikzpicture}
}

\newcommand{\updateLSigma}{
\begin{tikzpicture}[scale=1]
  
  \node at (-1.7,1.6) {(b)};
  \node[draw=\tensorborder, fill=\tensorcolor, line width=\linew, regular polygon, regular polygon sides=3, shape border rotate=90, minimum size=\rtensor,inner sep=0pt] (Ar) at (2, 0) [label=below:\tiny $A_{k}$]{};
  \node[draw=\tensorborder, fill=\tensorcolor, line width=\linew, regular polygon, regular polygon sides=3, shape border rotate=90, minimum size=\rtensor,inner sep=0pt] (Ar_dag) at (2, 1) [label=above:\tiny $A_{k}^*$]{};
\node[draw=\tensorborder, rectangle, rounded corners=\corner,  line width=\linew, fill=\tensorcolor, minimum height=1.25cm, minimum width=\rtensor] (L) at (-0.98, 0.5) {\tiny $L_{X(Y)}$};
\node[draw=\tensorborder, rectangle, rounded corners=\corner, line width=\linew, fill=\tensorcolor, minimum height=1.25cm, minimum width=\rtensor] (L2) at (1.12, 0.5) {\tiny $L_{X(Y)}$};
\draw[-{Latex[scale=1.0]},thick] (-0.1,0.5) -- (0.4,0.5);
	
\draw[thick] (1.625,0) -- (Ar);
\draw[thick]   (2.375,0) -- (Ar) ;
\draw[thick] (1.625,1) -- (Ar_dag);
\draw[thick]   (2.375,1) -- (Ar_dag) ;
\draw[thick] (Ar) -- (Ar_dag);

\draw[thick] (0.125-0.6,1) -- (0.375-0.6,1);
\draw[thick] (0.125-0.6,0) -- (0.375-0.6,0);
\end{tikzpicture}
}

\newcommand{\updateLZSigma}{
\begin{tikzpicture}[]

 \node at (-0.6,1.6) {(c)};
  \node[draw=\tensorborder, fill=\tensorcolor, line width=\linew, regular polygon, regular polygon sides=3, shape border rotate=90, minimum size=\rtensor,inner sep=0pt] (Ar) at (2, 0) [label=below:\tiny $A_{j}$]{};
\node[draw=\tensorborder, fill=\tensorcolor, line width=\linew, regular polygon, regular polygon sides=3, shape border rotate=90, minimum size=\rtensor, inner sep=0pt] (Ar_dag) at (2, 1) [label=above:\tiny $A_{j}^*$]{};
\node[draw=\opborder, rectangle,  fill=\opcolor,  line width=\linew, minimum width=\rtensor, minimum height=\rtensor] (Op) at (2, 0.5) {\tiny $Z$};
\node[draw=\tensorborder, rectangle, rounded corners=\corner,  line width=\linew, fill=\tensorcolor, minimum height=1.25cm, minimum width=\rtensor] (L) at (-0.15, 0.5) {\tiny $L_{Z}$};
\node[draw=\tensorborder, rectangle, rounded corners=\corner, line width=\linew, fill=\tensorcolor, minimum height=1.25cm, minimum width=\rtensor] (L2) at (1.35, 0.5) {\tiny $L_{Z}$};
\draw[-{Latex[scale=1.0]},thick] (0.4,0.5) -- (0.9,0.5);
	
\draw[thick] (1.625,0) -- (Ar);
\draw[thick]   (2.375,0) -- (Ar) ;
\draw[thick] (1.625,1) -- (Ar_dag);
\draw[thick]   (2.375,1) -- (Ar_dag) ;
\draw[thick] (Op) -- (Ar_dag);
\draw[thick] (Op) -- (Ar);

\draw[thick] (0.125,1) -- (0.375,1);
\draw[thick] (0.125,0) -- (0.375,0);

\end{tikzpicture}
}
\def\rtensor{12}
\def\linew{0.6pt}
\def\corner{0.25}
\def\tensorcolor{blue!30}
\def\tensorborder{black}
\def\opcolor{gray!50}
\def\opborder{black}
\def\Nq{5}

\def\doi{http://dx.doi.org/}

\newcommand{\be}{\begin{equation}}
\newcommand{\ee}{\end{equation}}
\newcommand{\bec}{\begin{equation*}}
\newcommand{\eec}{\end{equation*}}
\newcommand{\bea}{\begin{eqnarray}}
\newcommand{\eea}{\end{eqnarray}}

\usepackage{enumitem}

\usepackage{calc}

\usepackage{pifont}

\usepackage{dsfont}
\newcommand{\G}{\mathcal{G}}
\newcommand{\M}{\mathcal{M}}
\newcommand{\nGoccu}{\M_\text{occu}}
\newcommand{\nGOG}{\M_{\text{OG}}}
\newcommand{\nGNO}{\M_{\text{NO}}}
\newcommand{\nGR}{\M_R}

\newcommand{\ME}{E_{T}}
\newcommand{\avgnGoccu}{\overline{\mathcal{M}}_\textrm{occu}}
\newcommand{\avgnGNO}{\overline{\mathcal{M}}_\textrm{NO}}
\newcommand{\Id}{\mathbbm{1}}
\renewcommand{\d}{\text{d}}
\newcommand{\brap}[1]{(#1|}
\newcommand{\ketp}[1]{|#1)}
\newcommand{\braketp}[2]{(#1|#2)}
\newcommand{\cutoff}{R_c}

\begin{document}

\title{Computable fermionic non-Gaussianity from the covariance matrix}

\newcommand{\TUM}{\affiliation{Technical University of Munich, TUM School of Natural Sciences, Physics Department, James-Franck-Stra{\ss}e 1,
85748 Garching, Germany}}
\newcommand{\MCQST}{\affiliation{Munich Center for Quantum Science and Technology (MCQST), Schellingstra{\ss}e 4, 80799 M{\"u}nchen, Germany}}
\newcommand{\Quantinuum}{\affiliation{Quantinuum, Leopoldstra{\ss}e 180, 80804 M{\"u}nchen, Germany}}

\author{Poetri Sonya Tarabunga\,\orcidlink{0000-0001-8079-9040}}\email{poetri.tarabunga@tum.de} \TUM \MCQST
\author{Bernhard Jobst\,\orcidlink{0000-0001-7027-3918}} \TUM \MCQST
\author{Ra{\'u}l Morral-Yepes\,\orcidlink{0009-0007-4247-4772}} \TUM \MCQST 
\author{Marc~Langer\,\orcidlink{0009-0008-4515-7171}} \TUM \MCQST
\author{Barbara Kraus\,\orcidlink{0000-0001-7246-6385}} \TUM \MCQST
\author{Frank Pollmann\,\orcidlink{0000-0003-0320-9304}} \TUM \MCQST
\author{Sheng-Hsuan Lin\,\orcidlink{0000-0002-8599-0378}}\email{sheng-hsuan.lin@quantinuum.com}  \Quantinuum

\begin{abstract}
Fermionic non-Gaussianity, or fermionic magic, is a key resource underlying the computational complexity of fermionic quantum systems, yet tractable and operationally meaningful ways to quantify it remain limited. We address this challenge by developing a convex resource theory of fermionic non-Gaussianity and introducing two families of computable quantities for pure fermionic states, both derived from the Williamson normal form of the covariance matrix. The first family, the occupation number entropies, is defined as the Tsallis-$\alpha$ entropy of the occupation numbers. We prove that two members of this family are monotonic under Gaussian protocols, establishing them as computable convex resource monotones. They consequently lower-bound the number of non-Gaussian gates needed for state preparation. The second family, the natural-orbital participation entropies, is given by the R\'enyi-$\alpha$ entropy of the squared amplitudes of the state in the natural-orbital basis, defined by the eigenvectors of the covariance matrix. They quantify state compressibility in this basis and thus upper-bound the classical simulation cost in an orthonormal Gaussian basis. We analyze both families for stabilizer and translation-invariant states, where they simplify and reveal additional structure. We further study representative examples, including random SWAP-doped matchgate circuits and the bond-modulated XXZ model, highlighting the role of non-Gaussianity in many-body phenomena. Our work establishes a resource-theoretic framework for computable fermionic non-Gaussianity that unifies notions arising across quantum information, condensed-matter physics, and quantum chemistry, opening new directions for studying the complexity of quantum many-body systems and providing practical tools to assess the classical simulability of fermionic states relevant for quantum advantage.
\end{abstract}

\maketitle

\section{Introduction}

Quantum computers hold promise to provide significant computational advantage over classical computers~\cite{preskill2012quantum}. This expectation originates from the fact that the difficulty of representing generic quantum systems with a classical computer grows exponentially with the number of constituents, so that naively simulating it on a classical computer requires exponential resources. Nevertheless, not all quantum states are equally complex: several important families of states can be efficiently simulated classically. Understanding which features of a quantum state hinder such efficient classical simulation has thus become a central goal in identifying the true sources of quantum advantage~\cite{harrow2017quantum}.

One of the first properties recognized to characterize the complexity of quantum states is entanglement~\cite{horodecki2009quantum}.
However, it is now understood that entanglement, while necessary for classical intractability, is not sufficient. Highly entangled states can still be efficiently simulated when they belong to certain classes of states. For example, the stabilizer formalism identifies such a tractable class: Clifford circuits acting on stabilizer states can be simulated in polynomial time~\cite{gottesman1998theory, aaronson2004improved}. Universal quantum computation requires going beyond this class through the injection of nonstabilizer or magic states combined with feedforward~\cite{bravyi2005universal}. Another tractable class is that of fermionic Gaussian states (FGSs), also known as free-fermion states~\cite{surace2022fgs}. These are states generated by fermionic Gaussian unitaries (FGUs), i.e., unitaries generated by Hamiltonians quadratic in fermion operators, acting on a vacuum state. Their structure allows for efficient classical simulation via their covariance matrix.
To go beyond this class, one requires non-Gaussian operations or fermionic magic states~\cite{hebenstreit2019all} that depart from the classical simulation framework of free-fermion states, again combined with feedforward.
Together, entanglement, nonstabilizerness, and non-Gaussianity represent distinct types of quantum resources, each formalized within the framework of quantum resource theories~\cite{chitambar2019quantum, gour2024resources}.

A resource theory is built from two ingredients: a set of \emph{free states}, which can be prepared at no cost, and a set of \emph{free operations}, which cannot create the resource and hence map free states to free states. A central task is to quantify how resourceful a state is. The basic requirement on such a quantity is \emph{monotonicity} under the free operations: it cannot increase under free operations. We call a quantity satisfying this property a \emph{monotone}. A different (and not necessarily stronger~\footnote{It becomes a stronger requirement when the quantity is convex or in the pure-state setting, in which case strong monotonicity implies monotonicity.}) requirement is \emph{strong monotonicity}: the quantity does not increase on average under probabilistic free protocols. Throughout we follow the convention from Ref.~\cite{chitambar2019quantum}, reserving \emph{monotone} for the former and \emph{strong monotone} for the latter~\footnote{Terminology differs across communities. In entanglement theory, following Ref.~\cite{vidal2000entanglement}, the term \emph{monotone} is itself defined through the on-average condition under probabilistic protocols, so that a monotone corresponds to what we call a \emph{strong} monotone. A quantity that is monotonic only under deterministic protocols is instead called a \emph{measure}, which corresponds to what we call a monotone.}.

In the resource theory of fermionic non-Gaussianity, the free operations are the
Gaussian operations---for example Gaussian unitaries, the addition or removal of Gaussian
ancillas, or Gaussian measurements---which preserve the efficiently simulable class of
FGSs, and under which a proper monotone must be non-increasing. Moreover, to be practically relevant in a many-body setting, a monotone should be computationally tractable for a broad class of states, whether in experiments or in numerical simulations, so that it can be investigated systematically in large-scale systems of interest.

Recent progress in quantifying nonstabilizerness in qubit systems has highlighted the importance of computationally tractable resource monotones with clear operational meaning. In particular, the stabilizer R\'enyi entropy (SRE)~\cite{leone2022stabilizer}, shown to be a monotone under Clifford protocols~\cite{leone2024stabilizer} and to bound the resources needed to generate state designs~\cite{bittel2025operationalinterpretation}, has ignited extensive studies of nonstabilizerness in many-body systems, both in and out of equilibrium~\cite{haug2022quantifying, tarabunga2023many, tarabunga2023magic, falcao2024nonstabilizerness, ding2025evaluating, timsina2025robustnessmagic, lami2024quantum, haug2024probing, turkeshi2024magic, tirrito2024anticoncentration, falcao2025magicdynamics, dowling2025magicheisenbergpicture, tirrito2025universalspreading, fux2023entanglementmagic, bejan2023dynamical, tarabunga2024magictransition, frau2024non, frau2024stabilizer, tarabunga2024mps,nehra2025topol,sierant2026theory,xiao2026diffusive}.

The study of fermionic non-Gaussianity in many-body systems has an even longer history~\cite{gottlieb2007properties, gottlieb2014correlations, turner2017optimal, pachos2018quantifying, meichanetzidis2018free, patrick2019interaction}, motivated by the central role of free fermions in quantum many-body physics~\cite{surace2022fgs}.
Indeed, free-fermionic states serve as the natural starting point for a wide range of many-body models, accurately capturing ground-state correlations in systems from mean-field Hartree-Fock states~\cite{echenique2007a,BaLi94} to BCS superconductors~\cite{bardeen1957theory}.
Interactions, however, drive the ground state away from any free description, giving rise to exotic strongly correlated phases, such as high-temperature superconductivity~\cite{lee2006doping,Keimer2015} and the fractional quantum Hall effect~\cite{tsui1982fqh,laughlin1983,Jain2007}, %
that require approaches beyond mean-field and perturbation theory.
Fermionic non-Gaussianity quantifies this departure: it vanishes for free-fermion states and aims to quantify the correlations that interactions generate beyond any free-fermion description, making it a natural diagnostic of strongly interacting many-body physics.
Yet a major challenge has been to develop quantitative tools that are simultaneously computationally tractable, monotonic under physically motivated free operations (defined below), and operationally meaningful.
Non-Gaussianity monotones such as the Gaussian fidelity, Gaussian extent, or Gaussian rank~\cite{dias2024classical, cudby2023gaussian, ReardonSmith2024improved}, involve computationally demanding optimizations that preclude their evaluation in large systems, while the more recently proposed computable quantifiers~\cite{lumia2024measurement, lyu2024fermionicgaussiantesting, sierant2025fermionicmagicresourcesquantum, coffman2025measuringnongaussianmagicfermions,sierant2026theorymatchgatecommutant} have not been shown to be monotonic beyond unitary Gaussian operations and mostly lack a clear operational interpretation related to the complexity of the state. The field thus still lacks an operationally grounded and computationally tractable framework, in stark contrast to nonstabilizerness.

In this work, we address this gap by formulating a convex resource theory of fermionic non-Gaussianity, in which the free states are convex mixtures of fermionic Gaussian states and the free operations are Gaussian protocols with feedforward---adaptive Gaussian operations that are native to fermionic hardware yet still classically tractable. This setting is natural for magic-state injection in universal fermionic quantum computation~\cite{hebenstreit2019all}, yet it has not been fully developed in previous literature, which primarily considers only the more restrictive set of Gaussian operations without feedforward. Within this framework, we introduce two complementary families of computable quantities for pure fermionic states, both built from the Williamson normal form of the covariance matrix. The first family, the occupation number entropies, is solely defined from the Williamson eigenvalues, making them immediately applicable to large-scale numerical studies and to experimental settings where covariance matrix data are accessible. We prove that two members of this family are strong monotones under pure-state Gaussian protocols. This result directly yields lower bounds on the number of non-Gaussian gates needed to prepare a state. These entropies further upper-bound the deviation from Wick's theorem, which therefore still holds approximately for states with small occupation number entropy. The second family, the natural-orbital participation entropies, are defined as the R\'enyi participation entropies---the R\'enyi entropies of the squared wave-function amplitudes---of the state in the natural-orbital basis. These quantities combine information from both the covariance matrix and the quantum state itself, and quantify how well a state can be approximated by a sparse superposition of orthonormal Gaussian states in the natural-orbital basis, thereby upper-bounding the cost of classical simulation based on orthonormal Gaussian expansions. Beyond these properties, we show that both families also serve as diagnostics of quantum phase transitions in interacting many-body systems.
In a parallel work by one of the authors~\cite{tarabunga2026fermionic}, a different computable non-Gaussianity monotone was introduced, providing a complementary approach beyond the covariance-matrix-based characterizations.

Overall, our work provides a unified and computable framework for fermionic non-Gaussianity, built around the covariance matrix and the structure of the state in the natural-orbital basis. It connects resource monotones from quantum information, Wick's theorem and quantum phase transitions central to condensed-matter physics, and natural orbitals from quantum chemistry, making each connection concrete: the occupation number entropy (which has also been introduced in quantum chemistry under the name of correlation entropy~\cite{ziesche1995correlation}), which we prove to be a resource monotone, bounds the deviation from Wick's theorem and diagnoses quantum phase transitions. The companion natural-orbital participation entropy instead quantifies how broadly the state spreads over the natural-orbital basis, controlling its classical-simulation cost. Together with efficient algorithms to evaluate them, these quantities provide a common, computable toolbox across quantum information, many-body physics, and quantum chemistry.

The remainder of the paper is organized as follows.
We provide a short summary of the main results in Sec.~\ref{sec: summary}.
Sec.~\ref{sec: preliminaries} introduces the fermionic resource theory of non-Gaussianity. 
Sec.~\ref{sec: nG_measures} defines the occupation number entropies and natural-orbital participation entropies and establishes their main properties. 
Secs.~\ref{sec:pure_stab} and~\ref{sec:TI} provide discussion focusing on stabilizer states and translation-invariant systems. 
Sec.~\ref{sec: example} presents numerical applications of the introduced quantities in the setting of random circuits and a quantum phase transition in a many-body system, and Sec.~\ref{sec:concl} concludes.

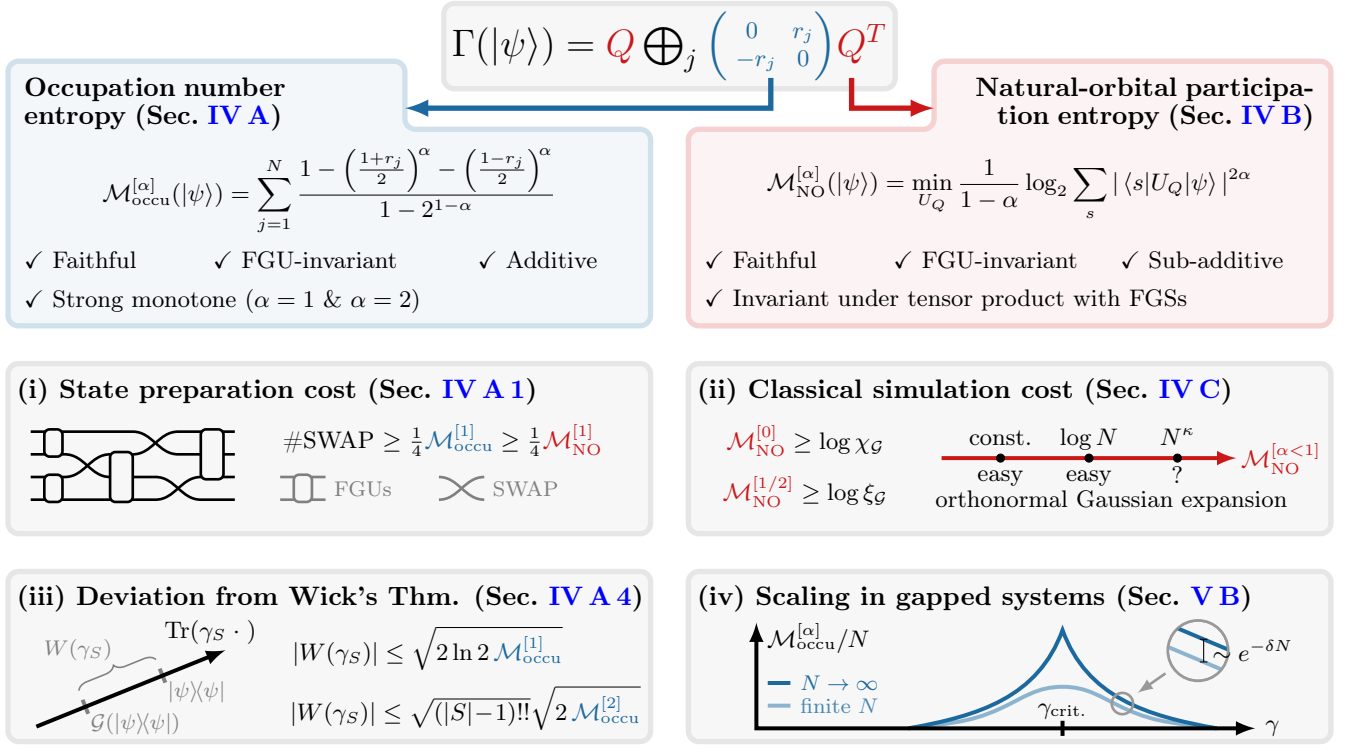
\begin{figure*}
	
	\newcommand{\maineqtextsize}{\Large}
	\newcommand{\headlinetext}[1]{\normalsize\textbf{#1}}
	\def\picheight{7}
	\def\maineqxpos{8.75} %
	\def\cutoutmaineqx{3.5}
	\def\cutoutmaineqy{0.9}
	\def\quantityboxheight{3.5}
	\def\boxdist{0.5}
	\def\roundedcorners{4pt}
	\def\drawlinewidth{2pt}
	\definecolor{moccu_in_maineq}{HTML}{1b699e}%
	\colorlet{moccu_box_fg}{moccu_in_maineq!20}
	\colorlet{moccu_box_bg_a}{moccu_in_maineq!5}
	\colorlet{moccu_box_bg_b}{moccu_in_maineq!5}
	\definecolor{mno_in_maineq}{HTML}{cc1719}%
	\colorlet{mno_box_fg}{mno_in_maineq!20}
	\colorlet{mno_box_bg_a}{mno_in_maineq!5}
	\colorlet{mno_box_bg_b}{mno_in_maineq!5}
	\colorlet{generic_gray_a}{black!3}
	\colorlet{generic_gray_b}{black!3}
	\colorlet{generic_gray_fg}{black!10}
	\centering
	\begin{tikzpicture}[remember picture]
		\node[draw = generic_gray_fg, fill, top color = generic_gray_a,bottom color = generic_gray_b, 
			line width=\drawlinewidth, rounded corners=\roundedcorners]
		 (CMequation) at (\maineqxpos,\picheight+0.2) { \maineqtextsize
			$\Gamma(\ket\psi) = 
				{\color{mno_in_maineq} Q} 
				\bigoplus_j \subnode{maineq_eigval}{{\color{moccu_in_maineq}
					{\small \begin{pmatrix} 0 & r_j \\ -r_j & 0 \end{pmatrix}}
					}} 
				{\color{mno_in_maineq} \subnode{maineq_rot}{Q}^T}$
		};

		\draw[draw=moccu_box_fg, bottom color=moccu_box_bg_b, top color=moccu_box_bg_a,
		 	line width=\drawlinewidth,rounded corners=\roundedcorners]
			(0,\picheight) coordinate(moccu_box_top_left) -- 
			($(\maineqxpos-\cutoutmaineqx,\picheight)$) coordinate(moccu_box_top_right) -- 
			($(\maineqxpos-\cutoutmaineqx,\picheight-\cutoutmaineqy)$) -- 
			($(\maineqxpos-0.5*\boxdist,\picheight-\cutoutmaineqy)$) coordinate(moccu_box_outer_corner) --
			($(\maineqxpos-0.5*\boxdist,\picheight - \quantityboxheight)$)--
			($(0,\picheight - \quantityboxheight)$) coordinate(moccu_box_bottom_corner) -- cycle;
		
		\draw[draw=mno_box_fg, bottom color=mno_box_bg_b, top color=mno_box_bg_a,
		 	line width=\drawlinewidth,rounded corners=\roundedcorners]
			(2*\maineqxpos,\picheight) coordinate(mno_box_top_right)-- 
			($(\maineqxpos+\cutoutmaineqx,\picheight)$) coordinate(mno_box_top_left)-- 
			($(\maineqxpos+\cutoutmaineqx,\picheight-\cutoutmaineqy)$) -- 
			($(\maineqxpos+0.5*\boxdist,\picheight-\cutoutmaineqy)$) coordinate(mno_box_outer_corner)--
			($(\maineqxpos+0.5*\boxdist,\picheight - \quantityboxheight)$)--
			($(2*\maineqxpos,\picheight - \quantityboxheight)$) coordinate(mno_box_bottom_corner) -- cycle;
			
		\coordinate(moccu_arrow_start) at (maineq_eigval.south |- maineq_rot.south);
		\coordinate (moccu_arrow_lower) at ($(moccu_arrow_start) - (0,0.4)$);
		\coordinate(mno_arrow_start) at (maineq_rot.south);
		\coordinate (mno_arrow_lower) at ($(mno_arrow_start) - (0,0.4)$);

		\draw[-latex, line width = \drawlinewidth, moccu_in_maineq] 
			(moccu_arrow_start) -- (moccu_arrow_lower) -- (moccu_arrow_lower -| moccu_box_top_right);
			
		\draw[-latex, line width = \drawlinewidth, mno_in_maineq] 
			(mno_arrow_start) -- (mno_arrow_lower) -- (mno_arrow_lower -| mno_box_top_left);

		\node[xshift=0.1cm,yshift =-0.1cm, anchor = north west, text width=4.8cm] at (moccu_box_top_left) 
				{\headlinetext{Occupation number\\ entropy (Sec.~\ref{subsec:nG_occu})}};
		
		\node[yshift = -0.1cm,anchor = north, text width=8cm]
			 at ($(moccu_box_top_left |- moccu_box_outer_corner)!0.5!(moccu_box_outer_corner)$)
			{ \[\nGoccu^{[\alpha]}(\ket{\psi}) = 
			\sum_{j=1}^N \frac{1-\left(\frac{1 + r_j}{2}\right)^{\!\alpha} - \left(\frac{1 - r_j}{2}\right)^{\!\alpha}}{1-2^{1-\alpha}}\]};

		\node[xshift=0.1cm, yshift = -1.5cm, anchor = north west] 
			at (moccu_box_top_left |- moccu_box_outer_corner) 
			{$\checkmark$ Faithful};
		\node[xshift=2.6cm, yshift = -1.5cm, anchor = north west] 
			at (moccu_box_top_left |- moccu_box_outer_corner) 
			{$\checkmark$ FGU-invariant};
		\node[xshift=6.1cm, yshift = -1.5cm, anchor = north west] 
			at (moccu_box_top_left |- moccu_box_outer_corner) 
			{$\checkmark$ Additive};
		\node[xshift=0.1cm, yshift = -2cm, anchor = north west] 
			at (moccu_box_top_left |- moccu_box_outer_corner) 
			{$\checkmark$ Strong monotone ($\alpha = 1$ \& $\alpha = 2$)};

		\node[xshift=-0.1cm,yshift =-0.1cm, anchor = north east, align=right,text width=4.8cm] 
			at (mno_box_top_right) 
				{\headlinetext{Natural-orbital participation entropy (Sec.~\ref{subsec:NOPE})}};
		
		\node[yshift = -0.25cm,anchor = north, text width=8cm]
			 at ($(mno_box_top_right |- mno_box_outer_corner)!0.5!(mno_box_outer_corner)$)
			{ \[\nGNO^{[\alpha]}(\ket{\psi}) = 
			\min_{U_Q} \frac{1}{1-\alpha} \log_2 \sum_s \vert \braket{s}{U_Q | \psi} \vert^{2\alpha}
			\]};
		
		\node[xshift=0.1cm, yshift = -1.5cm, anchor = north west] at (mno_box_outer_corner) 
			{$\checkmark$ Faithful};
		\node[xshift=2.6cm, yshift = -1.5cm, anchor = north west] at (mno_box_outer_corner) 
			{$\checkmark$ FGU-invariant};
		\node[xshift=5.6cm, yshift = -1.5cm, anchor = north west] at (mno_box_outer_corner)
			{$\checkmark$ Sub-additive};
		\node[xshift=0.1cm, yshift = -2cm, anchor = north west] at (mno_box_outer_corner)
			{$\checkmark$ Invariant under tensor product with FGSs};

		\def\totalwidth{2*\maineqxpos}
		\def\smboxdist{0.5}
		\def\boxwidth{(0.5*\totalwidth - 0.5*\smboxdist)}
		\def\smboxheight{2.25}

		\coordinate (application_start) at ($(moccu_box_bottom_corner) + (0,-\smboxdist)$);
		\coordinate (application_end) at ($(mno_box_bottom_corner) + (0,-\smboxdist)$);

		\draw[draw = generic_gray_fg, top color = generic_gray_a,bottom color = generic_gray_b,
		 	line width=\drawlinewidth,rounded corners=\roundedcorners]
			(application_start) coordinate (b1_left) 
				-- coordinate[midway] (b1_north)
			($(application_start) + ({\boxwidth},0)$) -- 
			($(application_start) +({\boxwidth},-\smboxheight) $)-- coordinate[midway] (b1_south) 
			($(application_start) +(0,-\smboxheight)$) coordinate (b1_botleft) -- cycle;
		
		\draw[draw = generic_gray_fg, top color = generic_gray_a,bottom color = generic_gray_b,
		 	line width=\drawlinewidth,rounded corners=\roundedcorners]
			($(application_start) + ({\boxwidth + \smboxdist},0)$) coordinate (b2_left)
				-- coordinate[midway] (b2_north)
			($(application_start) + ({2*\boxwidth + \smboxdist},0)$) -- 
			($(application_start) + ({2*\boxwidth + 1*\smboxdist},-\smboxheight)$) 
				-- coordinate[midway] (b2_south) 
			($(application_start) + ({1*\boxwidth + \smboxdist},-\smboxheight)$) 
				-- cycle;		
		
		\draw[draw = generic_gray_fg, top color = generic_gray_a,bottom color = generic_gray_b,
		 	line width=\drawlinewidth,rounded corners=\roundedcorners]
			($(application_start) + ({0*\boxwidth + 0*\smboxdist},-\smboxheight - \smboxdist)$) 
				coordinate (b4_left) -- coordinate[midway] (b4_north)
			($(application_start) + ({1*\boxwidth + 0*\smboxdist},-\smboxheight - \smboxdist)$) 
				-- coordinate[midway] (b4_east) 
			($(application_start) + ({1*\boxwidth + 0*\smboxdist} ,-2*\smboxheight - \smboxdist)$) 
				-- coordinate[midway] (b4_south) 
			($(application_start) + ({0*\boxwidth + 0*\smboxdist},-2*\smboxheight - \smboxdist)$) 
				-- coordinate[midway] (b4_west) cycle;
			
		\draw[draw = generic_gray_fg, top color = generic_gray_a,bottom color = generic_gray_b,
		 	line width=\drawlinewidth,rounded corners=\roundedcorners]
			($(application_start) + ({1*\boxwidth + 1*\smboxdist},-\smboxheight - \smboxdist)$)
				 coordinate (b3_left) -- coordinate[midway] (b3_north)
			($(application_start) + ({2*\boxwidth + 1*\smboxdist} , -\smboxheight - \smboxdist)$) 
				-- coordinate[midway] (b3_east) 
			($(application_start) + ({2*\boxwidth + 1*\smboxdist} ,-2*\smboxheight - \smboxdist)$)
				 -- coordinate[midway] (b3_south) 
			($(application_start) + ({1*\boxwidth + 1*\smboxdist},-2*\smboxheight - \smboxdist)$) 
				-- coordinate[midway] (b3_west) cycle;

		\node[xshift=0.0cm,yshift =-0.07cm, anchor = north west, align=left] 
			at (b1_left) 
				{\headlinetext{(i) State preparation cost (Sec.~\ref{subsec:Moccu alpha1})} };
		
		\node at ($(b1_north)!0.45!(b1_south) + (1.5,0)$) {\#SWAP $
			\geq \frac14{\color{moccu_in_maineq} \nGoccu^{[1]}}
			\geq \frac14{\color{mno_in_maineq} \nGNO^{[1]}} $};
		\begin{scope}[shift ={(b1_botleft)}, scale = 0.3]\begin{scope}[shift = {(1,1.5)}]
			\draw[line width = 1pt] (0,0) -- (5.5,0) to[out=0,in=180] (7.5,1) -- (9,1);
			\draw[line width = 1pt] (0,1) -- (1.5,1) 
					to[out=0,in=180] (3.5,2) -- (4.5,2) to[out=0,in=180] (6.5,3) -- (9,3);
			\draw[line width = 1pt] (0,2) -- (1.5,2) 
					to[out=0,in=180] (3.5,1)  -- (5.5,1) to[out=0,in=180] (7.5,0) -- (9,0);
			\draw[line width = 1pt] (0,3) -- (4.5,3) to[out=0,in=180] (6.5,2) -- (9,2);
			
			\draw[line width = 1pt, rounded corners = 2pt, fill=generic_gray_b]
				(0.6,-0.1) rectangle (1.4,1.1); 
			\draw[line width = 1pt, rounded corners = 2pt, fill=generic_gray_b]
				(0.6,1.9) rectangle (1.4,3.1); 
			\draw[line width = 1pt, rounded corners = 2pt, fill=generic_gray_b]
				(3.5,-0.1) rectangle (4.5,2.1); 
			\draw[line width = 1pt, rounded corners = 2pt, fill=generic_gray_b]
				(7.5,0.9) rectangle (8.5,3.1);
		
		\end{scope}\begin{scope}[shift = {(11,1.5)}, black!50]
			\begin{scope}[xshift=8cm]
			\draw[line width = 1pt] (0,0) to[out=0,in=180] (2,1);
			\draw[line width = 1pt] (0,1) to[out=0,in=180] (2,0);
			\node[anchor = west] at (2,0.5) {\footnotesize SWAP}; \end{scope}
			
			\begin{scope}[xshift=1cm,yshift=-2cm]
			\draw[line width = 1pt] (0,2) -- (2,2);
			\draw[line width = 1pt] (0,3) -- (2,3);
			\node[anchor = west] at (2,2.5) {\footnotesize FGUs}; 
			\draw[line width = 1pt, rounded corners = 2pt, fill=generic_gray_b]
				(0.6,1.9) rectangle (1.4,3.1); \end{scope}

		\end{scope}\end{scope}

		\node[xshift=0.0cm,yshift =-0.07cm, anchor = north west, align=left] 
			at (b2_left) 
				{\headlinetext{(ii) Classical simulation cost (Sec.~\ref{subsec: classical_sim})}};
		
		\node at ($(b2_north)!0.45!(b2_south) - (2.7,0)$) {$
			{\color{mno_in_maineq} \nGNO^{[0]}} \geq \log \chi_\mathcal{G}$};
		\node at ($(b2_north)!0.75!(b2_south) - (2.7,0)$) {$
			{\color{mno_in_maineq} \nGNO^{[1/2]}} \geq \log \xi_\mathcal{G}$};
		\coordinate (b2_line_start) at ($(b2_south) + (-0.9,1.0)$);
		\coordinate (b2_line_end) at ($(b2_south) + (3.0,1.0)$);
		
		\draw[-latex, line width=1.5pt,mno_in_maineq] (b2_line_start) --
				node[below, yshift=-0.3cm, xshift=0.3cm, black]{orthonormal Gaussian expansion}  (b2_line_end)
				node[right, xshift=-0.7mm] {$\nGNO^{[\alpha<1]}$};
		
		\draw[fill] ($(b2_line_start)!0.2!(b2_line_end)$) circle(1.5pt) 
			node[above] {\phantom{g}const.\phantom{g}} node[below, yshift=-0.5mm] {easy};
		\draw[fill] ($(b2_line_start)!0.5!(b2_line_end)$) circle(1.5pt) 
			node[above] {$\log N$} node[below, yshift=-0.5mm] {easy};
		\draw[fill] ($(b2_line_start)!0.8!(b2_line_end)$) circle(1.5pt)
        node[above] {\phantom{g}$N^\kappa$\phantom{g}} node[below] {?};

		\node[xshift=0.0cm,yshift =-0.07cm, anchor = north west, align=left] 
			at (b4_left) 
				{\headlinetext{(iii) Deviation from Wick's Thm.  (Sec.~\ref{subsubsec:wick})}};

		\node at ($(b4_north)!0.45!(b4_south) + (1.3,0)$) {$ %
			|W(\gamma_S)|\leq \sqrt{2\ln2\, {\color{moccu_in_maineq} \nGoccu^{[1]}}}$};

		\node at ($(b4_north)!0.8!(b4_south) + (1.8,0)$) {$
		|W(\gamma_S)| \leq \sqrt{(|S|\!-\!1)!!}\sqrt{2\,{\color{moccu_in_maineq} \nGoccu^{[2]}}}$};

		\begin{scope}[shift={(b4_west)}, rotate=22]
			\colorlet{wicknonemphcolor}{black!50}
		
			\draw[line width=1.5pt,wicknonemphcolor] (0.7,-1.1) 
				node[below right, yshift=0.15cm, xshift=-0.1cm] {\footnotesize $\mathcal{G}(\ketbra{\psi})$}
				 -- (0.7,-0.9) coordinate (wick_g);
			\draw[line width=1.5pt,wicknonemphcolor] (1.8,-1.1) 
				node[below right, yshift=0.15cm, xshift=-0.1cm] {\footnotesize $\ketbra{\psi}$}
				-- (1.8,-0.9) coordinate (wick_o);
			
			\draw[wicknonemphcolor, line width=0.6pt,
						decorate,decoration={brace,amplitude=4pt,raise=0.13cm}] (wick_g) 
				-- node[midway, anchor = south east, align=center, yshift=0.2cm] 
					{\footnotesize $W(\gamma_S)$} (wick_o);
					
			\draw[-latex, line width=1.5pt] (0.0,-1) -- (2.7,-1) 
				node[above, xshift=-0.2cm, yshift=-0.05cm]{$\Tr (\gamma_S \;\cdot\;)$};
		\end{scope}

		\node[xshift=0.0cm,yshift =-0.07cm, anchor = north west, align=left] 
			at (b3_left) 
				{\headlinetext{(iv) Scaling in gapped systems (Sec.~\ref{sec:TI})}}; 
		
		\coordinate (b3_imshift) at (0.7,-0.35);
				
		\coordinate (b3_xstart) at ($(b3_west)!0.2!(b3_east) - (0,0.6) + (b3_imshift)$);
		\coordinate (b3_xend) at ($(b3_west)!0.8!(b3_east) - (0,0.6) + (b3_imshift)$);
		\coordinate (b3_critl) at ($(b3_west)!0.5!(b3_east) - (0,0.6) + (b3_imshift)$);
		\coordinate (b3_crith) at ($(b3_west)!0.5!(b3_east) + (0,0.7) + (b3_imshift)$);
		
		\definecolor{moccu_lighter_line}{HTML}{ed6d6e}
		\draw[line width=1.5pt, moccu_in_maineq!50]%
			($(b3_xstart)!0.1!(b3_xend)$) to[out=8,in=205] 
			($(b3_xstart)!0.34!(b3_xend) + (0,0.265)$) to[out=25,in=180] 
			($(b3_crith) - (0,0.75)$) to[out=0,in=155]
			($(b3_xstart)!0.66!(b3_xend) + (0,0.265)$) to[out=-25,in=172] 
			($(b3_xstart)!0.9!(b3_xend)$);
			
		\draw[line width=1.5pt, moccu_in_maineq] 
			($(b3_xstart)!0.1!(b3_xend)$) to[out=10,in=-110] 
			(b3_crith) to[out=-70,in=170] coordinate[midway] (b3_im_midpoint2)
			($(b3_xstart)!0.9!(b3_xend)$);
			
		\draw[latex-latex, line width=1.5pt] 
			($(b3_xstart) + (-1.5,1.4)$) node[anchor=west, xshift=0.01cm, yshift=-0.2cm] 
					{\small $\nGoccu^{[\alpha]}/N$}--  
			($(b3_xstart) + (-1.5,0)$)-- (b3_xend) node[right] {\small $\gamma$};
		\draw[line width=1.5pt] ($(b3_critl) + (0,0.1)$) node[above, yshift=-0.1cm] {\small $\gamma_\text{crit.}$} -- 
			($(b3_critl) - (0,0.1)$) ;
		
		\colorlet{zoomcirclecolor}{black!40}
		\coordinate (b3_im_midpoint) at ($(b3_im_midpoint2) - (0,0.05)$);
		\coordinate (b3_center_inset) at ($(b3_im_midpoint) + (1.0,0.7)$);
		
		\draw[latex-, line width=1pt, zoomcirclecolor] 
			($(b3_im_midpoint)!0.2!(b3_center_inset)$) -- ($(b3_im_midpoint)!0.57!(b3_center_inset)$) ;
		\draw[line width=1pt, zoomcirclecolor](b3_im_midpoint) circle(4pt);
		
		\begin{scope}
			\path[clip, preaction={draw, line width=1pt, zoomcirclecolor}] (b3_center_inset) circle(0.4cm);
			\draw[line width=1.5pt, moccu_in_maineq!50]%
				($(b3_center_inset) + (-0.4,0.05)$) --coordinate[midway] (b3cil2m) +(0.95,-0.4);
				
			\draw[line width=1.5pt, moccu_in_maineq] 
				($(b3_center_inset) + (-0.4,0.35)$) --coordinate[midway] (b3cil1m) +(0.95,-0.4);
			\draw[line width=1pt, zoomcirclecolor] (b3_center_inset) circle(0.4cm);
			
			\draw ($(b3cil1m) - (0.05,0)$) -- +(0.1,0); 
			\draw ($(b3cil2m) - (0.05,0)$) -- +(0.1,0); 
			\draw (b3cil2m) -- coordinate[midway] (b3cilmm) (b3cil1m);
		\end{scope}
		
		\node[anchor=west] at (b3cilmm) {$\sim e^{-\delta N}$};
	
		\coordinate (b3legend) at ($(b3_xstart) + (-1.35,0.6)$);
		
		\draw[line width=1.5pt, moccu_in_maineq]
				($(b3legend) $) -- +(0.3,0) node[right] {\footnotesize $N\to\infty$};
				
		\draw[line width=1.5pt, moccu_in_maineq!50]%
				($(b3legend) -(0,0.3)$) -- +(0.3,0) node[right] {\footnotesize \textcolor{moccu_in_maineq!80}{finite $N$}};

	\end{tikzpicture}

    \caption{
    Overview of the two families of computable quantifiers of fermionic non-Gaussianity introduced in this work, both built from the Williamson normal form of the covariance matrix $\Gamma(\ket{\psi}) = Q \bigoplus_i \!\left(\begin{smallmatrix} 0 & -r_i \\ r_i & 0 \end{smallmatrix}\right)\! Q^T$. The \emph{occupation number entropies} $\nGoccu^{[\alpha]}$ (left, blue) depend only on the Williamson eigenvalues $\{r_i\}$ and are efficiently computable from the covariance matrix alone. The \emph{natural-orbital participation entropies} $\nGNO^{[\alpha]}$ (right, red) quantify how broadly the state spreads over the natural-orbital basis, optimized over degenerate bases when some $r_j$ coincide (typically $r_j$ are non-degenerate, in which case the natural-orbital basis is unique). The check marks indicate the properties that we prove for the corresponding quantity; notably, $\nGoccu^{[1]}$ and $\nGoccu^{[2]}$ are strong monotones under general Gaussian protocols, whereas monotonicity remains open for both $\nGoccu^{[\alpha\notin\{1,2\}]}$ and $\nGNO^{[\alpha]}$. However, we do establish it for $\nGNO^{[\alpha]}$ under the restricted stabilizer Gaussian protocols of the stabilizer non-Gaussianity resource theory. The rest of the boxes summarize the main operational implications: (i) both quantities lower-bound the number of SWAP gates required to prepare $\ket{\psi}$ from any Gaussian reference state via Gaussian protocols; (ii) $\nGNO^{[\alpha<1]}$ controls the cost of classical simulation by an expansion of the state in a particular orthonormal basis spanned by Gaussian states, with $\nGNO^{[0]}$ and $\nGNO^{[1/2]}$ bounding respectively the Gaussian rank $\chi_\mathcal{G}$ and the Gaussian extent $\xi_\mathcal{G}$; (iii) $\nGoccu^{[1]}$ and $\nGoccu^{[2]}$ can be used to bound $W(\gamma_S)=\Tr[(\rho-\G(\rho))\gamma_S]$, the deviation of every Majorana correlator $\Tr[\gamma_S\rho]$ from its corresponding expectation value $\Tr[\G(\rho)\gamma_S]$ with respect to the Gaussian state with the same covariance matrix, i.e., the expectation value predicted by Wick's theorem; and (iv) both quantities serve as diagnostics of quantum phase transitions, as illustrated by the schematic behavior of the density $\nGoccu^{[\alpha]}/N$ near a critical point $\gamma_{\text{crit.}}$.
    }
    \label{fig: overview}
\end{figure*}

\section{Summary of main results}
\label{sec: summary}

In this section, we present a summary of the main results of our work, which we also illustrate in Fig.~\ref{fig: overview}. This work establishes a rigorous and computationally tractable framework for quantifying the fermionic non-Gaussianity of pure states. To this end, we start by defining a convex resource theory of non-Gaussianity in Sec.~\ref{sec:convex_rt}, whose free states are convex mixtures of FGSs, which are efficiently classically simulable as long as one can efficiently sample their weights~\cite{melo2013}, and whose free operations are Gaussian protocols with feedforward. This is the natural setting in the context of universal quantum computation, since, without measurements, injecting magic states alone does not suffice to achieve universality~\cite{hebenstreit2019all}. %
Within this framework, a valid non-Gaussianity monotone must be non-increasing under Gaussian protocols. We then focus on the pure-state setting, where the allowed Gaussian protocols are those that map pure states to pure states (or, more generally, to ensembles of pure states~\footnote{%
    By an \emph{ensemble of pure states} we mean a collection $\{(p_k,\ket{\psi_k})\}$ of pure states $\ket{\psi_k}$ occurring with probabilities $p_k$; a Gaussian protocol with intermediate measurements produces such an ensemble, returning the pure state $\ket{\phi_k}$ conditioned on the measurement outcome $k$.
}).

Building on this foundation, we introduce two complementary families of computationally tractable quantifiers of fermionic non-Gaussianity in pure states in~Sec.~\ref{sec: nG_measures}, whose definitions and main properties we summarize below as well as in Fig.~\ref{fig: overview}. They are both constructed using information contained in the covariance matrix $\Gamma$ with elements
\begin{equation}
    \Gamma(\rho)\big|_{j,k}
    = -\frac{i}{2} \Tr([\gamma_j,\gamma_k] \rho).
\end{equation}
Note that the covariance matrix is well-defined for any state, Gaussian or not. Its eigenvalues come in pairs $\pm i r_j$ with $r_j\in[0,1]$; for a pure FGS they saturate at $r_j=1$ and, by Wick's theorem~\cite{Wi50, BaLi94}, $\Gamma$ fixes the state completely. For a pure non-Gaussian state, instead, some $r_j<1$ is a direct signature of non-Gaussianity. For a general state $\rho$, $\Gamma$ likewise defines a unique (generally mixed) FGS $\mathcal{G}(\rho)$ with the same covariance matrix.
Whereas the occupation number entropies depend only on $\Gamma$, and thus only depend on the Gaussian part of the state, the natural-orbital participation entropies depend on the full state, and thus also depend on its non-Gaussian correlations that are not captured by the covariance matrix.

\textit{Occupation number entropies} (Sec.~\ref{subsec:nG_occu})---The first family, $\nGoccu^{[\alpha]}$, is defined for an $N$-qubit state solely from the $N$ Williamson eigenvalues $\{r_j\}$ of the covariance matrix $\Gamma$, or equivalently from the occupation numbers $\lambda_j=(1+r_j)/2$ that lie in the interval $\big[\frac{1}{2},\,1\big]$. It is computed as
\begin{equation}
    \nGoccu^{[\alpha]}(\ket{\psi}) = \sum_{j=1}^N T_{\alpha}\!\left(\left\lbrace \lambda_j, 1-\lambda_j\right\rbrace\right),
\end{equation}
where $T_\alpha$ is the normalized $\alpha$-Tsallis entropy $T_\alpha(\left\lbrace p_i\right\rbrace) = \left(1 - \sum_i p_i^\alpha\right) / (1-{2^{1-\alpha}})$, and the normalization is chosen such that $0\leq\nGoccu^{[\alpha]}\leq N$. In words, this family computes the sum of Tsallis entropies of each occupation number pair, %
hence we call it \emph{occupation number entropy}. The basic idea of this family is that, as noted above, for a pure FGS all occupation numbers equal one, so that $\nGoccu^{[\alpha]}=0$, whereas any $\lambda_j<1$ yields a nonzero value. At the opposite extreme, a state with vanishing covariance matrix has $\lambda_j=\tfrac12$ for every mode, and hence the maximal value $\nGoccu^{[\alpha]}=N$.

It can be efficiently computed with widely used numerical methods, such as tensor-network methods, and includes as special cases the previously introduced relative entropy of non-Gaussianity~\cite{lyu2024fermionicgaussiantesting, lumia2024measurement} and the recently introduced fermionic antiflatness~\cite{sierant2025fermionicmagicresourcesquantum}.

In the von Neumann limit $\alpha\to1$, this reduces to
\begin{equation}
    \nGoccu^{[1]}(\ket{\psi}) = S\!\left(\mathcal{G}(\rho)\right),
\end{equation}
where $\mathcal{G}(\rho)$ is the unique FGS sharing the same covariance matrix as $\rho=\ketbra{\psi}{\psi}$, and $S$ is the von Neumann entropy. It can further be shown that $\nGoccu^{[1]}$ is the minimum quantum relative entropy to the set of FGSs---the relative entropy of fermionic non-Gaussianity~\cite{lyu2024fermionicgaussiantesting}. While $\nGoccu^{[\alpha]}$ is computable from the two-point Majorana correlators for any
$\alpha$, the case $\alpha=2$ stands out in admitting a closed form that avoids
diagonalizing the covariance matrix,
\begin{equation}
    \nGoccu^{[2]}(\ket{\psi}) = N - \sum_{j>k}\bra{\psi} i\gamma_j\gamma_k\ket{\psi}^2,
\end{equation}
making it especially convenient for numerical simulation and experimental measurement.

The central result here is that $\nGoccu^{[1]}$ and $\nGoccu^{[2]}$ are \emph{strong monotones} under pure-state Gaussian protocols, making them novel \emph{computable convex resource monotones for fermionic non-Gaussianity}. For $\alpha\neq1,2$, confirming or refuting monotonicity remains an open problem. Monotonicity yields a lower bound on the non-Gaussian gate count: we show that any preparation of $\ket{\psi}$ from a Gaussian reference state using Gaussian protocols and $n_s$ SWAP gates must satisfy
\begin{equation}
    \frac{1}{4}\nGoccu^{[\alpha]}(\ket{\psi})\leq n_s,
\end{equation}
for $\alpha\in\{1,2\}$.

The $\alpha=2$ case provides additional practical insights. It upper- and lower-bounds the Gaussian fidelity $F_\mathcal{G}$---the maximal overlap of a pure state with any FGS---as
\begin{equation}
    1-\frac{\nGoccu^{[2]}}{2}\leq F_\mathcal{G}\leq 1-\frac{1}{4}\!\left(1-\sqrt{1-\frac{\nGoccu^{[2]}}{N}}\,\right)^{\!2},
\end{equation}
proven in Appendix~\ref{sec:bounds_Fg}. Using these bounds, one can show that efficient Gaussian state testing%
---i.e., the task of deciding, from copies of an unknown pure state, whether it is close to or far from the set of fermionic Gaussian states---is possible with $\mathcal{O}(N^4)$ single-copy measurements, improving on the previous $\mathcal{O}(N^5\log N)$ state of the art~\cite{bittel2025optimal}. 

We further show that the occupation number entropies upper-bound the deviation from Wick's theorem: they bound the deviation $W(\gamma_S)=\Tr[(\rho-\G(\rho))\gamma_S]$ of every $2m$-point Majorana correlator $\Tr[\rho\gamma_S]$, with $S\subset[2N]$, from its Gaussian expectation value $\Tr[\G(\rho)\gamma_S]$---the Pfaffian of the covariance matrix as computed in Wick's theorem---as
\begin{equation}
\begin{aligned}
   |W(\gamma_S)| &\leq \sqrt{2\ln2\;\nGoccu^{[1]}(\ket{\psi})} \\
   \text{and}\qquad
   |W(\gamma_S)| &\leq \sqrt{(2m-1)!!}\;\sqrt{2\,\nGoccu^{[2]}(\ket{\psi})},
\end{aligned}
\end{equation}
so that $\nGoccu^{[1]}$ controls the deviation uniformly in the Majorana order, while the bound from $\nGoccu^{[2]}$ weakens rapidly with the Majorana order; for the lowest (four-point) order the latter tightens to $|W(\gamma_S)|\leq\sqrt{2\,\nGoccu^{[2]}(\ket{\psi})}$. A small occupation number entropy thus guarantees that the state remains accurately Gaussian, in the sense that Wick's theorem still holds approximately. 

The occupation number entropies also have a direct connection to (global) entanglement: they equal the residual entanglement (as quantified by the sum of single-mode Tsallis entanglement entropies) that remains after optimally disentangling the state by a Gaussian unitary. More precisely, we show that
\begin{equation}
    \nGoccu^{[\alpha]}(\ket{\psi}) = \min_{U\in\mathcal{G}_\text{U}}\ME^{[\alpha]}(U^\dagger\ket{\psi}),
\end{equation}
where $\ME^{[\alpha]}(\ket{\psi})=\sum_j T_\alpha(\rho_j)$ is a family of global entanglement monotones. For $\alpha=1$, %
it coincides with the total mutual information~\cite{modi2010unified}, while for $\alpha=2$, $\ME^{[2]}$ is proportional, up to the conventional normalization, to the Meyer--Wallach global entanglement measure~\cite{MeyerWallach2002}.

While $\nGoccu^{[\alpha]}$ quantifies the cost of preparing a state using non-Gaussian gates, it does not directly capture the cost of \emph{classically simulating} the state. This distinction is illustrated by the GHZ-like superposition $c_0 \ket{0}^{\otimes N} + c_1 \ket{1}^{\otimes N}$ (for an even number of qubits $N$), which is a superposition of only two FGSs and is therefore efficiently simulable via Gaussian expansion, yet still exhibits an extensive occupation number entropy $\nGoccu^{[\alpha]} \propto N$.

The quantity that does control the simulation cost is the number of FGSs needed to express the state, namely the Gaussian rank or Gaussian extent~\cite{dias2024classical,ReardonSmith2024improved,cudby2023gaussian}. However, these quantities require an intractable optimization over all superpositions into pure FGSs. This motivates the search for a computable proxy based on a fixed, physically motivated Gaussian basis, leading to the second family of non-Gaussianity quantifiers introduced now.

\textit{Natural-orbital participation entropies.} (Sec.~\ref{subsec:NOPE})---A natural choice of basis to represent a state in this context is the \emph{natural-orbital basis}~\cite{lowdin1955quantum}, defined by the covariance matrix via its Williamson normal form. This basis is efficiently accessible from $\Gamma$ and is widely used in quantum chemistry, where it often provides compact representations of correlated many-body states~\cite{RevModPhys.92.015003}. Importantly, in this basis FGSs become product states in the computational basis, so any deviation from product structure directly signals non-Gaussianity. For the example of the GHZ-like superposition above, this basis correctly resolves its underlying low-rank structure, as shown numerically in Sec.~\ref{subsec:superposition_FGS}.

We therefore define the \emph{natural-orbital participation entropy}
\begin{equation}
    \nGNO^{[\alpha]}(\ket{\psi}) = \min_{U_Q}\, S^{\mathrm{part}}_\alpha(U_Q^\dagger\ket{\psi}),
\end{equation}
where $S^{\mathrm{part}}_\alpha(\ket{\psi}) = \frac{1}{1-\alpha}\log_2 \sum_s |\langle s|\psi\rangle|^{2\alpha}$ and $\ket{s}$ denotes computational basis states. The minimization is over FGUs $U_Q$ that bring the covariance matrix into Williamson form, and is required only in the presence of degeneracies in the Williamson eigenvalues~\footnote{Note that generally the natural-orbital basis is not the optimal fermionic Gaussian basis that minimizes the participation entropy, as verified using an explicit example in Sec.~\ref{subsec:superposition_FGS}. }.

Operationally, $\nGNO^{[\alpha]}$ quantifies how delocalized the state is in the natural-orbital basis. It  upper-bounds the cost of classical simulation, as it bounds the number of (orthonormal)
FGSs needed to represent the state. As shown in Sec.~\ref{subsec: classical_sim}, at most logarithmic growth of $\nGNO^{[\alpha<1]}$ implies that polynomially many FGSs in the natural-orbital basis suffice for accurate approximation, whereas superlogarithmic growth for $\alpha>1$ (or linear growth for $\alpha=1$) might imply the need for exponentially many FGSs. Moreover, $\nGNO^{[\alpha]}$ upper-bounds the Gaussian rank and extent. In this sense, $\nGNO^{[\alpha]}$ complements $\nGoccu^{[\alpha]}$: the former quantifies representational complexity in classical simulation, while the latter quantifies non-Gaussian gate cost in state preparation.

Finally, monotonicity of $\nGNO^{[\alpha]}$ under general Gaussian protocols remains an open question. However, in Sec.~\ref{sec:pure_stab} we establish monotonicity within the restricted setting of stabilizer Gaussian protocols, i.e., Gaussian protocols that additionally preserve the stabilizer structure (thus defining a different resource theory), where both the occupation number entropies and the natural-orbital participation entropies are strong monotones for all $\alpha$.

\textit{Applications.} (Secs.~\ref{sec: special} and~\ref{sec: example})---Beyond establishing the resource-theoretic properties of these quantities, we apply them to various classes of states of broad interest. For stabilizer and translation-invariant states (Sec.~\ref{sec: special}) these quantities take a simpler form such that further analytical insights can be gained. We also numerically investigate these classes in representative physical settings (Sec.~\ref{sec: example}), developing efficient algorithms for evaluating both quantities across a range of state representations, including exact state vectors, stabilizer states~\cite{aaronson2004improved,gottesman1998theory}, and matrix product states (MPSs)~\cite{Fannes1992,Cirac2021}.

First, we consider (pure) fermionic states that are also stabilizer states. The covariance matrix of such states has a special structure, which causes the occupation number entropies to be independent of the R\'enyi index $\alpha$.  
Exploiting their efficient simulability, we track non-Gaussianity generation in SWAP-doped random matchgate circuits, finding rapid growth under global evolution and slower, diffusive growth under brick-wall circuits.

Second, for translation-invariant states, the exponential decay of correlators in gapped phases allows us to reliably extract the density $\nGoccu^{[\alpha]}/N$ in an infinite system from finite-range correlators. In a finite system it converges exponentially fast in system size to its value in the thermodynamic limit, under the standard exponential finite-size convergence assumption for local two-point functions.
As a concrete physical example, we demonstrate numerically that both families in the bond-modulated XXZ chain---dual under the Jordan--Wigner transformation~\cite{JoWi28} to spinless interacting fermions---exhibit a kink at the critical point; moreover, we find numerically that a constant subleading term in $\nGoccu^{[\alpha]}$ is absent in the gapped phases (consistent with the exponential convergence) but present at the critical point, providing a further diagnostic of the transition.

 \section{Preliminaries}
\label{sec: preliminaries}
This section introduces the background material underlying this work. Sec.~\ref{app:review_fgs} reviews fermionic Gaussian states, their covariance matrix description, and the Williamson eigenvalues that underlie the computation of our quantities. Sec.~\ref{sec:convex_rt} establishes the convex resource theory framework, where the free operations are Gaussian protocols with feedforward and the free states are convex mixtures of FGSs. We then focus on pure-state Gaussian protocols, i.e., those protocols whose input and outcome are pure states. Sec.~\ref{sec:ng_monotone} formalizes the properties expected from a valid non-Gaussianity monotone, and Sec.~\ref{subsec:survey_nG} surveys previously introduced quantifiers of fermionic non-Gaussianity, identifying the gap between computability and rigorous monotonicity that this work addresses.

\subsection{Review of fermionic Gaussian states}
\label{app:review_fgs}

We consider a set of $N$ qubits associated with the Hilbert space $\mathcal{H}_N = \bigotimes_{j=1}^N \mathcal{H}_j$, where $\mathcal{H}_j \simeq \mathbb{C}^2$. A system of $N$ qubits is equivalent to a system of $N$ fermionic modes by the Jordan--Wigner transformation~\cite{JoWi28}, where the Majorana operators are defined as
\begin{align}
    \gamma_{2j-1} &= \left( \prod_{k=1}^{j-1} Z_k \right) X_j  \\
    \gamma_{2j}   &= \left( \prod_{k=1}^{j-1} Z_k \right) Y_j,
\end{align}
for $j \in [N] \coloneqq \{1, 2, \dots, N\}$. Majorana operators are Hermitian, traceless, and satisfy the anticommutation relation $\{\gamma_i,\gamma_j\} =2\delta_{i,j} \Id$. Throughout this work, we formulate all operations in the qubit language when suited.

Given an ordered subset $A \subset [2N]$, we define the Majorana monomials $\gamma_A$ as the ordered product of the Majorana operators:
\begin{equation}
    \gamma_A =  \prod_{j \in A} \gamma_j,
\end{equation}
where the product is taken in increasing order, and  $\gamma_\varnothing\coloneqq\Id$. The $4^N$ operators $\gamma_A$ form an orthonormal basis in the space of linear operators, which we will refer to as the Majorana basis. Specifically, the expansion of a linear operator $O$ in the Majorana basis is 
\begin{equation}
    O = \sum_A O_A \gamma_A \quad\text{with}\quad O_A = 2^{-N}\Tr(O \gamma_A^\dagger).
\end{equation}
An operator is even if all coefficients $O_A=0$ for $|A|$ odd. A pure state is even if it is in the fixed parity subspace $P=\pm1$, where $P=\prod_i Z_i$. In this work, all states we consider are even states, which are the physical states in fermionic systems.

A fermionic Gaussian unitary (FGU) is a unitary operator whose action on Majorana operators is described by an orthogonal transformation $Q \in O(2N)$ acting as
\begin{equation} \label{eq:action_fgu}
    U \gamma_j U^\dagger = \sum_{k=1}^{2N} Q_{jk} \gamma_k,
\end{equation}
for all $1\leq j \leq 2N$. Such transformations are generated by quadratic fermionic Hamiltonians, which realize the $SO(2N)$ subgroup, together with parity-changing operators such as individual Majorana operators that extend the representation to the full orthogonal group $O(2N)$. The corresponding unitary $U_Q$ is uniquely determined by $Q$ up to a global phase.
We denote the set of FGUs as $\mathcal{G}_\text{U}$. Under the Jordan--Wigner transformation, FGUs are mapped to matchgate unitaries~\cite{Va01} in the qubit representation.

Fermionic Gaussian states (FGS) $\rho$ are even states which can be expressed as
\begin{equation}
    \rho = U \, \bigotimes_{j=1}^N\left( \frac{1+r_j Z_j}{2} \right) \, U^\dagger
\end{equation}
for an FGU $U$ and real numbers $|r_j| \leq 1$. A pure FGS has $r_j = \pm 1$ and can always be written as $\ket{\psi}=U \ket{0}^{\otimes N}$ with an FGU $U$. Any mixed FGS arises as a reduced state of a pure FGS. Furthermore, any FGS is uniquely determined by its covariance matrix (defined below), which encodes its complete information through Wick's theorem~\cite{Wi50, BaLi94}. We denote the set of all FGSs as $\mathcal{G}$ and the set of pure FGSs as $\mathcal{G}_\text{p}$.

Given a quantum state $\rho$, the covariance matrix $\Gamma(\rho)$ is defined as a $2N \times 2N$ matrix with elements
\begin{equation}
    \Gamma(\rho)\big|_{j,k}
    = -\frac{i}{2} \Tr([\gamma_j,\gamma_k] \rho),
\end{equation}
where $j,k \in [2N]$ and $[A, B] = AB - BA$ denotes the commutator of $A$ and $B$. Covariance matrices are real and antisymmetric, with (purely imaginary) eigenvalues $\pm i r_j$ which satisfy $|r_j| \leq 1$. Moreover, they can be brought into a normal form
\begin{equation}
    \label{eq:normal_form}
    \Gamma(\rho) = Q \, \bigoplus_{j=1}^N  \left(\!\begin{array}{cc}
        0 & r_j \\
        -r_j & 0 \\
    \end{array}\right)
    \, Q^T,
\end{equation}
where $Q$ is an orthogonal matrix in $O(2N)$ and $r_j \geq 0$ are called the Williamson eigenvalues of the covariance matrix~\cite{BoRe04}. The covariance matrix %
of $\rho$ evolved under the FGU $U_Q$ is given by $\Gamma(U_Q \, \rho \, U_Q^\dagger) = Q \, \Gamma(\rho) \, Q^T$.

At the heart of the efficient classical simulation of FGSs lies Wick's theorem~\cite{Wi50, BaLi94}, which expresses the expectation value of any Majorana string in an FGS through the covariance matrix: for an even number $2m$ of distinct indices,
\begin{equation} \label{eq:wick_pfaffian}
   \langle \gamma_{k_1}\gamma_{k_2}\cdots\gamma_{k_{2m}}\rangle = i^{m}\,\mathrm{Pf}\!\big(\Gamma|_{\{k_1,\dots,k_{2m}\}}\big),
\end{equation}
where $\mathrm{Pf}$ denotes the Pfaffian and $\Gamma|_{\{k_1,\dots,k_{2m}\}}$ is the principal submatrix of $\Gamma$ on the rows and columns $k_1,\dots,k_{2m}$ (correlators of an odd number of Majoranas vanish). Since the Pfaffian can be evaluated in $\mathcal{O}(N^3)$ time, all such correlators are efficiently obtained from $\Gamma$ alone.

A pure state $\ket{\psi}$ is known to be an FGS if and only if it satisfies~\cite{bravyi2005lagrangian, dias2024classical, melo2013}
\begin{equation} \label{eq:Lambda_condition}
    \Lambda \ket{\psi} \ket{\psi} = 0,
\end{equation}
where the operator $\Lambda$ is given by 
\begin{equation} \label{eq:Lambda_def}
    \Lambda = \sum_{j=1}^{2N} \gamma_j \otimes \gamma_j.
\end{equation}
More generally, a state $\rho$ is a mixed FGS if and only if it satisfies~\cite{bravyi2005lagrangian}
\begin{equation} \label{eq:Lambda_condition_mixed}
    [\Lambda, \rho \otimes \rho] = 0.
\end{equation}
A useful property of the operator $\Lambda$ is that it is invariant under $U\otimes U$ for any FGU $U$~\cite{bravyi2005lagrangian},
\begin{equation}
    (U \otimes U) \, \Lambda \, (U^\dagger \otimes U^\dagger) = \Lambda. 
\end{equation}
In other words, $\Lambda$ belongs to the second-order commutant of FGUs. More generally, $\Lambda$ generates the full commutant algebra of FGUs~\cite{sierant2026theorymatchgatecommutant, BrDi26, LasMou26}.

\subsection{Convex resource theories} \label{sec:convex_rt}
Quantum resource theories provide a systematic framework for quantifying the usefulness of quantum states for information-processing tasks~\cite{chitambar2019quantum, gour2024resources}. 
A resource theory is specified by two ingredients: a set of \emph{free states} and a set of \emph{free operations}. 
Free states are those that can be prepared without consuming the resource, while free operations are physical transformations that cannot generate the resource from free states. Various resource theories have been proposed in the literature, including for entanglement~\cite{vidal2000entanglement}, coherence~\cite{baumgratz2014quantifying}, and nonstabilizerness~\cite{veitch2014resource}.

In this work, we focus on the resource theory of (fermionic) non-Gaussianity. Previous works have discussed such a resource theory by considering the set of FGSs as the free states, and Gaussian operations---operations that map Gaussian states to Gaussian states---as the free operations~\cite{marian2013relative,lumia2024measurement,lyu2024fermionicgaussiantesting,gottlieb2007properties,gottlieb2014correlations,turner2017optimal}. The choice is motivated by the fact that Gaussian states can be efficiently simulated classically and arise as ground and thermal states of quadratic fermionic Hamiltonians---a choice that, however, treats as costly certain non-Gaussian operations that are in fact easy to implement~\cite{hebenstreit2019all}.

The most important such operations are \emph{adaptive measurements}---Gaussian operations conditioned on intermediate measurement outcomes---which require nothing beyond Gaussian operations and classical processing and are thus natively available on fermionic platforms. Taking them to be free yields a resource theory whose free set reflects the actually accessible operations while remaining classically tractable.

Accordingly, we adopt as free operations the class of Gaussian operations combined with feedforward, which we refer to as \emph{Gaussian protocols}, operationally defined as a finite sequence of the following elementary operations:
\begin{enumerate}
	\item evolution under any FGU;\label{op:FGU_evolution}
	\item tensor product with any FGS;\label{op:FGS_tensor_product}
	\item computational basis measurements;\label{op:measurements}
	\item discarding of qubits; and\label{op:discarding_qubits}
	\item the above operations conditioned on classical randomness or the outcomes of measurements.\label{op:conditioned_operations}
\end{enumerate}
Operations~\ref{op:FGU_evolution}--\ref{op:discarding_qubits} are standard in the quantification of fermionic non-Gaussianity~\cite{lumia2024measurement,lyu2024fermionicgaussiantesting,marian2013relative,gottlieb2007properties,gottlieb2014correlations}, whereas operation~\ref{op:conditioned_operations} is less standard and deserves justification. First, a basic requirement of any quantum resource theory is that classical randomness should not be regarded as a resource; conditioning subsequent operations on classical information---in particular, on the outcomes of intermediate measurements---should therefore be free, which is precisely what requirement~\ref{op:conditioned_operations} encodes. Second, the resulting Gaussian protocols are the fermionic analogue of the stabilizer protocols underlying the resource theory of nonstabilizerness~\cite{veitch2014resource} and of the bosonic Gaussian protocols of the convex resource theory of non-Gaussianity~\cite{takagi2018convex,albarelli2018}---both considerably more developed than the fermionic case---which similarly include operations conditioned on measurement outcomes.

In contrast to Gaussian operations only, Gaussian protocols can produce probabilistic mixtures of Gaussian states, which are generally non-Gaussian since the set of Gaussian states is not convex~\cite{SpSc18}. This gives rise to the convex resource theory framework, where the free states are the convex hull of FGSs, also called convex-Gaussian states, defined as
\begin{equation} \label{eq:convex_fgs}
    \mathrm{conv}(\mathcal{G})
    = \left\{ \int_\mu \!\d\mu\, p_\mu \sigma_\mu \,\middle|\,
    \sigma_\mu\!\in\!\mathcal{G},\,
    p_\mu\ge0,
    \int_\mu \!\d\mu\, p_\mu=1 \right\}.
\end{equation}
Note that this is equivalent to the convex hull of all pure FGSs, since any state $\sigma_\mu\in\mathcal{G}$ is a convex combination of pure FGSs. The set $\mathrm{conv}(\mathcal{G})$ is exactly the largest set of states that Gaussian protocols can generate from Gaussian inputs.

Beyond physical implementability, the resulting free states are \emph{weakly simulable}: convex-Gaussian states can be sampled efficiently on a classical computer~\cite{melo2013,dias2024classical,ReardonSmith2024improved,cudby2023gaussian}, provided one can efficiently sample the distribution $p_\mu$, since the (conditional) state is a Gaussian state which can be simulated through its covariance matrix~\cite{surace2022fgs}. Consequently, the free operations preserve the efficiently simulable class and cannot, on their own, yield a computational advantage, so states outside $\mathrm{conv}(\mathcal{G})$ are necessary for universal quantum computation. Furthermore, similarly to the case of Clifford circuits~\cite{bravyi2005universal}, Gaussian protocols can be promoted to universality by means of magic-state injection~\cite{hebenstreit2019all}. This choice of free states and operations is thus what makes the resource theory operationally relevant to universal quantum computation.%

In principle, one could adopt other sets of free operations, most notably the \emph{maximal resource-nongenerating operations}~\cite{brandao2015reversible,gour2017quantum}---in our setting, the complete set of operations that map convex-Gaussian states to convex-Gaussian states---which has more convenient mathematical properties but is not generally physically implementable. Our approach is instead operational, taking as free those operations that are actually accessible; we do not consider these maximal resource-nongenerating operations. We expect this maximal set to be strictly larger than the Gaussian protocols, as is the case in entanglement (separable maps vs.\ LOCC~\cite{bennet1999quantum}) and nonstabilizerness (completely stabilizer-preserving operations vs.\ stabilizer protocols~\cite{heimendahl2022axiomatic}).

In this work, we focus primarily on pure states. Therefore, the protocols that we consider are pure-state Gaussian protocols, for which 
\begin{equation} \label{eq:pure_state_protocols}
    \mathcal{E}(\ketbra{\psi}{\psi})=\ketbra{\phi}{\phi},
\end{equation}
where $\ket{\psi}, \ket{\phi}$ are pure states, and $\mathcal{E}$ denotes a Gaussian protocol. %
For pure states, the set of free states reduces from the convex hull of FGSs to the set of pure FGSs $\mathcal{G}_\mathrm{p}$.
Thus, within the pure-state setting the free states coincide with the free states of the non-convex resource theory. However, the set of free operations remains the larger class of Gaussian protocols as described above.

Within this convex resource theory framework, any valid non-Gaussianity monotone should be non-increasing under Gaussian protocols, which will be discussed in the following subsection.

\subsection{Non-Gaussianity monotones}
\label{sec:ng_monotone}

A central task in any resource theory is to quantify the amount of the resource contained in a given quantum state. This is achieved by introducing a \emph{resource monotone}, namely a function that assigns a nonnegative real number to each quantum state and does not increase under free operations.

In the present setting, a non-Gaussianity monotone is a function $\mathcal{M}$ that quantifies how far a state is from being convex-Gaussian, with the defining requirement of monotonicity under Gaussian protocols,
\begin{equation}
    \mathcal{M}\big(\mathcal{E}(\rho) \big) \leq \mathcal{M}(\rho),
\end{equation}
for every such protocol $\mathcal{E}$. %
In particular, a monotone for pure state needs to fulfill
\begin{equation}
    \mathcal{M}(\ket{\phi}) \leq \mathcal{M}(\ket{\psi}),
    \label{eq:monotonicity_condition}
\end{equation}
for any pair of pure states, $\ket{\phi},\ket{\psi}$, for which there exists a Gaussian protocol $\mathcal{E}$ which maps $\ket{\psi}$ to $\ket{\phi}$ (deterministically).

Below we list five elementary properties that are useful to have for a non-Gaussianity monotone $\M$ defined over the set of pure states:
\begin{enumerate} [label={P.\arabic*}]
    \item[\refstepcounter{enumi}\label{prop:faithful}\theenumi\phantom{*}]
    Faithfulness:\\
    \makebox[0.9\linewidth]{
        $\M(\ket{\psi}) = 0$ iff $\ket{\psi} \in \G_\text{p}$.}
    \item[\refstepcounter{enumi}\label{prop:stability}\theenumi\phantom{*}]
    Invariance under FGUs:\\
    \makebox[0.9\linewidth]{
        $\M(U \ket{\psi})=\M(\ket{\psi})$ for all FGUs $U\in \G_\text{U}$.}
    \item[\refstepcounter{enumi}\label{prop:sub-add}\theenumi\phantom{*}]
    Sub-additivity:\\
    \makebox[0.9\linewidth]{
        $\M(\ket{\psi} \otimes \ket{\phi})
        \leq \M(\ket{\psi}) + \M(\ket{\phi})$,}
\end{enumerate}
or the stronger condition
\begin{enumerate} [label={P.\arabic**}]
    \setcounter{enumi}{2}
    \item Additivity:\\
    \makebox[0.9\linewidth]{
        $\M(\ket{\psi} \otimes \ket{\phi})
        = \M(\ket{\psi}) + \M(\ket{\phi})$.
        \label{prop:add}}
\end{enumerate}
\begin{enumerate} [label={P.\arabic*}]
    \setcounter{enumi}{3}
    \item[\refstepcounter{enumi}\label{prop:strong}\theenumi\phantom{*}]
    Non-increasing, on average, under computational basis measurements.
    \item[\refstepcounter{enumi}\label{prop:composition_fgs}\theenumi\phantom{*}]
    Invariance under composition with an FGS:\\
    \makebox[0.9\linewidth]{
        $\M( \ket{\psi} \otimes \ket{\phi})=\M(\ket{\psi})$ for any FGS $\ket\phi$.}
\end{enumerate}
We note that only invariance under FGUs~(\ref{prop:stability}) and invariance under composition with an FGS~(\ref{prop:composition_fgs}) are strictly necessary properties for a non-Gaussianity monotone (though not sufficient). Both are invariances under free operations that admit a free inverse, so monotonicity in both directions implies equality. The remaining properties are desirable but not required; nevertheless, the property~(\ref{prop:strong}), composed with the two invariances above implies monotonicity on average under any probabilistic Gaussian protocol that maps $\ket{\psi}$ to an ensemble of states $\{p_k,\ket{\phi}_{k}\}$,
\begin{equation}
	\mathcal{M}(\ket{\psi}) \geq \sum_{k} p_{k} \, \mathcal{M}(\ket{\phi}_{k}),
	\label{eq:strong_monotonicity}
\end{equation}
a property known as \emph{strong monotonicity}~\cite{chitambar2019quantum}, and we accordingly call such a quantity a \emph{strong monotone}. For pure states and for convex functions, this property is stronger than the monotonicity condition in Eq.~\eqref{eq:monotonicity_condition}, and therefore provides a direct route for proving monotonicity under general Gaussian protocols, as will be discussed below.
We will also loosely refer to any quantity satisfying properties~\ref{prop:faithful}--\ref{prop:sub-add} as a \emph{quantifier} of fermionic non-Gaussianity, reserving the term \emph{monotone} for those that satisfy the full monotonicity condition, Eq.~\eqref{eq:monotonicity_condition}.%

With the help of the strong monotonicity property in Eq.~\eqref{eq:strong_monotonicity}, one can show that $\mathcal{M}$ satisfies monotonicity, Eq.~\eqref{eq:monotonicity_condition}, for pure states~\cite{TaFr25}. To see this, let $\mathcal{E}$ be a Gaussian protocol, consisting of a local measurement followed by conditioned Gaussian operations that are either unitary operators or the removal/addition of an ancilla. We denote by $M_k$ the Kraus operator encoding the update of the system's state corresponding to the measurement outcome $k$, so that
\begin{equation}
    \mathcal{E}(\,\cdot\,)= \sum_k M_k \,(\,\cdot\,)\, M^\dagger_k.
\end{equation}
Let $p_k = \Tr\!\big(M_k \, \rho \, M_k^\dagger\big)$ be the probability of obtaining the measurement outcome $k$ and $\rho_k = M_k \, \rho \, M_k^\dagger  \,/ \, p_k$ be the updated state. If the quantity $\mathcal{M}$ is non-increasing under measurements and the removal/addition of an ancilla, then
\begin{equation}
    \sum_k p_k \, \mathcal{M}(\rho_k)\leq \mathcal{M}(\ketbra{\psi}{\psi}).
\end{equation}
On the other hand, since $\mathcal{E}(\ketbra{\psi}{\psi})$ is pure, we must have $\rho_k = \ketbra{\phi}{\phi}$ for all $k$ (or $\rho_k=0$), so that $\sum_k p_k \, \mathcal{M}(\rho_k) = \mathcal{M}(\ketbra{\phi}{\phi})$, which proves monotonicity under deterministic operations. The same argument applies for arbitrary sequences of measurements followed by Gaussian operations.

We note that any strong monotone defined on pure states extends to a mixed-state monotone through the standard convex-roof construction~\cite{vidal2000entanglement}. However, this construction involves an optimization over all pure-state decompositions and is generally intractable. To construct computable mixed-state quantifiers, we therefore consider instead the smaller class of Gaussian operations that excludes the conditioned operations~(\ref{op:conditioned_operations}). This provides a trade-off between computational tractability and the generality of the free states and operations.

\subsection{Survey on quantifiers of non-Gaussianity \label{subsec:survey_nG}}

Various quantities have been proposed in the literature to quantify fermionic non-Gaussianity, motivated by operational, algebraic, or information-theoretic considerations. Only a subset of these quantities, however, are known to be monotones formulated within a rigorous resource-theoretic framework. Since the present work focuses primarily on pure states, below we restrict our discussion and review of prior works to this setting unless otherwise stated.

\paragraph*{Resource-theoretic monotones.}

The monotones most directly rooted in the resource theory of fermionic Gaussian states are the \emph{Gaussian fidelity}, the \emph{Gaussian rank}, and the \emph{Gaussian extent}~\cite{dias2024classical, cudby2023gaussian}.

The Gaussian fidelity is defined as
\begin{equation}
    F_{\G}(\ket{\psi}) 
    = \max_{\ket{\phi} \in \G_\text{p}} 
    |\langle\phi|\psi\rangle|^2 .
\end{equation}
Thus, $F_{\G}(\ket{\psi})$ quantifies the maximal overlap between $\ket{\psi}$ and any pure FGS, i.e., its proximity to the free set.
It is a monotone under pure-state Gaussian protocols and is super-multiplicative under tensor products. In other words, the quantity $-\log(F_{\G})$ is sub-additive under tensor products.

The Gaussian rank is defined as
\begin{equation}
\begin{aligned}
    \label{eq:grank_def}
    \chi_{\G}(\ket{\psi})
    = \min\!\big\{ \chi\in\mathbb{N} \,\big|\,
    &\exists\,\{\ket{\phi_j}\}_{j=1}^\chi\!\subset \G_\text{p},
    \; x_j\in \mathbb{C} \\
    &\text{s.t. } 
    \ket{\psi}=\sum_{j=1}^\chi x_j \ket{\phi_j} \big\}.
\end{aligned}
\end{equation}
It counts the minimal number of (not necessarily orthogonal) pure FGSs required to write $\ket{\psi}$ as their superposition, and upper-bounds the cost of exact classical simulations. For approximate simulations, the more relevant quantity is the $\delta$-approximate Gaussian rank,
\begin{equation}
\begin{aligned}
    \chi^\delta_\G(\psi) 
    = \min\!\big\{ 
    \chi_\G(\ket{\psi'}) \,\big|\, 
    \ket{\psi'}\!\in\!\mathcal{H},
    \lVert \ket{\psi} \!-\! \ket{\psi'} \rVert \!\le\! \delta 
    \big\},
\end{aligned}
\end{equation}
which captures the minimal Gaussian rank required to approximate $\ket{\psi}$ within error $\delta$ for some given state norm $\lVert \,\cdot\, \rVert$.

Closely related to the Gaussian rank is the Gaussian extent~\cite{dias2024classical, cudby2023gaussian}, defined as
\begin{equation}
    \label{eq:gextent_def}
    \xi_\G(\ket{\psi})
    = \!\inf_{M\in \mathbb{N}} \,
    \inf_{\{\ket{\phi_j}\}\subset\G_\text{p}}
    \!\Big\{\!
    \lVert x \rVert_1^2 
    \,\Big|\,  
    \ket{\psi} = \sum_{j=1}^M x_j \ket{\phi_j}
    \!\Big\},
\end{equation}
where $\| x \|_1 = \sum_{j=1}^M |x_j|$ denotes the $\ell^1$-norm of the coefficient vector. The Gaussian extent is the squared $\ell^1$-norm of the coefficients, minimized over all expressions of $\ket{\psi}$ as a finite superposition of FGSs. It upper-bounds the $\delta$-approximate Gaussian rank, which in turn upper-bounds the classical complexity of approximately simulating near-Gaussian fermionic circuits~\cite{dias2024classical, ReardonSmith2024improved}.

Although the monotonicity of the Gaussian rank $\chi_{\G}$ and the Gaussian extent $\xi_\G$ under Gaussian protocols was not explicitly noted previously, it can be shown to hold, as is true for similarly defined quantities in any convex resource theory. Moreover, %
both $\log{\chi_{\G}}$ and $\log{\xi_{\G}}$ are sub-additive under tensor products. The Gaussian extent is multiplicative for tensor products of $4$-qubit states~\cite{reardonSmith2025extentmultiplicative}.

Despite all above quantities being monotones, with the Gaussian extent and rank providing upper bounds on the classical simulation complexity, they are not practically applicable, since no efficient algorithm is known for computing them. The goal of the present work is to address this issue, providing computable quantifiers and monotones of non-Gaussianity
within a rigorous resource-theoretic framework.

\paragraph*{Algebraic and correlation-based quantifiers.}

Alternative ways of quantifying non-Gaussianity exploit the algebraic structure of fermionic systems and their correlations.
Proposed quantities include the nonfreeness~\cite{gottlieb2014correlations, gottlieb2007properties}, violations of Wick's theorem~\cite{pachos2022quantifying}, the fermionic nullity~\cite{MeHe25}, the twisted parity~\cite{semenyakin2025classifying}, the non-Gaussian entropy~\cite{lyu2024fermionicgaussiantesting, coffman2025measuringnongaussianmagicfermions}, and the fermionic anti-flatness~\cite{sierant2025fermionicmagicresourcesquantum}.

Among these quantities, we single out the Gaussian nullity~\cite{MeHe25}, owing to its computational simplicity and its connection to the quantities we introduce. It is defined as %
\begin{equation}
    \nu_\G = N - d_\G,
\end{equation}
where $N$ is the system size and $d_\G$ is the Gaussian dimension, which simply counts the number of eigenvalues of the covariance matrix that are equal to $+i$. While it is efficiently computable and has been shown to lower-bound the complexity of state preparation and upper-bound the complexity of classical simulation (in the unitary setting)~\cite{MeHe25}, it remains unknown whether the Gaussian nullity is a valid non-Gaussianity monotone in the convex resource theory framework. This restricts the application of the Gaussian nullity to unitary operations.

\paragraph*{Entanglement-based approaches.}

A complementary perspective comes from entanglement theory. The interaction distance~\cite{turner2017optimal} quantifies, via the trace distance between entanglement spectra, how far a state is from the closest free-fermion theory~\cite{turner2017optimal, pachos2018quantifying}, and has been applied to many-body models~\cite{meichanetzidis2018free,patrick2019interaction}. Related diagnostics instead build on the entropies of the single-particle reduced density matrices~\cite{gigena2015entanglement, liao2024quantum} or on the level statistics of the entanglement spectrum~\cite{projansky2024entanglement, carrasco2024entanglement}.

\paragraph*{Diagrammatic and circuit-based approaches.}

Recently, in an effort to unify and extend Clifford and matchgate circuits, the so-called Quon language was introduced, together with a quantity termed non-matchgateness, defined by counting irremovable holes in the associated diagrams~\cite{kang20252d}. This proposal establishes a structural connection to classical simulability. In a different line of work, the connection between circuit complexity and non-Gaussianity was studied~\cite{Molar}.

In summary, existing quantifiers capture distinct facets of non-Gaussianity---geometric distance to the free set, violations of the algebraic structure, entanglement signatures, or circuit complexity. Only a few are known to be resource monotones, and among those, computability remains a major obstacle. The aim of this work is to bridge this gap by introducing quantities that are both computationally tractable and provably monotone under Gaussian protocols.

\section{Quantifying non-Gaussianity}
\label{sec: nG_measures}
In this section we present two complementary families of computable quantifiers of fermionic non-Gaussianity, both built from the Williamson normal form of the covariance matrix. The first class, the occupation number entropies~$\nGoccu^{[\alpha]}$, is defined directly from the eigenvalues of the covariance matrix, while the second class, the natural-orbital participation entropies~$\nGNO^{[\alpha]}$, evaluates participation entropies in the natural-orbital basis. The two families probe complementary aspects of non-Gaussianity: the occupation number entropies quantify the non-Gaussian resources required to \emph{prepare} a state, whereas the natural-orbital participation entropies quantify the cost of classically \emph{representing} it as a superposition of orthonormal Gaussian states. Being built from the covariance matrix, both families are accessible not only in numerical simulations but also in experiment, where the covariance matrix can be measured directly, for instance in ultracold-atom platforms~\cite{ardila2018measuring}. %
A summary of the properties~\ref{prop:faithful}--\ref{prop:composition_fgs} for each of the quantities is shown in Tab.~\ref{tab:summary}.

\begin{table}[htbp]
\centering
\setlength{\tabcolsep}{4pt}
\renewcommand{\arraystretch}{1.15}
\begin{tabular}{lcccccc}
\hline\hline
\multicolumn{7}{c}{Gaussian protocols}\\
\hline
\rule[-1mm]{0pt}{4.5mm}Quantity
& \ref{prop:faithful} 
& \ref{prop:stability} 
& \ref{prop:sub-add} 
& \ref{prop:add} 
& \ref{prop:strong} 
& \ref{prop:composition_fgs} \\
\hline
\rule[0pt]{0pt}{11pt}%
$\nGoccu^{[\alpha]}$ 
& $\checkmark$ & $\checkmark$ & $\checkmark$
& $\checkmark$ & $\checkmark^{\,(*)}$
& $\checkmark$ \\

\rule[-5pt]{0pt}{11pt}%
$\nGNO^{[\alpha]}$ 
& $\checkmark$ & $\checkmark$ & $\checkmark$
& $?$ & $?$ 
& $\checkmark$ \\

\hline\hline
\multicolumn{7}{c}{Stabilizer Gaussian protocols}\\
\hline
\rule[-1mm]{0pt}{4.5mm}Quantity
& $\Tilde{P}.1$
& $\Tilde{P}.2$
& $\Tilde{P}.3$
& $\Tilde{P}.3^*$
& $\Tilde{P}.4$
& $\Tilde{P}.5$ \\
\hline
\rule[0pt]{0pt}{11pt}%
$\nGoccu^{\text{S-FGS}}(\mathrm{STAB})$ 
& $\checkmark$ & $\checkmark$ & $\checkmark$
& $\checkmark$ & $\checkmark$
& $\checkmark$ \\

\rule[-5pt]{0pt}{11pt}%
$\nGNO^{\text{S-FGS}}(\mathrm{STAB})$ 
& $\checkmark$ & $\checkmark$ & $\checkmark$
& $?$ & $\checkmark$
& $\checkmark$ \\
\hline\hline
\end{tabular}
\caption{
Summary of the properties satisfied by each quantity, for general states and for the special class of stabilizer states.
A checkmark ($\checkmark$) indicates the corresponding property holds; a question mark ($\,?\,$) denotes an open question.
$^{(*)}$Property~\ref{prop:strong} holds for $\alpha=1$ and $\alpha=2$; the case $\alpha\notin \{1,2\}$ remains open. The lower half of the table refers to the quantities computed for stabilizer states and where the free operations are restricted to stabilizer Gaussian protocols, see Sec.~\ref{sec:pure_stab}. A tilde over a property label ($\Tilde{P}.k$) indicates that the property is taken with respect to the restricted stabilizer Gaussian protocols. All properties hold for $\nGoccu^{[1]}$ in the general setting, making it a true non-Gaussianity monotone. Similarly, all properties hold for $\nGoccu^{[\alpha]}$ and $\nGNO^{[\alpha]}$ in the stabilizer setting, making them true monotones in the corresponding resource theory. %
}
\label{tab:summary}
\end{table}

\subsection{Occupation number entropy
\label{subsec:nG_occu}}

Consider the set of $N$ Williamson eigenvalues of the covariance matrix $\{ r_j\}$ such that $r_j \in [0, 1]$ for $j\in[N]$. They define the real and positive occupation numbers as $\lambda_j = (1+r_j)/2$. Our first family quantifies fermionic non-Gaussianity by exploiting the observation that a pure FGS has occupation numbers $\lambda_j=1$ so that any $\lambda_j<1$ is a signature of non-Gaussianity, which we quantify by its entropy. %

Concretely, we define the $\alpha$-occupation number entropy as the sum of $\alpha$-Tsallis entropies evaluated on the binary distributions formed by the pairs of occupation numbers
\begin{equation}
\label{eq: Moccu}
    \nGoccu^{[\alpha]}(\ket{\psi}) = \sum_{j=1}^N T_{\alpha}\left(\left\lbrace \lambda_j, 1-\lambda_j\right\rbrace\right),
\end{equation}
where $T_\alpha$ is the (normalized) $\alpha$-Tsallis entropy
\begin{equation} \label{eq:tsallis_prob}
T_\alpha(\left\lbrace p_i\right\rbrace) = \frac{1}{1-{2^{1-\alpha}}} \left(1 - \sum_i p_i^\alpha\right),
\end{equation}
for a discrete probability distribution $p_i$ with $\sum_i p_i=1$ and $p_i \geq 0$. The normalization has been chosen such that $0 \leq \nGoccu^{[\alpha]}(\ket{\psi}) \leq N$. Indeed, $T_{\alpha} \left( \left\{ \lambda_j, 1-\lambda_j \right\} \right)$ is maximized by $\lambda_j = 1/2$, where $T_{\alpha} \left(\left\{ 1/2, 1/2 \right\}\right)=1$ for any $\alpha$.
Defining a Hermitian matrix $H(\ket{\psi}) = \frac{1}{2} \big(\Id + i\Gamma(\ket{\psi})\big)$, where $\Gamma(\ket{\psi})$ is the covariance matrix of the state $\ket{\psi}$, the occupation number entropy can also be written as
\begin{equation}
    \nGoccu^{[\alpha]}(\ket{\psi}) = \frac{1}{1-{2^{1-\alpha}}} \left(N - \Tr[H(\ket{\psi})^\alpha]\right).
\end{equation}

The $\alpha$-Tsallis entropy is a generalization of classical entropy measures: it reduces to the Shannon entropy in the limit $\alpha \to 1$, and to the linear entropy for $\alpha = 2$. One can easily show that $\nGoccu^{[\alpha]}(\ket{\psi})$ satisfies properties~\ref{prop:faithful}--\ref{prop:sub-add} and \ref{prop:composition_fgs} for any $\alpha\geq0$, as presented in Appendix~\ref{sec:occu_properties}. Below, we focus the discussion of the occupation number entropy $\nGoccu^{[\alpha]}(\ket{\psi})$ on the cases $\alpha\to1$ and $\alpha=2$.  In particular, we will prove that $\nGoccu^{[1]}(\ket{\psi})$ and $\nGoccu^{[2]}(\ket{\psi})$ satisfy \emph{strong monotonicity}. As a result, we will show that these quantities can be used to bound the number of SWAP gates required to prepare a state. In the following we will also write $\nGoccu$ instead of $\nGoccu^{[1]}$ and drop the explicit dependence on $\ket{\psi}$ when this does not generate confusion.

Notably, since $\nGoccu^{[\alpha]}$ is defined in terms of expectation values of $\mathcal{O}(N^2)$ Pauli operators, it is efficiently computable in numerical simulations with, e.g., exact state vectors, MPSs, and stabilizer states. The overall cost can be reduced by a further factor of $N$ by constructing the covariance matrix one row at a time, so that consecutive entries $i\gamma_j\gamma_k\ket{\psi}$ are obtained from one another by a constant-support operator rather than by rebuilding the full Pauli string (Appendix~\ref{sec:cov_mat_construction}).
Leveraging this structure, the construction of the covariance matrix can be performed with cost $\mathcal{O}(N^2 2^N)$ for exact state vectors, $\mathcal{O}(N^{2}\chi^{3})$ for MPSs (where $\chi$ is the bond dimension), and $\mathcal{O}(N^{3})$ for stabilizer states (see Appendix~\ref{sec:cov_mat_construction}). For MPSs, an alternative method for $\alpha=2$ involves a matrix product operator (MPO) formulation, by representing the operator $\Lambda^2$ as an MPO with bond dimension $3$. This approach evaluates $\nGoccu^{[2]}$ at a cost $\mathcal{O}(N\chi^{5})$, thereby offering a trade-off between system size and bond dimension. These algorithms are discussed in Appendix~\ref{sec:mps_algos}.

\subsubsection{The limit \texorpdfstring{$\alpha\to1$: $\nGoccu^{[1]}$}{alpha to 1: nGoccu1}}
\label{subsec:Moccu alpha1}
In the von Neumann limit $\alpha\to1$, we have
\begin{equation}
    \nGoccu \!=\! \sum_{j=1}^N \left[ -\lambda_j \log_2(\lambda_j) - (1 \!-\! \lambda_j) \log_2(1 \!-\! \lambda_j) \right].
\end{equation}
This expression is equivalent to 
\begin{equation}
    \nGoccu = S(\G(\rho)),
\end{equation}
where $\G(\rho)$ is the unique (but generically mixed) FGS with the same covariance matrix as the given pure state $\rho=\ketbra{\psi}{\psi}$ and $S(\rho)=-\Tr(\rho \log_2 \rho)$ is the von Neumann entropy.
It has been shown
that $\G(\rho)$ is the closest FGS to $\rho$ in terms of the quantum relative entropy~\cite{lyu2024fermionicgaussiantesting}\footnote{For the particle-preserving case, the statement had been previously proved in Ref.~\cite{gottlieb2014correlations}.}:
\begin{equation}
S(\rho \Vert \sigma) = -\Tr(\rho \log_2 \sigma) - S(\rho).
\end{equation}
Thus, $\nGoccu$ can be interpreted as the minimum quantum relative entropy to the set of FGSs, which is referred to as the relative entropy of fermionic non-Gaussianity in Ref.~\cite{lyu2024fermionicgaussiantesting}. %

So far the occupation number entropies have been defined only for pure states. A mixed-state extension can be constructed following the standard convex-roof construction~\cite{vidal2000entanglement}, which retains the resource-theoretic properties---including, for $\alpha=1$, monotonicity under the full set of Gaussian protocols---but, like any convex-roof monotone, requires an optimization over all pure-state decompositions and is generally intractable. We instead adopt a different mixed-state extension that is computationally tractable, at the cost of being valid only under the smaller class of Gaussian operations without feedforward.
In particular, for $\alpha=1$, it is the relative entropy of non-Gaussianity,
\begin{equation}
\begin{aligned} \label{eq:Mr_def}
    \nGR(\rho)
    &= \min_{\sigma \in \G} S(\rho \Vert \sigma) = S(\rho \Vert \G(\rho))\\
    &= S(\G(\rho)) - S(\rho),
\end{aligned}
\end{equation}
which reduces to $\nGoccu^{[1]}(\ket{\psi})$ for pure states $\rho=\ketbra{\psi}{\psi}$.
While the elementary properties of \emph{faithfulness}~(\ref{prop:faithful}), \emph{stability}~(\ref{prop:stability}), and \emph{additivity}~(\ref{prop:add}, which implies property~\ref{prop:composition_fgs} as a corollary) have been discussed in Ref.~\cite{lyu2024fermionicgaussiantesting}, the monotonicity property remained unknown.
We prove that it additionally satisfies \emph{strong monotonicity}~(\ref{prop:strong}) and monotonicity under partial tracing in Appendix~\ref{sec:Mr_properties}. The proof relies on the relative entropy definition in Eq.~\eqref{eq:Mr_def} and the monotonicity of relative entropy, on average, under measurements~\cite{vedral1998entanglement}.
By the argument in Sec.~\ref{sec:ng_monotone}, it then follows that $\nGoccu$---that is, $\nGR$ restricted to pure states---is a strong monotone under pure-state Gaussian protocols. This establishes $\nGoccu$ as a proper monotone in the convex resource theory of non-Gaussianity.

The monotonicity property is crucial to provide a non-trivial lower bound on the number of non-Gaussian operations, such as the SWAP gates, required to prepare a given state. Note that the SWAP gate promotes matchgate circuits to universal quantum computation~\cite{jozsa2008matchgates, BrGa11}. To obtain the lower bound, we note that a SWAP gate can be implemented by a gadget consisting of matchgates and feedforward~\cite{hebenstreit2019all}, consuming the magic state
\begin{equation} \label{eq:magic_state}
    \ket{M} = |\Phi^+\rangle_{13} \, |\Phi^+\rangle_{24},
\end{equation}
where $\ket{\Phi^+}=(\ket{00}+\ket{11})/\sqrt{2}$.
From monotonicity under Gaussian protocols, it follows that 
\begin{align}
    &\nGoccu(\text{SWAP} \ket{\psi})\leq \nGoccu(\ket{\psi} \otimes \ket{M}) \nonumber \\
    &\qquad= \nGoccu(\ket{\psi}) + \nGoccu(\ket{M}),
\end{align}
 for any state $\ket{\psi}$. By direct computation, we find that all elements of the covariance matrix of $\ket{M}$ vanish, so that the occupation number entropy is maximal, $\nGoccu(\ket{M})=4$. Therefore, if a state $\ket{\psi}$ can be prepared by a circuit composed of Gaussian protocols and $n_s$ SWAP gates acting on $\ket{0}^{\otimes N}$, then
\begin{equation} \label{eq:bound_swap}
    \frac{1}{4}\nGoccu(\ket{\psi}) \leq n_s.
\end{equation}
A similar bound has previously been obtained using the Gaussian nullity~\cite{MeHe25}. However, the latter bound is only valid in a more restricted setting of matchgates doped with SWAP gates. Instead, Eq.~\eqref{eq:bound_swap} holds more generally for any Gaussian protocol doped with SWAP gates.

As an example, consider the state
\begin{equation}
    \ket{M_3} = |\Phi^+\rangle_{13} \, |\Phi^+\rangle_{25} \, |\Phi^+\rangle_{46}.
\end{equation}
For the qubit ordering and partition $124|635$ it is the Choi state corresponding to the gate $M_3=\sum_{a_1,a_2,a_3\in\{0,1\}}\ket{a_1,a_2,a_3}\bra{a_2,a_3,a_1}$. Hence, $\ket{M_3}$ can be used to deterministically implement the gate $M_3$ using matchgates and feedforward~\cite{hebenstreit2019all}. To prepare $\ket{M_3}$, one can first prepare the Gaussian state $\ket{\Phi^+}^{\otimes 3}$ and apply a SWAP gate on qubits $2$ and $3$, and another SWAP gate on qubits $4$ and $5$. Direct computation shows that $ \nGoccu(\ket{M_3})=6$, which implies that at least two SWAP gates are required to prepare $\ket{M_3}$ by Eq.~\eqref{eq:bound_swap}. This establishes that the construction with two SWAP gates above is optimal for any Gaussian protocol doped with SWAP gates.

It is worth noting that $\nGoccu$ has appeared previously in different contexts under different names.
In the particle-number-preserving setting, it was introduced as \emph{correlation entropy} in Ref.~\cite{ziesche1995correlation} and \emph{nonfreeness} in Refs.~\cite{gottlieb2007properties, gottlieb2014correlations}.
It has since been used to characterize many-body localization transitions~\cite{bera2015mbl} and measurement-induced transitions~\cite{lumia2024measurement}, as well as to study the typical scaling of fermionic non-Gaussianity in random states~\cite{ares2026nongaussianityrandomquantumstates}.
Recently it has been shown that it is lower-bounded by the particle-number asymmetry~\cite{ares2026asymmetrylowerboundfermionic}.

\subsubsection{The case \texorpdfstring{$\alpha=2$: $\nGoccu^{[2]}$}{alpha = 2: nGoccu2}}
For $\alpha=2$, we have 
\begin{equation} \label{eq:ngoccu2_main}
    \nGoccu^{[2]}(\ket{\psi}) = 4  \sum_{j=1}^{N} \lambda_j (1 - \lambda_j )  = N-\sum_{j=1}^{N}r_j^2.
\end{equation}
This can equivalently be written as
\begin{equation}
\label{eq:occu_Lambda}
    \nGoccu^{[2]}(\ket{\psi}) = \frac{1}{2} \| \Lambda |\psi \rangle \otimes |\psi \rangle \|^2_2,
\end{equation}
where $\Lambda$ is defined in Eq.~\eqref{eq:Lambda_def}. This is shown in Appendix~\ref{sec:property_mlin}.
It thus links directly to the Gaussianity condition in Eq.~\eqref{eq:Lambda_condition}. %

Importantly, like $\nGoccu^{[1]}$, the case of $\alpha=2$, $\nGoccu^{[2]}$, is a \emph{strong monotone} under pure-state Gaussian protocols, as shown in Appendix~\ref{sec:strong_mono_occu2}, and therefore qualifies as a genuine monotone of fermionic non-Gaussianity. As a direct consequence, by the same argument as for $\nGoccu^{[1]}$, it likewise lower-bounds the number of SWAP gates required to prepare $\ket{\psi}$: since the magic state $\ket{M}$ of Eq.~\eqref{eq:magic_state} also satisfies $\nGoccu^{[2]}(\ket{M})=4$, monotonicity together with additivity yields 
\begin{equation}
    \tfrac{1}{4}\nGoccu^{[2]}(\ket{\psi})\leq n_s,
\end{equation}
whenever $\ket{\psi}$ is preparable using $n_s$ SWAP gates, in direct analogy with Eq.~\eqref{eq:bound_swap}. Although $\nGoccu^{[1]}(\ket{\psi})\geq\nGoccu^{[2]}(\ket{\psi})$ for every state~\footnote{%
    Writing $\nGoccu^{[1]}=\sum_j H_{\text{b}}(\lambda_j)$ with the binary entropy $H_{\text{b}}(\lambda)=-\lambda\log_2\lambda-(1-\lambda)\log_2(1-\lambda)$ and $\nGoccu^{[2]}=\sum_j 4\lambda_j(1-\lambda_j)$, the two are related by the elementary inequality $H_{\text{b}}(\lambda)\geq 4\lambda(1-\lambda)$, which holds for all $\lambda\in[0,1]$ with equality iff $\lambda\in\{0,\tfrac12,1\}$.
}%
---so the $\alpha=1$ bound is always tighter---the $\alpha=2$ bound has the practical advantage of following directly from the two-point Majorana correlators (as discussed below), without diagonalizing the covariance matrix, making it the more convenient estimator in both numerics and experiment.

Furthermore, $\nGoccu^{[2]}$ can be simply expressed in terms of expectation values as
\begin{equation} \label{eq:occu_exp}
    \nGoccu^{[2]}(\ket{\psi}) = N - \sum_{j>k} \,  \bra{\psi} i \gamma_j \gamma_k \ket{\psi} ^2.
\end{equation}
It is equivalent to the recently introduced \emph{fermionic antiflatness} with $k=1$~\cite{sierant2025fermionicmagicresourcesquantum}. This expression makes it highly convenient for both numerical simulation and experiment. For its experimental evaluation, we show in Appendix~\ref{sec:exp_meas} that $\nGoccu^{[2]}$ can be measured within $\epsilon$ accuracy via single-copy measurements with $\mathcal{O}(N^4\epsilon^{-2})$ copies of the state.

Apart from sampling errors, in experiments it is important to verify that $\nGoccu^{[2]}$ is robust to noise: Consider the situation where, instead of preparing the target pure state $\ket{\psi}$, one actually prepares the (possibly mixed) state $\rho$ due to noise or imperfections of the device. If the realized state remains close to $\ket{\psi}$, the experimentally measured value should approximate that of the target state. Indeed, we show that $\nGoccu^{[2]}$ is Lipschitz continuous with respect to the trace distance,
\begin{equation} \label{eq:bound_w_distance}
    \!\!|\nGoccu^{[2]}\!\;\!(\rho) - \nGoccu^{[2]}\!\;\!(\ket{\psi}) | \leq 2N (2N\!-\!1) \|\rho - \ketbra{\psi}{\psi} \!\|_1,\!\!
\end{equation}
for any pure state $\ket{\psi}$ and a (possibly mixed) state $\rho$. This is proven in Appendix~\ref{app:concent_ineq}. Here, $\nGoccu^{[2]}(\rho)$ for mixed states is defined as a natural extension of Eq.~\eqref{eq:occu_exp}:
\begin{equation} \label{eq:occu_exp_mixed}
    \nGoccu^{[2]}(\rho) = N - \sum_{j>k} \, |\Tr(i \gamma_j \gamma_k \rho) \;\! |^2.
\end{equation}  
Note, however, that this is not a faithful quantifier of non-Gaussianity (\ref{prop:faithful}) for mixed states. Combined with $\|\rho - \ketbra{\psi}{\psi}\|_1 \le 2\sqrt{1-F(\rho,\ketbra{\psi}{\psi})}$, the bound turns an available fidelity estimate---e.g.\ from cross-entropy benchmarking~\cite{arute2019quantum,Boixo2018,Wu2021,Morvan2024} or classical-simulation extrapolation~\cite{shaw2024benchmarking}---into a concrete bound on the deviation from the target value.

In addition, $\nGoccu^{[2]}$ provides both a lower and an upper bound on the Gaussian fidelity $F_\mathcal{G}$ as 
\begin{equation}
    \label{eq:occu_bound_Fg}
    1-\frac{\nGoccu^{[2]}}{2} \leq F_\mathcal{G} \leq  1 - \frac{1}{4}\left(\! 1-\sqrt{1-\frac{\nGoccu^{[2]}}{N}}\,\right)^{\!\!2}\!.
\end{equation}
The proof is given in Appendix~\ref{sec:bounds_Fg}.

As a byproduct, these bounds yield an efficient Gaussian-testing algorithm for pure states~\cite{buhrman2008quantum}: given the promise that either $F_\mathcal{G}\geq1-\epsilon_1$ or $F_\mathcal{G}\leq1-\epsilon_2$ (with $0<\epsilon_1<\epsilon_2$ satisfying suitable conditions), measuring $\nGoccu^{[2]}$ and applying Eq.~\eqref{eq:occu_bound_Fg} distinguishes the two cases with high probability using $\mathcal{O}(N^4)$ single-copy measurements, improving on the previous best bound of $\mathcal{O}(N^5\log N)$ copies~\cite{bittel2025optimal}. The improvement stems from measuring a low-degree polynomial of the covariance-matrix elements rather than estimating the full covariance matrix; the algorithm is detailed in Appendix~\ref{sec:testing}.

As a final note on the special case $\alpha=2$, we point out that it also allows a straightforward generalization to a computable mixed-state non-Gaussianity quantifier, valid under Gaussian operations without feedforward.
Using the Gaussianity condition for mixed states in Eq.~\eqref{eq:Lambda_condition_mixed} and with a similar idea as in Eq.~\eqref{eq:occu_Lambda}, we construct a mixed-state non-Gaussianity quantifier by considering the Hilbert-Schmidt norm
\begin{equation}
    \mathcal{M}_\Lambda(\rho) = \frac{1}{4 \Tr(\rho^2)^2} \lVert [\Lambda, \rho \otimes \rho] \rVert_2^2,  
\end{equation}
which can be simplified to 
\begin{equation}
\begin{split}
    \mathcal{M}_\Lambda(\rho) = &N - \frac{1}{\Tr(\rho^2)^2} \sum_{j>k}  |\Tr(\gamma_j \gamma_k \rho^2)|^2  \\
    &-\frac{1}{2\Tr(\rho^2)^2} \sum_{j,k} \Tr(\rho \gamma_j \rho \gamma_k)^2,
\end{split}
\end{equation}
as shown in Appendix~\ref{sec:M_Lambda_properties}.
The condition in Eq.~\eqref{eq:Lambda_condition_mixed} directly results in the faithfulness of $\mathcal{M}_\Lambda(\rho)$. It satisfies properties~\ref{prop:faithful},~\ref{prop:stability}, and~\ref{prop:add}, and $\mathcal{M}_\Lambda(\ketbra{\psi}{\psi})=\nGoccu^{[2]}(\ket{\psi})$ for pure states,  as shown in Appendix~\ref{sec:M_Lambda_properties}. More generally, any norm of the commutator $[\Lambda,\rho\otimes\rho]$, such as the trace norm, defines a faithful non-Gaussianity quantifier; we adopt the Hilbert--Schmidt norm because it yields the closed, efficiently computable form above.

\subsubsection{Connection to Gaussian nullity} \label{subsubsec:ngoccu_nullity}

The occupation number entropy is also related to the Gaussian nullity $\nu_\G$ in the limit $\alpha \to 0$ and  $\alpha \to \infty$ separately as
\begin{equation}
    \nu_\G = \lim_{\alpha \to 0}  \nGoccu^{[\alpha]},
\quad\text{and}\quad
    \nu_\G = \lim_{\alpha \to \infty}  \nGoccu^{[\alpha]}.
\end{equation}
We also have the bound
\begin{equation}
    \nGoccu^{[\alpha]} \leq \nu_\G,
\end{equation}
To see this, note that any state with Gaussian nullity $\nu_\G$ can be brought, by an FGU $U$, to the form $U\ket{\psi} = \ket{0}^{N-\nu_\G} \ket{\phi}$, where $\ket{\phi}$ is a pure state of $\nu_\G$ qubits~\cite{MeHe25}. The inequality follows from invariance under FGUs, additivity, faithfulness, and the bound $\nGoccu^{[\alpha]} \leq N$.

\subsubsection{Connection to the deviation from Wick's theorem}
\label{subsubsec:wick}

Wick's theorem, Eq.~\eqref{eq:wick_pfaffian}, expresses every Majorana correlator $\Tr[\rho \gamma_S]$ of an FGS, with $S\subset[2N]$, as a Pfaffian of the covariance matrix. %
Non-Gaussian states do not satisfy these relations, so non-Gaussianity is naturally diagnosed by the deviation from Wick's theorem. Here we show how the occupation number entropies explicitly upper-bound this deviation. The bounds have a direct operational meaning: when the occupation number entropies are small, Wick's theorem still holds approximately---every higher correlator stays close to the Pfaffian of its two-point functions---so that the state remains accurately described by a Gaussian approximation, with the occupation number entropies controlling the approximation error. This regime is physically relevant: small occupation number entropies arise, for example, in quantum impurity models~\cite{sierant2025fermionicmagicresourcesquantum,Bravyi17impurity}, in which a Gaussian system is perturbed by a few localized impurities, or in many-body localized systems~\cite{bera2015mbl,falcao2026fermionic}.

We write 
\begin{equation}
    W(\gamma_S)=\Tr[(\rho-\G(\rho))\gamma_S]
\end{equation}
for the deviation of a correlator $\Tr[\rho\gamma_S]$ from its Gaussian value $\Tr[\G(\rho)\gamma_S]$ (the Pfaffian of the covariance matrix as given by Wick's theorem), which vanishes for all $S$ if and only if $\rho$ is an FGS. The relative entropy of non-Gaussianity bounds it uniformly in the Majorana order: H\"older's and Pinsker's inequalities~\cite{Wilde2013} give
\begin{equation} \label{eq:wick_relent}
   |W(\gamma_S)| \leq \big\|\rho-\G(\rho)\big\|_1 \leq \sqrt{2\ln2\;\nGR(\rho)},
\end{equation}
where we used $\|\gamma_S\|_\infty=1$. A single quantity thus bounds the deviation at \emph{every} order.

The $\alpha=2$ case provides a complementary bound, more efficiently accessible from low-order Majorana data, at the cost of giving looser bounds for higher-order Majorana correlators. As we show in Appendix~\ref{app:wick}, the deviation of each single Wick contraction can be expressed in terms of the commutator $[\Lambda,\rho^{\otimes2}]$, thus directly connecting the Gaussianity condition $[\Lambda,\rho^{\otimes2}]=0$ [Eq.~\eqref{eq:Lambda_condition_mixed}] to the emergence of Wick's theorem. For a pure state and a $2m$-point correlator, this relation yields a bound,
\begin{equation} \label{eq:wick_higher}
   |W(\gamma_S)| \leq \sqrt{(2m-1)!!}\;\sqrt{2\,\nGoccu^{[2]}(\ket\psi)} ,
\end{equation}
as shown in Appendix~\ref{app:wick}.  Here the double factorial is defined as $(2m-1)!!=1\cdot3\cdot5\cdots(2m-1)$. For the lowest (four-point) order, the bound tightens to $|W(\gamma_S)|\leq\sqrt{2\,\nGoccu^{[2]}(\ket\psi)}$. The prefactor $\sqrt{(2m-1)!!}$ grows super-exponentially, so the bound weakens rapidly with the Majorana order, in contrast to the relative-entropy bound in Eq.~\eqref{eq:wick_relent}, which is uniform in the order.

\subsubsection{Connection to entanglement}

An active line of research in resource theories is to understand how different quantum resources are related to one another. For example, nonstabilizerness has been linked to entanglement~\cite{tirrito2023quantifying, frau2024nonstabilizerness, dowling2025bridging, Gu2025} and coherence~\cite{turkeshi2023measuring, tirrito2025universalspreading, tarabunga2025magictransition}. Here, we establish that the occupation number entropies have a direct connection to entanglement. In particular, we define a family of global entanglement monotones, the total Tsallis entropies, as
\begin{equation} \label{eq:me_def_main}
    \ME^{[\alpha]}(\ket{\psi}) = \sum_{j=1}^N T_{\alpha}\left(\rho_j\right),
\end{equation}
where $\rho_j$ is the single-qubit reduced density matrix at site $j$ and $T_\alpha(\rho)$ is defined as
\begin{equation}
T_\alpha(\rho) = \frac{1}{1-{2^{1-\alpha}}} \left(1 - \Tr( \rho^\alpha)\right) ,
\end{equation}
analogously to $T_\alpha(\{p_i\})$ in Eq.~\eqref{eq:tsallis_prob} for a discrete probability distribution $p_i$. Note that $\ME^{[\alpha]}$ can be viewed as the sum of the bipartite entanglement between each site $j$ and the rest of the system, where the bipartite entanglement is quantified by a family of entanglement monotones known as the Tsallis-$\alpha$ entanglement~\cite{kim2010tsallis}. Consequently, the monotonicity of $\ME^{[\alpha]}$ under LOCC is directly inherited from the bipartite entanglement measures~\cite{ma2024multipartite}. For $\alpha=2$, $\ME^{[2]}$ is equivalent to the Meyer--Wallach measure~\cite{MeyerWallach2002, Brennen2003}, while for $\alpha=1$, $\ME^{[1]}$ is equivalent to the total mutual information~\cite{modi2010unified}. 

Working in the natural-orbital basis, i.e., the basis in which the covariance matrix is in the normal form, any FGS becomes a product state in the computational basis, where the state is completely disentangled.
We show that the family $\ME^{[\alpha]}$, when evaluated in this natural-orbital basis, coincide exactly with the occupation number entropies.
Explicitly,
\begin{equation} \label{eq:me_natural}
    \nGoccu^{[\alpha]} (\ket{\psi})=  \ME^{[\alpha]}(U_Q^\dagger | \psi \rangle) 
\end{equation}
where $U_Q$ is the FGU associated to the orthogonal matrix $Q$ such that $\Gamma(U_Q^\dagger | \psi \rangle)$ is in a normal form. While $U_Q$ may not be unique (see Sec.~\ref{subsec:NOPE}), Eq.~\eqref{eq:me_natural} holds true for any choice $U_Q$ satisfying the condition above. More strongly, we show that the natural-orbital basis is precisely the Gaussian basis 
that minimizes the global entanglement monotones in Eq.~\eqref{eq:me_def_main}. That is,
\begin{equation}
\label{eq:me_min_natural}
    \nGoccu^{[\alpha]}(\ket{\psi}) = \min_{U \in \G_\text{U}} \ME^{[\alpha]}(U^\dagger \ket{\psi}).
\end{equation}
This is proven in Appendix~\ref{sec:connection_multipartite}. Note that the case $\alpha=1$ was previously shown in Ref.~\cite{gigena2015entanglement}. Interestingly, the relations in Eq.~\eqref{eq:me_min_natural} bear resemblance to the entanglement R\'enyi entropies, which minimize the participation entropies [defined below in Eq.~\eqref{eq:pe}] over local unitaries~\cite{Spekkens2001}.

Despite their simplicity and efficient computability, because they depend only on the covariance matrix, the occupation number entropies assign the same value to all states with identical covariance matrices, regardless of their higher-order correlations. This leads to a coarse quantification that mirrors the known limitation of single-qubit-based global entanglement measures such as the Meyer--Wallach measure~\cite{MeyerWallach2002, Brennen2003}. This motivates more refined quantities which, following the same guiding idea, vanish on computational basis states and are evaluated in the natural-orbital basis, as we discuss next.

\subsection{Natural-orbital participation entropy \label{subsec:NOPE}}

We motivate the second family from the perspective of classical simulation. The cost of simulating a state via a Gaussian expansion is governed by the number of Gaussian states the expansion required: minimizing over all Gaussian expansions defines the Gaussian rank $\chi_\mathcal{G}$, while the amplitude-weighted analogue yields the Gaussian extent $\xi_\mathcal{G}$~\cite{dias2024classical,ReardonSmith2024improved,cudby2023gaussian}. Although these quantities are the most natural targets, their computation requires an optimization over all Gaussian expansions, for which there is currently no known efficient algorithm.
To obtain a more practical alternative, we restrict attention to expansions in an \emph{orthonormal Gaussian basis}, defined as
$|\tilde{i}\rangle = U|i\rangle,\, U \in \mathcal{G}_\mathrm{U}$,
where $U$ is a fermionic Gaussian unitary acting on the computational basis. In particular, we use a single canonical orthonormal Gaussian basis: the \emph{natural-orbital basis}~\cite{lowdin1955quantum}, the (not necessarily unique) basis in which $\Gamma$ takes its normal form. This yields quantities efficiently accessible in state-vector simulations that upper-bound $\chi_\mathcal{G}$ and $\xi_\mathcal{G}$. Its relevance is that a pure FGS, rotated into its natural-orbital basis, becomes a product state in the computational basis, with vanishing participation entropy. The participation entropy in this basis therefore quantifies non-Gaussianity, with a nonzero value
showing deviation from Gaussianity. It also reflects classical simulability: a small value means the weight concentrates on a few natural-orbital states, so the state is well approximated by retaining only those, whereas a large value means the weight is spread over many states, requiring a correspondingly large expansion. We also illustrate in Sec.~\ref{subsec:superposition_FGS} through a simple example that the natural-orbital basis captures the relevant low-rank Gaussian structure of the state, even when it is not necessarily the optimal Gaussian basis.

Accordingly, we define the natural-orbital participation entropy as
\begin{equation}
    \label{eq:M_no}
    \nGNO^{[\alpha]} (\ket{\psi}) =
    \min_{U_Q} S^{\text{part}}_\alpha(U_Q^\dagger \ket{\psi}),
\end{equation}
where $S^{\text{part}}_\alpha (\ket{\psi})$ is the participation entropy
\begin{equation}
    \label{eq:pe}
    S^{\text{part}}_\alpha (\ket{\psi}) = \frac{1}{1-\alpha} \log_2 \sum_s |\langle s | \psi \rangle|^{2\alpha},
\end{equation}
that is, the
$\alpha$-R\'enyi entropy of the discrete probability distribution $\{p_s\}$ defined from the squared
amplitudes in the computational basis, $p_s = |\langle s | \psi \rangle|^2$. The minimization in Eq.~\eqref{eq:M_no} is performed over those $U_Q \in \G_\text{U}$ for which
the rotated covariance matrix
\begin{equation}
    \Gamma(U_Q^\dagger | \psi \rangle)
    = Q^T \, \Gamma(| \psi \rangle) \, Q
    = \bigoplus_{j=1}^N
    \begin{pmatrix}
        0 & r_j \\
        -r_j & 0 \\
    \end{pmatrix}
\end{equation}
is in normal form. 
We show in Appendix~\ref{sec:no_properties} that $\nGNO^{[\alpha]}$ satisfies
properties~\ref{prop:faithful},~\ref{prop:stability},~\ref{prop:sub-add}
and~\ref{prop:composition_fgs}.

Generically, the normal form uniquely fixes $Q$, so that no further minimization is required.
Minimization only becomes necessary in fine-tuned cases %
where either (i) one or more Williamson eigenvalues vanish, $r_j=0$, or (ii) there exist degenerate nonzero Williamson eigenvalues. Note, however, that case (i) generically occurs in stabilizer states (discussed in Sec.~\ref{sec:pure_stab}), while case (ii) arises in the presence of non-Abelian symmetries, such as the combination of a $U(1)$ $z$-rotation and spin-flip symmetry, which will appear in our numerical examples in Sec.~\ref{sec: example}.

In case (i), the blocks corresponding to $r_j=0$ are invariant under arbitrary orthogonal rotations that act non-trivially only on those blocks. In case (ii), the blocks take the form $\lambda \Omega$ where $\Omega=\bigoplus_{j=1}^{N_\lambda} \begin{pmatrix} 0 & 1 \\ -1 & 0 \\ \end{pmatrix}$, %
which are invariant under symplectic matrices $ S\Omega S^T =\Omega$ for $S \in \text{Sp}(2N_\lambda)$. Therefore, the minimization is performed for orthogonal symplectic matrices $Q \in O(2N_\lambda)\, \cap\, \text{Sp}(2N_\lambda)$ acting on the degenerate blocks. %

Experimentally, $\nGNO^{[\alpha]}$ can be measured by first applying an FGU that transforms the state into its natural-orbital basis, followed by a measurement of the participation entropy. The latter can be estimated from computational-basis measurements either by reconstructing wavefunction amplitudes directly or, for integer R\'enyi index $\alpha>1$, by estimating $\sum_s p_s^\alpha$ from correlations between independent measurement outcomes~\cite{luitz2014universal}.

The FGU $U_Q$ that brings the state into the natural-orbital basis does not, in general, minimize the participation entropy over all FGUs (an explicit example is given in Sec.~\ref{subsec:superposition_FGS}); minimizing over all FGUs would identify an even better Gaussian basis for a truncated expansion, at the cost of a harder computation.
This defines the orthogonal Gaussian participation entropy
\begin{equation} 
    \label{eq:M_og}
    \nGOG^{[\alpha]}(\ket{\psi} ) = \min_{U \in \G_\text{U}} S^{\text{part}}_\alpha(U^\dagger\ket{\psi}),
\end{equation}
where the minimization is over all FGUs. However, due to the need of optimization for any state, this quantity is generally difficult to compute.  Notice that this definition bears resemblance to the Gaussian rank [cf. Eq.~\eqref{eq:grank_def}]~\cite{dias2024classical, cudby2023gaussian} and the Gaussian extent [cf. Eq.~\eqref{eq:gextent_def}]~\cite{dias2024classical, cudby2023gaussian} for $\alpha=0$ and $\alpha=1/2$, respectively. In particular, while the Gaussian extent and Gaussian rank are defined via an optimal superposition of arbitrary (not necessarily orthogonal) Gaussian states, the expansion implicit in $\nGOG^{[\alpha]}$ is restricted to orthonormal Gaussian states. This immediately implies
\begin{equation} \label{eq:bound_og_extent_rank}
    \nGOG^{[1/2]} \geq \log_2 \xi_\G
\quad \text{and}\quad
    \nGOG^{[0]} \geq \log_2 \chi_\G.
\end{equation}
Further properties are discussed in Appendix~\ref{sec:NGOG_app}.

Since the natural-orbital basis is a particular orthonormal Gaussian basis, we have
\begin{equation}
 \nGNO^{[\alpha]} \geq \nGOG^{[\alpha]},   
\end{equation}
which also shows that $\nGNO^{[\alpha]}$ upper-bounds the Gaussian extent and the Gaussian rank, for R\'enyi index $\alpha=1/2$ and $\alpha=0$, respectively.
In particular, it has been conjectured that in the case of two fermions at fixed particle number, the natural-orbital basis minimizes the participation entropy among all Gaussian bases~\cite{aliverti2024can}.
One can also show the inequality
    \begin{equation}
        \nGoccu^{[1]} \geq \nGNO^{[1]},
    \end{equation}
which is proven in Appendix~\ref{sec:relation_occu}.
Furthermore, we have
\begin{equation} \label{eq:lower_fid_nore}
    -\log_2 F_{\G}  \leq \nGNO^{[\alpha]}%
\end{equation}
for any $\alpha\geq0$. The proof is given in Appendix~\ref{sec:bounds_Fg_ogre}.

Finally, these quantities are also related to the Gaussian nullity $\nu_\G$ as
\begin{equation}
    \nGNO^{[\alpha]} \leq \nu_\G
    \quad\text{and}\quad
    \nGOG^{[\alpha]} \leq \nu_\G,
\end{equation} 
for any $\alpha\geq0$. The proof is similar to the one mentioned in Sec.~\ref{subsubsec:ngoccu_nullity}.

\subsection{Connection to the classical simulation cost of fermionic circuits}
\label{subsec: classical_sim}

Here, we discuss the connection between the two quantities $\nGOG^{[\alpha]}$ and $\nGNO^{[\alpha]}$ and the computational cost of classical simulations. We recall that these quantities correspond to the participation entropy in a particular orthonormal Gaussian basis, which represents a favorable choice for simulation. Although the
simulation strategy described below may not be practical, in particular compared to recently developed methods~\cite{dias2024classical, cudby2023gaussian, ReardonSmith2024improved,dias2026optimalimprovedgatedecompositions},
it serves to provide a conceptual connection between our quantities and classical
simulation complexity.%

Let us consider the simulation of a matchgate circuit doped by SWAP gates. In order to simulate the circuit classically, we consider the approach of finding an efficient classical representation of the state throughout the evolution with the gates.
We choose to represent the state in an orthonormal Gaussian basis as $\ket{\psi} = \sum_j c_j \ket{g_j}$, where $\ket{g_j}$ are orthonormal Gaussian states. Equivalently, we can express the state as $\ket{\psi} = U \ket{\psi'}$, where $U$ is a matchgate unitary and 
\begin{equation}
    \ket{\psi'}=\sum_j c_j \ket{j}.
\end{equation}
 Therefore, it is sufficient to store a matchgate unitary $U$, which defines a Gaussian basis, and the set of coefficients $\{c_j\}$ to represent the state. If a matchgate unitary $U'$ is applied to the state, one can simply update $U \to U' U$. We then consider the application of a generic non-Gaussian gate $R=\exp\!\big(i \theta  \gamma_{p} \gamma_{q} \gamma_{r} \gamma_{s} \big)$. Note that non-Gaussian gates that are matchgate-equivalent to this form, such as the $\text{SWAP}$ gate, can be treated in an analogous way. We have $R \ket{\psi} = R U \ket{\psi'} = U (U^\dagger R U) \ket{\psi'}$, where $\tilde{R} = U^\dagger R U$ can be efficiently computed and takes the form $\tilde{R} = \exp\!\big(i \theta \sum_{p,q,r,s} \tilde{h}_{pqrs} \gamma_{p} \gamma_{q} \gamma_{r} \gamma_{s} \big)$, which satisfies $\left(\sum_{p,q,r,s} \tilde{h}_{pqrs} \gamma_{p} \gamma_{q} \gamma_{r} \gamma_{s} \right)^2=\Id$. The unitary $\tilde{R}$ can then be applied to update the coefficients $\{c_j\}$. To find the resulting state $\ket{\psi''} = \tilde{R} \ket{\psi'}= \sum_j c'_j \ket{j}$, we can determine the coefficients $c'_j$ as
\begin{equation}
\begin{aligned}
    c'_j &= \braket{j}{\psi''}
    = \bra{j} \tilde{R} \ket{\psi'}
    = \sum_k  c_k \bra{j} \tilde{R} \ket{k} \\
    &= \sum_k c_k\Big(  \cos(\theta) \, \delta_{jk}\\[-0.28cm]
    &\quad\qquad+ i \sum_{p,q,r,s} \sin(\theta) \, \tilde{h}_{pqrs} \bra{j} \gamma_{p} \gamma_{q} \gamma_{r} \gamma_{s} \ket{k} \Big),
\end{aligned}
\end{equation}
which can be computed efficiently, assuming the number of nonzero coefficients $c_j$ is polynomial. It can be verified that each gate would increase the number of nonzero coefficients at most by a polynomial factor of $\mathcal{O}(N^4)$. The final state $U \ket{\psi''}$ can then be written in the original Gaussian basis with updated coefficients $\{c'_j\}$.
In the next step, one can transform the basis to a different Gaussian basis, chosen according to some criteria, e.g., by minimizing the participation entropy.

This procedure enables a classical simulation of the fermionic circuits by keeping track of the coefficients in a Gaussian basis. The simulation can be done approximately by truncating all coefficients below a given error threshold $|c_j|<\epsilon$ to control the computational cost.
Related approaches are considered in the literature~\cite{mullinax2025large}. These coefficients play an analogous role to Schmidt values in the truncation of MPSs, and we can directly recover approximability results for non-Gaussian states in terms of their participation entropy in a specific basis from analogous results for MPSs in terms of entanglement R\'enyi entropies of the target state. In both settings, the question of how many terms must be retained for an accurate truncation reduces to how broad a single classical probability distribution is---the squared Schmidt coefficients for an MPS, and the squared amplitudes $\{|c_j|^2\}$ in the chosen Gaussian basis for our quantities. It therefore suffices to analyze this classical distribution, for which the relevant approximability results have been established for MPS in Refs.~\cite{Verstraete2006, schuch2008entropy, audenaert2007sharp}, and we now translate to our setting. The quantities $\nGOG^{[\alpha]}$ and $\nGNO^{[\alpha]}$ with $\alpha<1$ indicate when an accurate truncation is possible: following Ref.~\cite{Verstraete2006}, if these grow at most logarithmically in system size, only a polynomial number of coefficients need to be retained. The quantities $\nGOG^{[\alpha]}$ and $\nGNO^{[\alpha]}$ with $\alpha\geq1$, conversely, indicate when an accurate truncation is impossible: following Ref.~\cite{schuch2008entropy}, if the quantities with $\alpha>1$ grow at least as fast as some (positive) power of the system size, or the quantities with $\alpha=1$ grow linearly with system size, an accurate representation needs to retain exponentially many coefficients. These results can be shown for $\alpha\neq1$ using majorizing distributions and Schur convexity of R\'enyi entropies~\cite{Verstraete2006, schuch2008entropy}, and for $\alpha=1$ using Fannes' inequality~\cite{schuch2008entropy, audenaert2007sharp}.
Note that $\nGOG^{[\alpha]}$ is the relevant quantity if the state is represented in the optimized Gaussian basis, and $\nGNO^{[\alpha]}$ for the natural-orbital basis.

By similar reasoning, if the coefficients $\{c_j\}$ are represented by an MPS, the corresponding simulation cost is dictated by the entanglement minimized over Gaussian bases or the entanglement in the natural-orbital basis. While we do not discuss these quantities further, we note that they are related to $\nGOG^{[\alpha]}$ and $\nGNO^{[\alpha]}$, since the participation entropy provides an upper bound on the R\'enyi entanglement entropy for any corresponding R\'enyi index~\cite{Spekkens2001}. Moreover, we note the connection to entanglement-minimized orbitals as discussed in Ref.~\cite{li2025entanglement}. Likewise, one could investigate the simulation of the states in the natural-orbital basis via near-Clifford methods, whose cost is governed by the nonstabilizerness of the state in the corresponding basis, which has recently been considered in Ref.~\cite{grabarits2026universal}.

\section{Special cases}
\label{sec: special}

In this section we study the non-Gaussianity of two important classes of many-body states. We begin with stabilizer states (Sec.~\ref{sec:pure_stab}), for which we introduce a new resource theory---the resource theory of \emph{stabilizer non-Gaussianity}---whose free states are stabilizer FGSs and whose free operations are those preserving this structure. The covariance matrix of a stabilizer state is so strongly constrained that $\nGoccu^{[\alpha]}$ reduces to the Gaussian nullity for every $\alpha$, while $\nGNO^{[\alpha]}$ is likewise $\alpha$-independent and can be evaluated through a discrete optimization. Afterwards, we turn to translation-invariant systems (Sec.~\ref{sec:TI}), where momentum-space block-diagonalization simplifies the computation of the quantities defined in Sec.~\ref{sec: nG_measures} and yields exponential finite-size convergence in gapped phases.

\subsection{Pure stabilizer states}
\label{sec:pure_stab}

Stabilizer states constitute an important set of quantum states due to their efficient classical simulability~\cite{aaronson2004improved} and their diverse applications including quantum error correction~\cite{Go97} and measurement-based computation~\cite{RaBr01}.
In the resource theory of stabilizer non-Gaussianity, the free states are stabilizer fermionic Gaussian states (S-FGSs), while the free operations are Gaussian protocols that additionally preserve the stabilizer structure, which we refer to as \emph{stabilizer Gaussian protocols}.
Since stabilizer Gaussian protocols map pure S-FGSs to pure S-FGSs, the free operations cannot generate the resource out of free states, so this choice of free states and free operations defines a valid resource theory.
This is a restriction of the general resource theory of non-Gaussianity, in the sense that both its free states and its free operations are subsets of the corresponding free sets of the general theory.
This restricted free set is physically well motivated by the central role of stabilizer states.
Throughout this subsection, all properties discussed concern only the resource theory of stabilizer non-Gaussianity. The additional structure of stabilizer states strongly constrains the covariance matrix and allows us to establish properties that were not proved in the general setting: the occupation number entropy coincides with the Gaussian nullity for any R\'enyi index and is computable from the covariance matrix without diagonalization, while the natural-orbital participation entropy is a strong monotone under stabilizer Gaussian protocols, and its exact evaluation requiring a discrete optimization for which we provide an efficiently computable upper bound.

Consider a stabilizer state $\ket{\psi}$ on $N$ qubits, specified by a set of $N$ linearly independent Pauli strings $\{g_1,\dots, g_N\}$ that stabilize the state, i.e., $g_j\ket{\psi} = \ket{\psi}$ for all $j$.
For such states, expectation values of Pauli strings can only take the values $0$ or $\pm1$~\cite{Go97}, which imposes strong constraints on the structure of the covariance matrix: for each $\gamma_j$, at most one index $k \neq j$ satisfies $\langle \gamma_j \gamma_k \rangle = \pm 1$, while all remaining correlators vanish~\footnote{This results from the fact that $\gamma_j\gamma_k$ and $\gamma_j\gamma_{\ell}$ anticommute for any $k\neq\ell$ and $k,\ell \neq j$, along with the fact that all Pauli strings with nonzero expectation value must commute for a stabilizer state.}. 
Consequently, each row and column of the covariance matrix contains at most one nonzero off-diagonal entry of magnitude $1$, so a permutation of the modes brings $\Gamma$ into Williamson normal form. The Williamson eigenvalues are thus restricted to $0$ or $1$ and the occupation numbers to $0$, $1/2$, or $1$. The occupation numbers $0$ and $1$ correspond to Gaussian modes and $1/2$ to non-Gaussian modes, so $\Gamma$ gives direct access to the Gaussian nullity,
\begin{equation}
    \nu_\mathcal{G} = N-\frac{1}{2}\sum_{j,k}\abs{\Gamma_{jk}}.
\end{equation}
The covariance matrix itself can be efficiently computed in $\mathcal{O}(N^3)$ time, as described in Appendix~\ref{subapp: construction covmat}, and therefore $\nu_\mathcal{G}$ is efficiently computable. 
A direct evaluation of Eq.~\eqref{eq: Moccu} using this structure of the occupation numbers shows that the occupation number entropy of a stabilizer state coincides with the Gaussian nullity,
\begin{equation}
\label{eq:ngoccu_stab}
    \nGoccu^{[\alpha]} = \nu_{\mathcal{G}}.
\end{equation}
Since this value is independent of the R\'enyi index---as is the participation entropy below---we drop the superscript $\alpha$ for stabilizer states and write $\nGoccu^{\text{S-FGS}}$ and $\nGNO^{\text{S-FGS}}$. Moreover, the results of the previous sections show that $\nGoccu^{[1]}$ is a non-Gaussianity monotone; this monotonicity thus extends to $\nGoccu^{\text{S-FGS}}$ on this class of states~\footnote{Notice also that the unique FGS $\mathcal{G}(\ket{\psi})$ with the same covariance matrix as a stabilizer state is a (possibly mixed) stabilizer state. Specifically, the stabilizer state $\mathcal{G}(\ket{\psi})$ is stabilized by the set of Pauli operators in the stabilizer group of $\ket{\psi}$ which are quadratic in the Majorana operators. This set consists of $N-\nu_\mathcal{G}$ operators, and it generates the stabilizer group of $\mathcal{G}(\ket{\psi})$. As a consequence, we have $\nGoccu^{\text{S-FGS}} = S(\mathcal{G}(\ket{\psi})) = \nu_\mathcal{G}$, which gives an alternative derivation of Eq.~\eqref{eq:ngoccu_stab}.}. We further show in Appendix~\ref{sec:ng_stabilizer_mixed_state_monotone} that $\M_\Lambda^{\mathrm{S-FGS}}$ obeys monotonicity for (mixed) stabilizer states under stabilizer Gaussian operations (where we do not consider feedforward).

We now turn to the natural-orbital participation entropy.
Within the stabilizer non-Gaussianity resource theory we define $\nGNO^{\text{S-FGS}}$ via Eq.~\eqref{eq:M_no}, restricting the minimization to braiding gates---Clifford unitaries that are also matchgates~\cite{Br06}---which upper-bounds $\nGNO^{[\alpha]}$ for stabilizer states. Because every non-Gaussian mode of a stabilizer state has occupation $1/2$,
these modes are degenerate, and the minimization in Eq.~\eqref{eq:M_no} becomes nontrivial, since one must determine the basis that minimizes the participation entropy.
We prove that $\nGNO^{\text{S-FGS}}$ satisfies monotonicity under stabilizer Gaussian protocols in Appendix~\ref{sec:monotone_OG_stab}. 

To carry out the minimization, we use that the participation entropy of a stabilizer state is efficiently computable~\cite{SiTu22} and express the stabilizer table in the Majorana basis~\cite{Br06, BeSw25}. Writing each generator as $s_j \propto \gamma_1^{n_{j,1}} \dots \gamma_{2N}^{n_{j,2N}}$ (we omit the phase, which does not affect the non-Gaussianity), we define the stabilizer tableau in the Majorana representation as the binary matrix $(M_{\text{maj}})_{jk}=n_{j,k}$ with entries given by the exponent of the $k$th Majorana operator in the $j$th generating stabilizer. In this representation, the columns $2n-1$ and $2n$ of $M_{\text{maj}}$ correspond to qubit $n$. A braiding unitary then simply permutes the columns of $M_{\text{maj}}$, so the minimization amounts to finding the lowest participation entropy reachable by column permutations. However, there are $(2N)!$ such permutations, making the brute-force optimization unfeasible for large system sizes.

Equivalently, the minimization can be recast as finding the stabilizer FGS $\ket{\text{S-FGS}}$ whose stabilizer group $\mathcal{S}_{\text{S-FGS}}$ maximizes the intersection $\mathcal{S}\cap\mathcal{S}_{\text{S-FGS}}$ with the stabilizer group $\mathcal{S}$ of $\ket\psi$; for any candidate, this intersection is computed efficiently via the Zassenhaus algorithm~\cite{LuRa97}. This equivalence is proved in Appendix~\ref{app: equivalence optimizations}. The hardness of computing $\nGNO^{\text{S-FGS}}$ therefore lies in searching over the exponentially large space of candidates, since the total number of S-FGSs for $N$ qubits (excluding the irrelevant signs in each stabilizer) is $(2N-1)!!$, which corresponds to the number of pairs of Majorana operators that can be formed. To address this, in Appendix~\ref{app:nGNO approximate stabilizers} we provide an efficient stochastic algorithm for finding an approximate solution to the optimization problem, which then yields an upper bound on $\nGNO^{\text{S-FGS}}$. 
The optimizing state also yields an explicit expansion of $\ket\psi$ as a superposition of $2^{\nGNO^{\text{S-FGS}}}$ stabilizer Gaussian states (Appendix~\ref{sec:ng_stabilizer_superposition}). Thus $\nGNO^{\text{S-FGS}}$ quantifies the cost of simulating a stabilizer state in a stabilizer Gaussian basis, mirroring the role of $\nGNO^{[\alpha]}$ for general states; this is advantageous because stabilizer Gaussian states are more efficiently simulable than generic stabilizer states, so a stabilizer state with $\nGNO^{\text{S-FGS}}$ independent of system size can be simulated more efficiently than a generic one.

\subsection{Translation-invariant states}
\label{sec:TI}

In this section, we discuss the computation of the occupation number entropy $\nGoccu^{[\alpha]}$ in translation-invariant fermionic systems.
Such systems are ubiquitous in condensed matter physics and thus constitute an important class of states.
Throughout this section, we will mostly focus on one-dimensional (1D) systems for notational simplicity, though our results equally apply to translation-invariant systems in arbitrary spatial dimension $D$ as we detail in Appendix~\ref{sec:ti_appendix}.
We first consider finite translation-invariant systems in Sec.~\ref{sec:ngoccu_finite_TI}, and show that translation symmetry allows the covariance matrix to be block-diagonalized in momentum space, greatly simplifying the computation of the occupation number entropy.
We then analyze the thermodynamic limit in Sec.~\ref{sec:ngoccu_thermo}, where the momentum becomes continuous, and we derive expressions for the non-Gaussianity density. If the correlations decay sufficiently fast as a power law, we establish that the occupation number entropy in the thermodynamic limit can be reconstructed from finite-range correlations with an error that decreases polynomially with the cutoff range. We consider ground states in gapped phases of matter in Sec.~\ref{sec:ngoccu_gapped_main}. Due to the exponential decay of correlations in these states, the finite-range approximation can be improved to converge exponentially fast in the cutoff. Moreover, in this setting, the two-point correlators in a finite system are expected to quickly converge to their expectation value in the thermodynamic limit. We use this to show that the occupation number entropy density $\nGoccu^{[\alpha]} / N$ in a finite system converges exponentially fast with system size to its value in the thermodynamic limit.
Finally, we discuss translation-invariant (TI) MPSs in Sec.~\ref{sec:ti_mps_main}, which provide an explicit setting of gapped systems with translation invariance in 1D. In contrast to the finite-range approximations discussed in the preceding subsections, we show that the structure of TI MPSs allows us to obtain the covariance matrix in momentum space in closed form even in the thermodynamic limit.

\subsubsection{Simplification in finite systems}
\label{sec:ngoccu_finite_TI}

First, we show that the computation of $\nGoccu^{[\alpha]}$ can be considerably simplified in the presence of translation symmetry in finite systems. We consider a system of Majorana fermions with $L$ unit cells, each containing $2B$ Majorana operators. The total number of sites is $N=LB$. We will adapt our notation in this section to highlight the index corresponding to translation symmetry and separate it from other degrees of freedom: we label the Majorana operators as $\gamma_{j,\alpha}$, with the site index $j\in \{ -\lfloor L/2 \rfloor, - \lfloor L/2 \rfloor +1, \dots, \lceil L/2 \rceil -1 \}$ and an internal index $\alpha \in [2B]$.

Invariance under translations imposes that the system is invariant under the transformation $\gamma_{j,\alpha} \to \gamma_{j+1,\alpha}$ for bulk sites $j \in \{ -\lfloor L/2 \rfloor, - \lfloor L/2 \rfloor +1, \dots, \lceil L/2 \rceil -2 \}$, and $\gamma_{\lceil L/2 \rceil-1,\alpha} \to e^{i2\pi \theta}\gamma_{-\lfloor L/2 \rfloor,\alpha}$ at the boundary. We will particularly consider $\theta=0$ and $\theta=1/2$, which correspond to periodic and antiperiodic boundary conditions, respectively.

We write the elements of the covariance matrix as
\begin{equation}
    \Gamma_{j,\alpha;k,\beta} = -\frac{i}{2} \langle [\gamma_{j,\alpha}, \gamma_{k,\beta}] \rangle.
\end{equation}
Due to translation symmetry, the covariance matrix can be written in matrix form as 
\begin{equation}
    \Gamma = \sum_{\ell=-\lfloor L/2 \rfloor}^{\lceil L/2 \rceil -1} J^\ell \otimes t(\ell),
\end{equation}
where $t(\ell)$ are $2B \times 2B$ matrices and $J$ is a $L \times L$ matrix with nonzero elements $J_{j,j+1}=1$ for $1 \leq j \leq L-1$ and $J_{L,1}=e^{i2 \pi\theta}$. In other words, the matrix $\Gamma$ is block circulant~\cite{Davis1979} for $\theta=0$ and is block skew-circulant for $\theta=1/2$. We thus only need to evaluate $4B^2L$ instead of all $4N^2 = 4B^2L^2$ correlators to specify the covariance matrix. Note that the anticommutation relation of the Majorana operators implies that the matrix $t(\ell)$ satisfies
\begin{equation} \label{eq:t_symmetry}
    t(\ell)^T = -t(-\ell).
\end{equation}

Since $\Gamma$ is block (skew-)circulant, we can block-diagonalize it using a standard Fourier transform over the lattice indices~\cite{Davis1979}. The eigenvalues of $J$ are given by
\begin{equation}
    \omega_q = e^{-i\frac{2\pi}{L}(q-\theta)}, 
\end{equation}
for $q\in \{ -\lfloor L/2 \rfloor, - \lfloor L/2 \rfloor +1, \dots, \lceil L/2 \rceil -1 \}$.
In the eigenbasis of $J$, the covariance matrix $\Gamma$ takes the block-diagonal form ${\Gamma = \bigoplus_q \tilde{t}\big(\frac{q-\theta}{L}\big)}$, where we define 
\begin{equation} \label{eq:ti_momentum}
    \tilde{t}\!\left(\tfrac{q-\theta}{L}\right)
    = \sum_{\ell=-\lfloor L/2 \rfloor}^{\lceil L/2 \rceil -1}
    \! \omega_q^\ell \ t(\ell).
\end{equation}
These are the covariance matrix elements in momentum space for $q\in \{ -\lfloor L/2 \rfloor, - \lfloor L/2 \rfloor +1, \dots, \lceil L/2 \rceil -1 \}$.
The computation of all $\tilde{t}\big(\frac{q-\theta}{L}\big)$ can be carried out efficiently via a fast Fourier transform, with a cost of $\mathcal{O}(B^2 L \log L)$. Note that each $\tilde{t}\big(\frac{q-\theta}{L}\big)$ is skew-Hermitian, as can be shown using the relation in Eq.~\eqref{eq:t_symmetry}. This implies that the eigenvalues of $\tilde{t}\big(\frac{q-\theta}{L}\big)$ are purely imaginary. The eigenvalues of the set $\big\{\tilde{t}\big(\frac{q-\theta}{L}\big)\big\}_{q}$ give the full spectrum of $\Gamma$, which also has purely imaginary eigenvalues as expected.

Defining $H\big(\frac{q-\theta}{L}\big) = \frac{1}{2} \big( \Id_{2B} + i \tilde{t}\big( \frac{q-\theta}{L} \big) \big)$, we can express the occupation number entropy as
\begin{equation} \label{eq:ti_ngoccu}
    \nGoccu^{[\alpha]} = \frac{1}{1-2^{1-\alpha}}
    \left( N - \hspace{-1em}
    \sum_{q=-\lfloor L/2 \rfloor}^{\lceil L/2 \rceil -1} \!\!\!\!
    \Tr\!\left[H\big(\tfrac{q-\theta}{L}\big)^{\!\alpha}\right]
    \right)\!.
\end{equation}
In the limit $\alpha\to1$, this becomes
\begin{equation}
    \label{eq:ti_ngoccu_vne}
    \nGoccu = - \hspace{-1em}
    \sum_{q=-\lfloor L/2 \rfloor}^{\lceil L/2 \rceil -1} \!\!\!\!
    \Tr\!\left[H\big(\tfrac{q-\theta}{L}\big)
    \log_2 H\big(\tfrac{q-\theta}{L}\big)\right] .
\end{equation}
The case of $\nGoccu^{[2]}$ can be further simplified, as we can directly write
\begin{equation} \label{eq:ti_ngoccu2}
\begin{aligned}
    \nGoccu^{[2]} &= N - \frac{L}{2} \sum_{\ell=-\lfloor L/2 \rfloor}^{\lceil L/2 \rceil -1} \Tr[t(\ell) \, t(\ell)^\dagger] \\
    &= N\left( 1- \frac{1}{2B}
    \sum_{\alpha,\beta=1}^{2B} \,
    \sum_{j=-\lfloor L/2 \rfloor}^{\lceil L/2 \rceil -1}
    |\Gamma_{j,\alpha;0,\beta}|^2\right)\!,
\end{aligned}
\end{equation}
which only sums over entries contained in one column of blocks in the covariance matrix.

\subsubsection{Occupation number entropy in the thermodynamic limit}
\label{sec:ngoccu_thermo}

Here, we discuss the computation of $\nGoccu^{[\alpha]}/N$ for translation-invariant systems in the thermodynamic limit, taking the number of unit cells $L\to\infty$ and keeping their size $B$ fixed. Taking this limit in Eq.~\eqref{eq:ti_momentum} yields
\begin{equation} \label{eq:ti_momentum_inf}
    \tilde{t}(p) =  \sum_{\ell\in \mathbb{Z}} e^{-i2\pi p\ell} \, t(\ell),
\end{equation}
where the momentum $p$ now takes continuous values in the Brillouin zone $2\pi p\in [-\pi, \pi)$.

In the limit $L\to \infty$, the non-Gaussianity density $m_{(\infty)}^{[\alpha]} = \lim_{N\to\infty} \nGoccu^{[\alpha]}(N)/N$ becomes
\begin{equation}  \label{eq:ti_ngoccu_inf_main}
    m_{(\infty)}^{[\alpha]}
    = \frac{1}{1 - 2^{1-\alpha}}
    \left(
    1 - \frac{1}{B} \int_{-1/2}^{1/2} \hspace{-.66em} \d p \, \Tr[H(p)^{\alpha}]
    \right)
\end{equation}
for general $\alpha\neq1$, and in the limit $\alpha\to1$ becomes
\begin{equation} \label{eq:ti_ngoccu_vne_inf_main}
    m_{(\infty)}^{[1]}
    = - \frac{1}{B} \int_{-1/2}^{1/2} \hspace{-.66em} \d p \, \Tr[H(p)\log_2 H(p)].
\end{equation}
These expressions are obtained from taking the limit $L\to\infty$ in Eqs.~\eqref{eq:ti_ngoccu} and~\eqref{eq:ti_ngoccu_vne}, and we analogously define ${H(p) = \frac{1}{2}\big(\Id_{2B}+i\tilde{t}(p)\big)}$.

Many translation-invariant states of interest arise as low-energy states of local Hamiltonians, and therefore the two-point correlators appearing in their covariance matrix decay with their separation distance.
For systems where these correlation functions decay sufficiently fast, we can compute the occupation number entropy density $m_{(\infty)}^{[\alpha]}$ in the thermodynamic limit by only considering correlations within a certain cutoff-range $\cutoff$. If the correlators in a $D$-dimensional translation-invariant system decay as $\mathcal{O}(1/r^{D+\eta})$ with the distance $r$ for some $\eta>0$, the error of this finite-range approximation of $m_{(\infty)}^{[\alpha]}$ scales as $\mathcal{O}(\cutoff^{-\eta})$ with the size of the cutoff.
The idea is that the cutoff leads to a truncation of the Fourier transform of the covariance matrix in Eq.~\eqref{eq:ti_momentum_inf}. As the summand decays sufficiently fast, the error of this truncation is algebraically small with respect to the cutoff. To ensure this ultimately leads to a small approximation error for $m_{(\infty)}^{[\alpha]}$, we need the integrands in Eqs.~\eqref{eq:ti_ngoccu_inf_main} and~\eqref{eq:ti_ngoccu_vne_inf_main} to be stable under small perturbations of $H(k)$. To this end, we make use of the Lipschitz continuity of the Tsallis entropies in certain regimes. For $\alpha>1$ they are Lipschitz continuous, while for $\alpha\leq1$ they are Lipschitz continuous under the additional assumption that no eigenvalue of $H(k)$ becomes zero.
A zero eigenvalue of $H(k)$ corresponds to an eigenvalue $-i$ in the covariance matrix and thus to a Gaussian mode; we can thus equivalently interpret this as the condition that no Gaussian modes appear in the covariance matrix. The existence of such modes in translation-invariant interacting states is a fine-tuned condition that holds only on a measure-zero subset of states. The detailed derivation is presented in Appendix~\ref{sec:ngoccu_critical}.

\subsubsection{Occupation number entropy in gapped phases}
\label{sec:ngoccu_gapped_main}

In addition to translation invariance and locality, many physically relevant systems additionally possess a finite spectral gap. A key consequence of the gap is the exponential decay of correlations in the bulk for the ground states~\cite{hastings2006spectralgap, nachtergaele2006liebrobinson}, which implies that the real-space covariance matrix elements (for systems preserving fermionic parity) decay as $\sim e^{-|\ell|/\xi}$, where $\xi$ is the correlation length.

As a consequence, we can improve the approximation error of the finite-range approximation of $m_{(\infty)}^{[\alpha]}$ above. In particular, if we choose a cutoff $\cutoff\gg\xi$ to only retain correlators with distances $|\ell|\leq\cutoff$ in the Fourier transform in Eq.~\eqref{eq:ti_momentum_inf}, this leads to an approximation error for $m_{(\infty)}^{[\alpha]}$ that is bounded by $\mathcal{O}(e^{-\cutoff/\xi})$. This holds under the same assumptions on the spectrum of $H(k)$ as for the algebraic decay. The detailed derivation is presented in Appendix~\ref{sec:ngoccu_approx_infinite}.

Similarly, we can consider how fast finite-size values of the occupation number entropy density $\nGoccu^{[\alpha]}/N$ in a system of size $L$ converge to their thermodynamic-limit value $m_{(\infty)}^{[\alpha]}$. To establish that these also converge exponentially fast, we need to make an additional assumption on how fast the two-point correlators in a finite system converge to their value in the thermodynamic limit. In gapped systems the correlations decay exponentially fast and we expect that distant parts do not influence local quantities~\cite{hastings2006spectralgap, nachtergaele2006liebrobinson}. For systems much larger than the correlation length $L\gg2\xi$, we therefore expect local operators, such as two-point correlators with distance $\ell\leq\xi$, to be already converged up to exponentially small corrections. Conversely, one expects that the two-point correlators with distance $\ell>\xi$ are exponentially small in a finite system. Under this assumption, after taking the Fourier transform of the finite-system covariance matrix, its blocks are exponentially close in system size to those of the infinite system in Eq.~\eqref{eq:ti_momentum_inf} with the same momentum. Thus, up to those exponentially small corrections, the finite-size $\nGoccu^{[\alpha]}/N$ are given by Eqs.~\eqref{eq:ti_ngoccu_inf_main} and~\eqref{eq:ti_ngoccu_vne_inf_main} with the integral replaced by sums over the finite-size momenta. We can interpret the sums over the finite-size momenta as approximating the integral by a sum of $L$ discrete slices, i.e., by the trapezoidal rule. The error of such an approximation has been shown to decrease exponentially with $L$ if the integrand is a periodic analytic function integrated over one period~\cite{trefethen2014exponentiallyconvergent}.
Since the correlators decay exponentially fast, the momentum blocks of the infinite-system covariance matrix given in Eq.~\eqref{eq:ti_momentum_inf} are an analytic function of the momentum. This carries over to the integrands of Eqs.~\eqref{eq:ti_ngoccu_inf_main} and~\eqref{eq:ti_ngoccu_vne_inf_main}, if $\alpha>1$ and an integer or if no Gaussian modes remain in the covariance matrix. In this case, we can show that
\begin{equation}
    \frac{\nGoccu^{[\alpha]}(N)}{N}
    = m_{(\infty)}^{[\alpha]}
    + \mathcal{O}(e^{-\delta N}),
\end{equation}
where $\delta>0$ is a constant independent of $N$. %
All these statements are shown in more detail in Appendix~\ref{sec:ngoccu_gapped}.

This establishes that, in gapped, translation-invariant systems, the occupation number entropy density is a robust bulk quantity that can be
reliably extracted either from finite-range correlators in an infinite system or from  finite-size scaling in finite systems.

\subsubsection{Translation-invariant MPSs} \label{sec:ti_mps_main}

In this section, we explain how the occupation number entropies can be evaluated in TI MPSs using the momentum-space formalism developed above. TI MPSs are a physically important and versatile class: they faithfully and efficiently represent the ground states of gapped, translation-invariant local Hamiltonians in one dimension~\cite{Verstraete2006, Cirac2021}. Beyond their physical relevance, a key feature in the present context is that the momentum-space covariance matrix blocks $\tilde{t}(p)$ can be obtained analytically without introducing an approximation due to a real-space cutoff---in contrast to the cases we discussed before. This is because all covariance matrix elements can be obtained in closed form from the MPS transfer matrix as a finite sum of elementary functions. In practice, this allows the evaluation of the occupation number entropies by numerical integration of Eqs.~\eqref{eq:ti_ngoccu_inf_main} and~\eqref{eq:ti_ngoccu_vne_inf_main}, or by analytical integration if their structure remains simple enough.

We consider a TI MPS defined on $N$ qubits with bond dimension $\chi$~\cite{Fannes1992, Cirac2021}:
\begin{equation} \label{eq:ti_mps}
    \ket{\psi}
    = \! \sum_{\{s_j\}_{j=1}^{N}} \!
    \Tr[A_{s_1} A_{s_2} \cdots A_{s_N}] \,
    \ket{s_1, s_2, \dots, s_N}
\end{equation}
with $A_{s}$ being a $\chi \times \chi$ matrix and $s_j \in \{0,1\}$. We assume that the MPS is injective, i.e., that the matrices $A_s$ do not have a non-trivial invariant subspace~\cite{cirac2017mpdo}. This also implies that the transfer matrix has a unique dominant eigenvalue. We work in the thermodynamic limit $N\to\infty$; the formalism extends straightforwardly to finite systems and larger unit cells.

We further impose that the MPS has a definite parity, which implies the existence of a unitary matrix $U_P \in U(\chi)$ such that~\cite{perez-garcia2007mps}
\begin{equation} \label{eq:local_sym_mps}
    (-1)^s A_s = e^{i \phi_P} U_P A_s U_P^\dagger,
\end{equation}
with $\phi_P=0$ or $\phi_P=\pi$ and $U_P^2= \Id_\chi$~\cite{pollmann2012spt,chen2011classification}. Note that, assuming even parity, the periodic boundary conditions imposed here correspond to antiperiodic boundary conditions in the fermionic representation under the Jordan--Wigner transformation.

 The transfer matrix $E = \sum_s A_s \otimes \overline{A}_s$ and the mixed transfer matrices $E_O = \sum_{s, s'} O_{s,s'} A_s \otimes \overline{A}_{s'}$ (for $O\in\{X,Y,Z\}$) act on the virtual space $\mathcal{H}_V \simeq\mathbb{C}^{\chi^2}$ and standard MPS calculations yield all required correlators as $\langle Z_j \rangle = \brap{L_1} E_Z \ketp{R_1}$, $\langle O_j O'_{j+1} \rangle = \brap{L_1} E_O E_{O'} \ketp{R_1}$, and
\begin{equation}
    \langle O_j Z_{j+1} \cdots Z_{j+\ell-1} O'_{j+\ell} \rangle
    = \brap{L_1} E_O \, E_Z^{\ell-1} \, E_{O'} \ketp{R_1},
\end{equation}
with $O,O'\in\{X,Y\}$, where $\brap{L_j}$ and $\ketp{R_j}$ are the left and right eigenvectors of $E$ (eigenvalues $|\lambda_1|\geq|\lambda_2|\geq\cdots$, with the dominant eigenvalue normalized to $\lambda_1=1$ and $\braketp{L_j}{R_j}=1$ as normalization).

What is crucial to the present setting is the $\mathbb{Z}_2$ parity symmetry. The relation in Eq.~\eqref{eq:local_sym_mps} renders $E$ invariant under $\mathcal{U} = U_P \otimes \overline{U}_P$ (with $\mathcal{U}^2=\Id_{\chi^2}$), splitting the virtual space into parity sectors $\mathcal{H}_V = \mathcal{H}_V^0 \oplus \mathcal{H}_V^1$, so that each eigenvector $\ketp{R_j}$ carries a definite parity $\nu_j\in\{0,1\}$. The Pauli-$Z$ transfer matrix is likewise invariant, $E_Z = e^{i \phi_P} (U_P \otimes \Id_\chi) \, E \, (U_P^\dagger \otimes \Id_\chi)$, with spectrum $\lambda_j^Z = e^{i \phi_P}\lambda_j$ and eigenvectors $\brap{L_j^Z} = \brap{L_j} U_P \otimes \Id_\chi$, $\ketp{R_j^Z} = U_P^\dagger \otimes \Id_\chi \ketp{R_j}$ of unchanged parity $\nu_j$, whereas $E_X$ and $E_Y$ are \emph{charged} under $\mathcal{U}$. Expanding $E_Z^{\ell-1}$ in its eigenbasis therefore gives
\begin{equation}
\begin{aligned}
    &\brap{L_1} E_O \, E_Z^{\ell-1} E_{O'} \ketp{R_1}
    = e^{i \phi_P(\ell-1)} \sum_j \delta_{\nu_1,1-\nu_j} \, \lambda_j^{\ell-1}
    \\[-0.15cm]
    &\hspace{8.5em}\times\brap{L_1} E_O \ketp{R_j^Z} \brap{L_j^Z} E_{O'} \ketp{R_1}.
\end{aligned}
\end{equation}
The parity selection rule $\nu_j\neq\nu_1$ is the crucial feature here, as it excludes the dominant eigenvalue $\lambda_1$ from the sum, so the string correlators decay with $\ell$ and the resulting geometric series in momentum space converge.

Consequently, the real-space correlators $t(\ell)$ take the form of a finite sum of geometric series, each of which can be Fourier-transformed over $\ell$ in closed form. The momentum-space covariance matrix $\tilde{t}(p)$ is thereby obtained exactly, and the occupation number entropy density follows from the remaining one-dimensional momentum integral, which can be evaluated numerically. The details are given in Appendix~\ref{sec:ti_mps}; the dominant cost is the $\mathcal{O}(\chi^6)$ diagonalization of the transfer matrix.

As an alternative discussed previously and shown in Appendix~\ref{sec:ngoccu_approx_infinite}, the occupation number entropy can also be approximated from a finite range of correlations: TI MPSs are gapped, with exponentially decaying correlations~\cite{Cirac2021}, so computing the elements $t(\ell)$ for $|\ell|\leq\cutoff$ cost $\mathcal{O}(\cutoff\chi^3)$, and choosing $\cutoff\gg\xi=-1/\ln|\lambda_2|$ yields the occupation number entropy up to a correction exponentially small in $\cutoff$.

An example of explicit calculation for the MPS skeleton~\cite{skeleton2021, PhysRevLett.97.110403, PhysRevResearch.4.L022020} can be found in Appendix~\ref{sec:ti_mps_example}.

\section{Examples}
\label{sec: example}

In this section, we consider a range of representative examples and examine the behavior of the introduced quantifiers of fermionic non-Gaussianity. We begin with a simple, controlled example that illustrates that the natural-orbital participation entropy $\nGNO^{[\alpha]}$ works as anticipated: low-rank superpositions of a few orthonormal Gaussian states. $\nGNO^{[\alpha]}$ faithfully reveals the underlying low-rank Gaussian structure, whereas the occupation number entropy $\nGoccu^{[\alpha]}$ instead bounds the non-Gaussian gate count, so the two quantities exhibit distinct scaling behavior by construction. This example sets the stage for the two subsequent more physically relevant settings, both in and out-of-equilibrium.
First, we analyze SWAP-doped matchgate circuits, where occupation number entropies quantify the growth of non-Gaussianity.
Second, we consider interacting many-body ground states, showing that the two families can capture phase transitions in many-body systems. We also verify the statements of the previous section that the occupation number entropy converges exponentially fast with system size in gapped phases.

\subsection{Superposition of few orthogonal FGSs}
\label{subsec:superposition_FGS}

As a first example, we consider coherent superpositions of a small number of mutually orthogonal FGSs, which we refer to as low-rank states.
For this class of states, the two families introduced above exhibit distinct scaling behaviors by construction.
As a result, this setting provides a simple but transparent illustration of the differences between the two families.
In particular, we see that $\nGNO^{[\alpha]}$ is sensitive to the low-rank structure if the underlying state is a low-rank superposition of orthogonal FGSs.

\subsubsection{Rank-2 orthogonal FGSs}

\begin{figure}
    \centering
    \includegraphics[width=\linewidth]{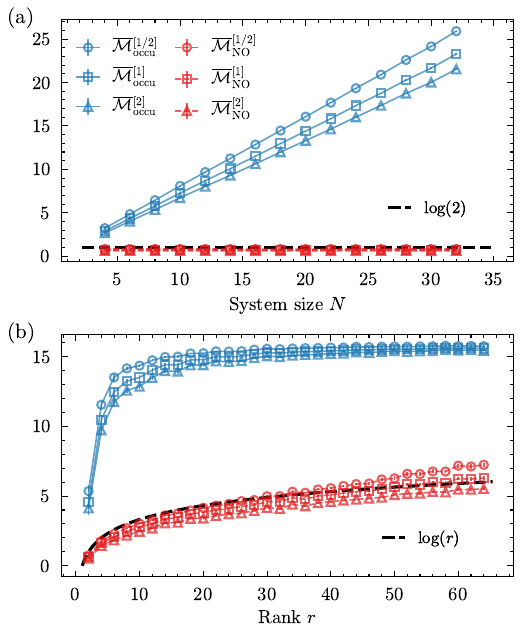}
    \caption{
    Comparison of the occupation number entropy $\nGoccu^{[\alpha]}$ (blue) and the natural-orbital participation entropy $\nGNO^{[\alpha]}$ (red) for $\alpha=1/2$ (circles), $\alpha=1$ (squares), and $\alpha=2$ (triangles) for random low-rank FGSs. Each data point shows the average of $10000$ realizations.
    (a) The non-Gaussianity of rank-2 superpositions of orthogonal FGSs against different system sizes $N$.
    The black dashed line indicates the upper bound $\log_2(2) = 1$ of the participation entropy of a rank-2 state.
    (b) The non-Gaussianity of rank-$r$ superpositions of orthogonal FGSs in a system of size $N=16$.
    The black dashed line indicates the upper bound $\log_2(r)$ of the participation entropy of a rank-$r$ state.
    In both scenarios, the natural-orbital participation entropy is able to correctly reveal the rank of the underlying state.
    }
    \label{fig: low-rank}
\end{figure}

We first consider rank-2 superpositions of two orthogonal FGSs, specifically two computational basis states with opposite occupation number,
\begin{equation}
    \ket{\psi} = c_1 \ket{0}^{\otimes N} + c_2 \ket{1}^{\otimes N} 
    \label{eq:rank2_GHZ_like}
\end{equation}
with $|c_1|^2 + |c_2|^2=1$. Note that taking the components to be computational-basis configurations is without loss of generality, since any other Gaussian basis can be brought to the computational basis by an FGU.
The real and imaginary parts of the random complex numbers $c_1, c_2$ are independently drawn from a standard normal distribution centered around zero and then normalized.
Equivalently, the coefficients $(c_1,c_2)$ may be viewed as a random state in the two-dimensional subspace spanned by
$\ket{0}^{\otimes N}$ and $\ket{1}^{\otimes N}$.
We restrict to even system sizes so that the fermionic parity constraint is satisfied and also consider only the system sizes $N\geq 4$ when non-Gaussian states appear~\cite{bravyi2005classical,hebenstreit2019all}.
For each system size, we draw $10000$ independent realizations and compute the average $\avgnGoccu^{[\alpha]}$ and $\avgnGNO^{[\alpha]}$.
The results are shown in Fig.~\ref{fig: low-rank}a.

We observe that $\avgnGoccu^{[\alpha]}$ grows linearly with system size, whereas $\avgnGNO^{[\alpha]}$ remains bounded above by $\log_2(2)=1$.
The crucial point to understand this behavior is to know that the state in Eq.~\eqref{eq:rank2_GHZ_like} is already expressed in its natural-orbital basis.
Indeed, for $N\geq 4$, the covariance matrix of
$c_1 \ket{0}^{\otimes N}+c_2 \ket{1}^{\otimes N}$
is already in the normal form in the original computational basis.
The corresponding occupation numbers are $N$-fold degenerate,
\begin{equation}
    \lambda_i = |c_2|^2,
    \qquad
    1-\lambda_i = |c_1|^2,
    \qquad
    i=1,\ldots,N .
\end{equation}
Therefore, the occupation number entropy is extensive, $\nGoccu^{[\alpha]} \sim \mathcal{O}(N)$,
whenever both amplitudes have nonzero weight.

On the other hand, the state has exactly two nonzero components in the natural-orbital basis.
Consequently, $\nGNO^{[\alpha]}$ depends only on the probability distribution
\begin{equation}
    (p_1,p_2) = \bigl(|c_1|^2, |c_2|^2\bigr),
\end{equation}
and is bounded by
\begin{equation}
    \nGNO^{[\alpha]} \leq \log_2 2 = 1 .
\end{equation}
This explains the separation observed in Fig.~\ref{fig: low-rank}a:
$\nGoccu^{[\alpha]}$ grows linearly with system size, while $\nGNO^{[\alpha]}$ remains bounded.
The bounded value of $\nGNO^{[\alpha]}$ reflects the fact that the GHZ-like state in Eq.~\eqref{eq:rank2_GHZ_like} is already a rank-2 state in the natural-orbital basis.

\subsubsection{Rank-\texorpdfstring{$r$}{r} orthogonal FGSs}

We next consider whether $\nGNO^{[\alpha]}$ continues to diagnose low-rank Gaussian structure when the number of components is increased.
We construct random rank-$r$ superpositions of orthogonal FGSs directly in the computational occupation basis,
\begin{equation}
    \ket{\psi}
    =
    \sum_{a=1}^{r} c_a \ket{\mathbf{n}_a},
    \label{eq:rankr_orthogonal_FGS}
\end{equation}
where each $\ket{\mathbf{n}_a}$ is an occupation-number product state with even fermionic parity, and
\begin{equation}
    \braket{\mathbf{n}_a}{\mathbf{n}_b}=\delta_{ab}.
\end{equation}
The coefficients satisfy $\sum_a |c_a|^2=1$.
In practice, the configurations $\{\mathbf{n}_a\}_{a=1}^r$ are drawn randomly with equal probability from the even-parity occupation basis, and the complex coefficients are drawn randomly from standard normal distribution and then normalized.

This construction differs from the rank-2 example in an important way.
For small $r<N$, the randomly selected occupation configurations are typically well separated.
More precisely, pairs of occupation configurations often have Hamming distance larger than four.
In this regime, the covariance matrix typically receives no cross-configuration contributions and remains in normal form, and the computational occupation basis is the natural-orbital basis.
As a result, the state in Eq.~\eqref{eq:rankr_orthogonal_FGS} is effectively a rank-$r$ state in the natural-orbital basis.
One therefore expects
\begin{equation}
    \nGNO^{[\alpha]} \leq \log_2 r ,
\end{equation}
with the precise value determined by the probability distribution $\{|c_a|^2\}$.

As $r$ is increased, this simple picture is no longer expected to hold.
The set of sampled configurations becomes denser in occupation space, and pairs of configurations with Hamming distance smaller than or equal to four become increasingly likely to appear.
Such nearby configurations can take the covariance matrix out of the normal form.
Consequently, the natural-orbital basis need no longer coincide with the original computational occupation basis.
After transforming to the natural-orbital basis, a state that was initialized with only $r$ computational-basis components can acquire more than $r$ nonzero components.
In this regime, $\nGNO^{[\alpha]}$ can exceed the naive $\log_2 r$ line because the rank in the natural-orbital basis can become larger than the rank in the computational basis.

We consider a fixed system size $N=16$ and vary the initial rank from $r=2$ to $r=4N$.
The results are shown in Fig.~\ref{fig: low-rank}b.
For small $r$, the average $\avgnGNO^{[\alpha]}$ lies below, or close to, the reference curve $\log_2 r$, consistent with the sampled configurations typically being separated by large Hamming distances, so that the covariance matrix remains in normal form and the computational basis is already the natural-orbital basis.
For larger $r$, the sampled configurations increasingly overlap in the sense described above, in which case the covariance matrix can depart from normal form and the transformation to the natural-orbital basis can spread the state over more Gaussian basis states.
This accounts for the observed small deviation from the $\log_2 r$ bound for the expansion in the computational basis, and provides an explicit example in which the natural-orbital basis is not the optimal orthonormal Gaussian basis. Crucially, however, the overshoot remains small: $\avgnGNO^{[\alpha]}$ stays close to $\log_2 r$ across the entire range. This is also the practically relevant regime---states of interest are typically not sparse in any a priori known basis, so the optimal Gaussian basis is unknown---and it shows that, although generally not optimal, the natural-orbital basis remains a compact, near-optimal basis for representing the state.
On the other hand, we observe that the values of the average $\avgnGoccu^{[\alpha]}$ quickly saturate to their maximal values at small $r$, indicating the distinct behavior of the two families.

\subsubsection{Remarks}

The examples above demonstrate a systematic difference between $\nGoccu^{[\alpha]}$ and $\nGNO^{[\alpha]}$. This reflects an intrinsic difference in what the two quantities probe.
On one hand, we observe that $\nGoccu^{[1]} \sim N$ has a linear scaling in the system size in the rank-2 GHZ-like example. In Sec.~\ref{subsec:nG_occu}, we show $\nGoccu^{[1]}$ provides a lower bound on the number of SWAP gates required for state preparation. The rank-2 states studied here exhibit a GHZ-type structure, which requires $\mathcal{O}(N)$ SWAP gates to prepare. Therefore, the observation is consistent and further indicates $\nGoccu^{[1]}$ captures closely the \emph{SWAP gate complexity}.

On the other hand, $\nGNO^{[\alpha]}$ is sensitive to a different structural property: the sparsity of the wavefunction in the natural-orbital basis.
For the rank-2 example, the state has exactly two components and is in the natural-orbital basis, so $\nGNO^{[\alpha]}$ remains bounded.
For the rank-$r$ example, the state has $r$ or close to $r$ components in the natural-orbital basis as the initialization basis is close to the natural-orbital basis.
When the covariance matrix is taken out of the normal form, the transformation to the natural-orbital basis can increase the rank, and $\nGNO^{[\alpha]}$ correspondingly grows slightly beyond $\log{(r)}$.

This feature may be particularly relevant in quantum chemistry~\cite{RevModPhys.92.015003}, where the distinction between single-reference and multi-reference states plays a central role. Single-reference methods, such as coupled-cluster approaches built on a single Slater determinant as a reference frame, perform well when the wavefunction is dominated by one Gaussian component. In contrast, strongly correlated (multi-reference) systems require coherent superpositions of several determinants with sizable amplitudes.
In this context, $\nGNO^{[\alpha]}$ quantifies the effective number of Gaussian components and may therefore serve as an indicator of multi-reference character~\cite{liao2024quantum}.
This distinction between $\nGoccu^{[\alpha]}$ and $\nGNO^{[\alpha]}$ therefore highlights their complementary utility, with each quantity being well-suited to probe different notions of complexity of quantum states.

We conclude with two caveats.
First, the exact GHZ state represents a fine-tuned point at which the natural-orbital basis becomes ill-defined due to the large degeneracy. At this singular point, evaluating $\nGNO^{[\alpha]}$ requires the additional full optimization over all natural-orbital bases, which for this case corresponds to all Gaussian bases and thus reduces to $\nGOG^{[\alpha]}$. This singular behavior does not diminish the utility of the proposed quantities, as an arbitrarily small $\epsilon$-perturbation away from the exact GHZ point lifts the degeneracy and restores the standard behavior of $\nGNO^{[\alpha]}$.
Second, a more substantial open issue arises for non-orthonormal low-rank states. Numerically, we observe that random non-orthonormal rank-2 states, when expanded in an orthonormal Gaussian basis, require a rank that grows exponentially with the system size.
At present, we lack a deeper theoretical understanding of this apparent rank proliferation.
Clarifying its origin and determining whether it reflects a generic geometric feature of non-orthogonal states remains an important direction for future work.

\subsection{SWAP-doped circuits} \label{subsec:example_swap_doped}

In this section, we study the behavior of non-Gaussianity in random matchgate circuits doped with SWAP gates.
Random circuits are widely used as minimal models of unitary time evolution and thermalization in chaotic quantum many-body systems, as they capture generic scrambling behavior without relying on microscopic structure~\cite{Fisher2023RandomCircuits}. 
Thus, our setup can be regarded as a noninteracting background, modeled by the random matchgate circuit, with inserted SWAP gates playing the role of interacting impurities embedded in an otherwise free system.

\begin{figure}[h!]
    \centering
    \includegraphics{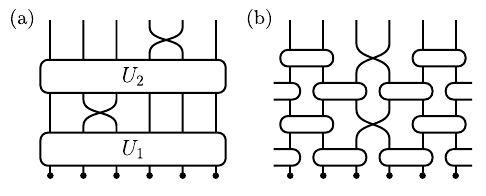}
    \caption{Schematic evolution of an $N$-qubit system where every time step consists of (a) a global random FGU followed by a SWAP operation at a random position, and (b) one brickwall layer of local matchgates, with the central one being replaced by a SWAP gate. The number of time steps $t$ in both cases corresponds to the total number of SWAP gates applied.}
    \label{fig:circuit_evolution}
\end{figure}

The complexity of simulating the evolution of such systems for large system sizes grows exponentially with the number $t$ of SWAP gates applied to the system~\cite{HeJo20, hakkaku2022, mocherla2024extendingmatchgatesimulationmethods, miller2025simulationfermioniccircuitsusing, dias2024classical, ReardonSmith2024improved}.
We consider two distinct setups, illustrated in Fig.~\ref{fig:circuit_evolution}. In the first setup, shown in Fig.~\ref{fig:circuit_evolution}a, each time step consists of applying a random FGU to the entire system, followed by a SWAP gate. While this global setup is less physical than a local circuit, it serves as an analytically tractable model which provides a first, illustrative example of how non-Gaussianity grows in random circuits. The second setup, depicted in Fig.~\ref{fig:circuit_evolution}b, involves a brickwall evolution composed of matchgates interspersed with SWAP gates. Local brickwall circuits of this kind are a standard minimal model of local many-body dynamics representing Hamiltonian evolution with time-dependent noise, with a local source of impurity~\cite{thoenniss2023nonequilibrium, yahui2025dynamics, Bravyi17impurity}.
Here, we focus on the behavior of the occupation number entropy, which we can characterize analytically and numerically utilizing Clifford and state-vector simulations. For the natural-orbital participation entropy, the achievable system sizes with state-vector simulations are insufficient to determine a scaling behavior. 

There are two limits of the occupation number entropies $\nGoccu^{[\alpha]}$ that are simple to study in our model.
For $\alpha=0$ (or equivalently $\alpha\to\infty$), the occupation number entropy corresponds to the Gaussian nullity, i.e., the number of Williamson eigenvalues that are not 0 or 1.
At each step, a SWAP gate acts nontrivially in at most four Gaussian modes, as described in Ref.~\cite{MeHe25}.
Since the random evolution with matchgates mixes the modes at every single time step, the nullity grows maximally at each time step almost surely, until reaching saturation at the value $N$:
\begin{equation}
    \nGoccu^{[0]}(t) = \nu_{\mathcal{G}}(t) = 4t.
\end{equation}
Note that this holds for both circuit layouts considered.

Secondly, we study in detail the case $\alpha=2$, where the occupation number entropy is quadratic in the Williamson eigenvalues.
Since we are interested in the average behavior, we can simulate the system with braiding gates, i.e., those unitaries that are both Clifford and matchgates.
The average coincides given that braiding gates form a matchgate 3-design~\cite{wan2023matchgateshadow}, implying that averaged quadratic quantities such as $\nGoccu^{[2]}$ coincide with those obtained for generic, randomly sampled FGUs.
Furthermore, since the SWAP gate is a Clifford unitary, the simulations can be performed efficiently using the stabilizer formalism~\cite{Go97, aaronson2004improved}, as discussed in Sec.~\ref{sec:pure_stab}.

\begin{figure}
    \centering
    \includegraphics{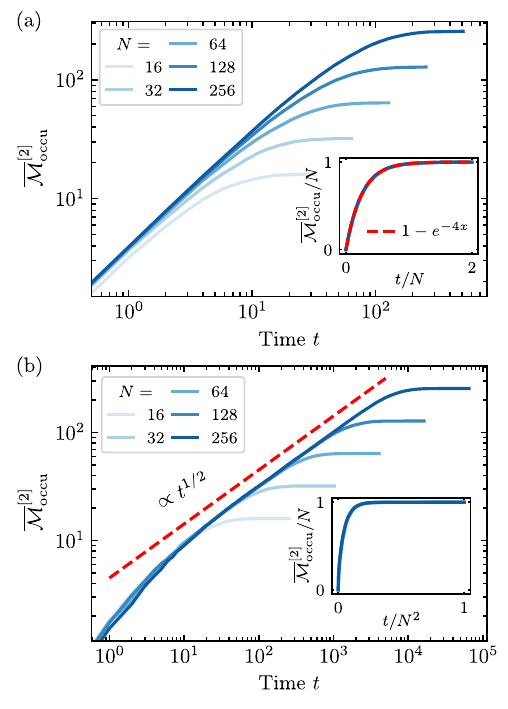}
    \caption{Time evolution of the averaged occupation number entropy $\avgnGoccu^{[2]}$ in fermionic Gaussian circuits doped with SWAP gates, as depicted in Fig.~\ref{fig:circuit_evolution}. Each quantity is averaged over 100 realizations of the circuit. (a)~Global unitary evolution as a function of system size $N$. The inset shows the collapse of the curve when rescaling with system size. (b)~Evolution from the brickwall circuit evolution, growing diffusively with the number of SWAP gates, $\propto t^{1/2}$. The inset shows the collapse when rescaling time by $N^2$. Note that the insets in both~(a) and~(b) include results for all system sizes considered, but the curves coincide and therefore appear as a single line due to the data collapse.}
    \label{fig: SWAP doped nG}
\end{figure}

The numerical results of the evolution for global random FGUs doped with SWAP gates are shown in Fig.~\ref{fig: SWAP doped nG}a, with each evolution averaged over $100$ distinct realizations.
We observe that it grows rapidly and saturates to $\avgnGoccu^{[2]}\rightarrow N$ in the long time limit (see inset).
As the dashed-red line in the inset shows, the dynamics are described by the equation
\begin{equation}
    \label{eq:evolution_swapdoped}
    \avgnGoccu^{[2]}(t)=N\left(1-\exp\!\left(-\frac{4t}{N}\right)\right).
\end{equation}
This behavior can be analytically derived by analyzing how braiding and SWAP gates modify the stabilizers of the state, which we approximately map to a classical destruction model in the large-$N$ limit. We present the details in Appendix~\ref{app:evolution occu doped}. Note that this result coincides with the average nullity in the random braiding circuits.

Now we turn to the brickwall evolution.
The numerical results are shown in Fig.~\ref{fig: SWAP doped nG}b.
As in the previous case, the occupation number entropy saturates to $N$ for long enough times.
However, here the thermalization is slower: it grows diffusively with time, $\avgnGoccu^{[2]}(t)\propto t^{1/2}$, consistent with the diffusive transport of random matchgate circuits without impurities~\cite{dias2021diffusiveoperatorspreadingrandom}. 
The thermalization time is, therefore, proportional to $N^2$. 
This is similar to the behavior of entanglement in the same setup on the absence of impurities~\cite{Nahum_2017, nahum_freefermions25}.
Further details on the entanglement dynamics in doped braiding circuits can be found in Ref.~\cite{PaLu26}.

Finally, we turn to the evolution of other quantities.
For that, we resort to simulating the exact evolution of the wavefunction, hence limiting the system sizes we can reach.
The occupation number entropies for different values of $\alpha$ look similar to that observed for $\alpha=2$: an exponential convergence to the steady-state value for the global evolution and a diffusive growth in the brickwall evolution---so we do not show this data.
For the natural-orbital participation entropy, instead, the system sizes simulable with exact state-vector methods are not enough to determine a scaling behavior in the two setups. Nevertheless, when the circuit is restricted to braiding gates (so that the state remains a stabilizer state), an upper bound on $\nGNO^{\text{S-FGS}}$ can be evaluated at much larger sizes via the stochastic optimization presented in Appendix~\ref{app:nGNO approximate stabilizers}.
The analysis of the behavior of such quantities is beyond the scope of this paper.

In summary, this example demonstrates the role of locality in the generation of non-Gaussianity resources. It grows monotonically with the number of SWAP gates and saturates near its maximal value, with a rate set by the circuit architecture: it saturates on timescale linear with system size for the global evolution, whereas in the local brickwall circuit the growth is limited by diffusion ($\sim t^{1/2}$), so that the state becomes maximally complex only on the longer timescale $t\sim N^2$. Relatedly, this locality-controlled buildup of non-Gaussianity governs the emergence of unitary designs in doped matchgate circuits, where local architectures likewise reach designs more slowly through the same diffusive mechanism~\cite{trigueros2026unitarydesignsdopedmatchgate}. This highlights the role of $\nGoccu^{[\alpha]}$ as a resource monotone that tracks the non-Gaussian resources generated along a circuit, thereby capturing the buildup of complexity in this minimal model of doped free-fermion dynamics.

\subsection{Bond-modulated XXZ model}

An important class of translation-invariant fermionic systems are given by interacting fermionic many-body systems. Here we want to investigate how our proposed quantities behave in such a setting. In particular, we want to see whether they capture the intuitive notion that critical ground states of quantum phase transitions should be more complex than those in gapped phases. While we generically expect the quantifiers of fermionic non-Gaussianity to be extensive due to the extensive number of interaction terms present in the Hamiltonian, their density can still show distinct behavior. Specifically, in Sec.~\ref{sec:TI}, we have argued that the subleading term of the occupation number entropy density should vanish (exponentially) within gapped phases, which we investigate numerically here.

As a concrete example of a many-body system, we consider the one-dimensional XXZ model with a bond modulation. The Hamiltonian is given by
\begin{equation}
    H \hspace{-0.14em}=\hspace{-0.14em} \sum_{j=1}^N \hspace{-0.14em} {\textstyle\frac{1+(-1)^j\gamma}{2}} \hspace{-0.14em}
    \left( X_jX_{j+1} \hspace{-0.14em}+\hspace{-0.14em} Y_jY_{j+1} \hspace{-0.14em}+\hspace{-0.14em} {\textstyle\frac{1+\gamma}{2}} Z_jZ_{j+1} \right)\hspace{-0.14em},
    \label{eq:XXZ_Hamiltonian}
\end{equation}
and we choose periodic boundary conditions. Under a Jordan--Wigner transformation, the term $\frac{1}{2}(X_jX_{j+1} + Y_jY_{j+1})$ corresponds to a fermionic nearest-neighbor hopping, while the term $Z_jZ_{j+1}$ maps to a repulsive nearest-neighbor density-density interaction. In the following, we will be interested in ground state properties of this model. As the bond-modulation strength $\gamma\in[-1,+1]$ is tuned, the ground state interpolates from singlets on odd bonds at $\gamma=-1$ to singlets on even bonds at $\gamma=+1$. Even though the interaction strength in the Hamiltonian increases monotonically with increasing $\gamma$, the two end-points $\gamma=\pm1$ thus correspond to Gaussian states. For $\gamma<0$, the ground state is in a symmetric and topologically trivial phase and transitions to a distinct symmetry-protected topological (SPT) phase for $\gamma>0$, with a quantum phase transition at the translation-invariant point $\gamma=0$~\cite{Verresen2017}.
The critical point at $\gamma=0$ is the usual XXZ model, which describes a Luttinger liquid~\cite{Giamarchi2004} and has central charge $c=1$~\cite{DiFrancesco1997}.

The model has both a $\mathbb{Z}_2$ symmetry consisting of $\pi$-rotations around the $x$-axis and a $U(1)$ symmetry consisting of arbitrary rotations around the $z$-axis, which in general do not commute.
The unitary representations $U_Q$ and $U_R$ of both symmetries can be written as FGUs, and thus have associated orthogonal matrices $Q$ and $R$ acting on Majorana modes---cf. Sec.~\ref{app:review_fgs}. Since the ground state preserves both symmetries, its covariance matrix $\Gamma$ will commute with both $Q$ and $R$, meaning we can simultaneously diagonalize $\Gamma$ and $Q$, or $\Gamma$ and $R$. If there are no degeneracies in the spectrum of $\Gamma$, its eigenbasis is unique, and both $Q$ and $R$ are diagonal in that basis. However, since $Q$ and $R$ do not mutually commute, they cannot be simultaneously diagonalized, which implies that $\Gamma$ must have some degeneracies in its spectrum to allow a nonunique eigenbasis. Similarly, the model has both a two-site translation and lattice inversion symmetry, that do not mutually commute and therefore also enforce degeneracies in the covariance matrix spectrum. Thus, to select a specific natural-orbital basis to compute the natural-orbital participation entropy in the following, we bring both the covariance matrix and simultaneously two commuting symmetries into (block-diagonal) normal form, with the normal form defined as in Eq.~\eqref{eq:normal_form} in Sec.~\ref{app:review_fgs}. We can choose one of either the $U(1)$ $z$-axis rotation or the $\mathbb{Z}_2$ $x$-axis $\pi$-rotation symmetry, and one of either the translation or the inversion symmetry. This means the natural-orbital basis inherits the symmetries brought into normal form (if all eigenvalues are nonzero)~\footnote{%
    In normal form, the covariance matrix is block diagonal $\Gamma = \bigoplus_j \alpha_j \, iY$ with general $\alpha_j\geq0$. If all $\alpha_j>0$, then also the symmetry is block diagonal $Q = \bigoplus_j \exp(-i\theta_j \, Y_j)$. In that case, the covariance matrices of the Gaussian basis states are just obtained by setting $\alpha_j=\pm1$ in the block-diagonal covariance matrix, and they all still commute with the symmetry $Q$. Thus we see that the Gaussian basis defined by simultaneously bringing $\Gamma$ and $Q$ into normal form has basis states symmetric under $Q$.},
such that when the original state is written in this Gaussian basis, it excludes certain basis states based on their symmetry eigenvalues.
In practice, we then take the minimal participation entropy over these four distinct combinations of symmetries, which most often leads to the combination of $U(1)$ and translation symmetry. We found this to be more successful than a numerical optimization of the participation entropy within the degenerate blocks.

\begin{figure}
    \centering
    \includegraphics[width=\linewidth]{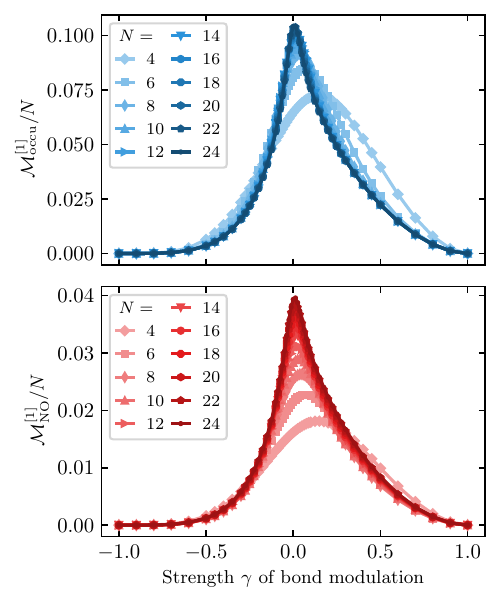}
    \caption{Occupation number entropy density $\nGoccu^{[1]}/N$ (top panel) and an upper bound on the natural-orbital participation entropy density $\nGNO^{[1]}/N$ (bottom panel) in the bond-modulated XXZ model [see Eq.~\eqref{eq:XXZ_Hamiltonian}] plotted against the strength $\gamma$ of the bond modulation. The two end-points $\gamma=\pm1$ correspond to Gaussian states where both quantities go to zero, and both peak at the critical point $\gamma=0$ without bond modulation. Instead of the costly direct minimization of the natural-orbital participation entropy within the degenerate subspaces, we compute it by taking the minimimum over the four bases preserving some of the symmetries (see text for details); we found this to yield smaller values than a direct brute-force optimization.}
    \label{fig:XXZ_nG}
\end{figure}

Fig.~\ref{fig:XXZ_nG} shows the results for the occupation number entropy density $\nGoccu^{[1]}/N$ (top panel) and the natural-orbital participation entropy density $\nGNO^{[1]}/N$ (bottom panel) as a function of $\gamma$ for different system sizes $N$. Starting from the Gaussian state at $\gamma=-1$, both quantities increase as we increase $\gamma$ up to the critical point at $\gamma=0$. From thereon, even though the strength of the interaction term in the Hamiltonian continues to grow, both quantities decrease as the ground state approaches another Gaussian state at $\gamma=+1$. With increasing system size, the curves begin to fall on top of each other, suggesting that they converge to the occupation number entropy density or natural-orbital participation entropy density, respectively, of the state in the thermodynamic limit. This convergence is slowest around the critical point, and seems generally slower for the natural-orbital participation entropy than the occupation number entropy.

\begin{figure}
    \centering
    \includegraphics[width=\linewidth]{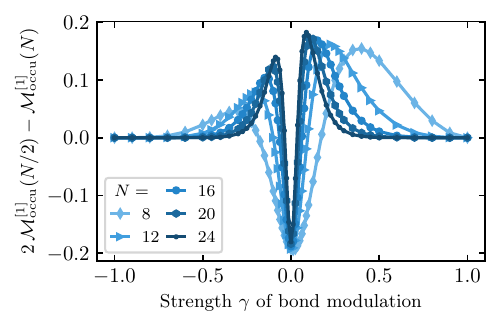}
    \caption{The subleading term of the occupation number entropy in the bond-modulated XXZ model [see Eq.~\eqref{eq:XXZ_Hamiltonian}] plotted against the strength $\gamma$ of the bond modulation. It is extracted by subtracting the occupation number entropy at a given system size $N$ from twice the value at half the system size. In the gapped phases where the correlation length is smaller than the system size it goes to zero, and there are positive finite-size corrections around the critical point where the correlation length is larger than the system. At the critical point, the subleading contribution is negative and mostly independent of system size.}
    \label{fig:XXZ_nG_subleading}
\end{figure}

In particular, for the occupation number entropy we have argued in Sec.~\ref{sec:TI} that in gapped phases of translation invariant models the subleading term of the non-Gaussianity should vanish exponentially fast once the system size is larger than the correlation length. We can extract the subleading term of the occupation number entropy by subtracting its value at a given system size $N$ from twice its value at half the system size. This is shown in Fig.~\ref{fig:XXZ_nG_subleading} as a function of $\gamma$. In the gapped phases, it quickly goes to zero, while it is nonzero around the critical point. We expect the nonzero regions around the critical point to correspond to those regions where the correlation length is comparable to or larger than the system size. Curiously, finite-size effects around the critical point lead to a positive subleading term, while the subleading term at the critical point is negative and seems largely system-size independent. Whether this negative offset at the phase transition can be related to some universal features of the critical point remains an open question. Note that in the XXZ model it can be continuously tuned by changing the strength of the interaction---in particular, it can be tuned to zero by setting the interaction strength to zero.

\begin{figure}
    \centering
    \includegraphics[width=\linewidth]{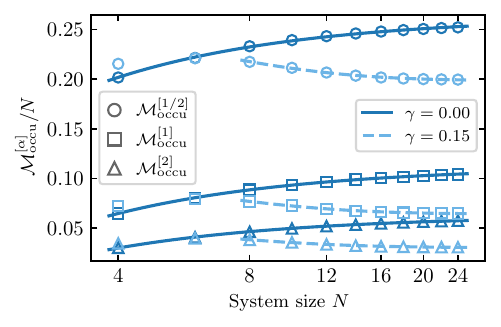}
    \caption{Convergence of the occupation number entropy density $\nGoccu^{[\alpha]}/N$ in the bond-modulated XXZ model [see Eq.~\eqref{eq:XXZ_Hamiltonian}] with system size $N$. The symbols show the occupation number entropy density for $\alpha=1/2$ (circles), for $\alpha=1$ (squares) and for $\alpha=2$ (triangles), both at the critical point $\gamma=0$ (darker blue) and at one representative example in a gapped phase at $\gamma=0.15$. Solid lines show fits of the form $f(N) = a - b/N$ to the data at the critical point, dashed lines show fits of the form $f(N) = a + b \, e^{-c \, N}$ to the data in the gapped phase. The exponential convergence to a constant with system size in the gapped phase is consistent with the result derived in Sec.~\ref{sec:TI}.}
    \label{fig:XXZ_nG_convergence}
\end{figure}

To check that the convergence is indeed exponential, in Fig.~\ref{fig:XXZ_nG_convergence} we compare the convergence with system size at the critical point $\gamma=0$ (dark blue) with a representative example in the gapped phase at $\gamma=0.15$ (light blue). The different marker shapes correspond to different values of $\alpha$ in the definition of the occupation number entropy; circles correspond to $\alpha=1/2$, squares to $\alpha=1$, and triangles to $\alpha=2$. At the critical point ($\gamma=0$), the occupation number entropy is consistent with a constant subleading term,
\begin{equation*}
    \nGoccu^{[\alpha]}/N = m_{(\infty)}^{[\alpha]} + c/N,
\end{equation*}
as indicated by the solid lines which show a fit of this form to the data. In the gapped phase ($\gamma=0.15$), we expect an exponential convergence of the occupation number entropy,
\begin{equation*}
    \nGoccu^{[\alpha]}/N = m_{(\infty)}^{[\alpha]} + c \, e^{-\delta N}.
\end{equation*}
The dashed lines show the result of fitting a function of this form to the data, where we have excluded $N\in\{4,6\}$ from the fit due to strong finite-size effects. For any $\alpha$, the data are consistent with the predicted exponential convergence.

As a second example, we consider an interacting quantum Ising model in Appendix~\ref{app:Ising} and find similar results.

\section{Conclusion}
\label{sec:concl}

We have introduced and analyzed two families of computable quantifiers of fermionic non-Gaussianity---the occupation number entropies $\nGoccu^{[\alpha]}$ and the natural-orbital participation entropies $\nGNO^{[\alpha]}$---establishing their resource-theoretic properties, operational implications, and practical utility across a range of physically relevant settings.

Our work raises several open questions. %
A central one is when the natural-orbital basis is the optimal basis for the participation entropy: although $\nGNO^{[\alpha]}$ is defined through this basis, it need not minimize the participation entropy over all Gaussian bases, and the true optimal Gaussian expansion may even be non-orthogonal. It would be valuable to characterize the states---e.g. ground states of local Hamiltonians---for which the natural-orbital basis is optimal, or near-optimal, and to identify the property responsible for this optimality. Concretely, a recent numerical study found that the entanglement-minimized-orbital basis~\cite{li2025entanglement} can give a lower ($\alpha\to\infty$) participation entropy than the natural-orbital basis, so relating natural orbitals, entanglement-minimized orbitals, and the true optimum is an intriguing problem.%

In the context of quantum computation, fermionic non-Gaussianity is the analogue of nonstabilizerness within the fermionic (matchgate) computational model. Just as nonstabilizerness is the resource injected via magic states to promote Clifford circuits to universality---thereby setting the $T$-gate cost of fault-tolerant computation---fermionic non-Gaussianity is the resource injected via magic states to promote matchgate circuits to universal fermionic quantum computation~\cite{hebenstreit2019all}, and our monotone $\nGoccu^{[1]}$ lower-bounds the corresponding number of non-Gaussian (e.g.\ SWAP) gates required to prepare a state.
The two are nonetheless distinct, complementary resources: as we have seen, stabilizer states can exhibit volume-law fermionic non-Gaussianity, while fermionic Gaussian states can carry volume-law nonstabilizerness.
Many physically relevant systems---particularly strongly correlated interacting fermions---exhibit volume-law non-Gaussianity, as observed in this work and in Ref.~\cite{sierant2025fermionicmagicresourcesquantum}; volume-law non-Gaussianity thus appears to be a common feature of such states. This ubiquity in correlated matter underscores the value of the efficiently computable quantities developed here for quantifying the resource content of fermionic many-body states. %

We also introduced two computable quantifiers of fermionic non-Gaussianity in mixed states, which are valid under Gaussian operations without feedforward, whose systematic study and physical applications we leave for future work. Whether a computable mixed-state \emph{monotone} in the full convex resource theory exists remains open, and may even be ruled out by the hardness of the convex-Gaussian membership problem, as recently shown for convex mixtures of stabilizer states~\cite{leone2026unbearable}. A promising route forward is instead to construct computable quantitative non-Gaussianity witnesses for mixed states~\cite{tang2026witness}, following the recent approach for nonstabilizerness~\cite{haug2025efficientwitnessingtestingmagic} and entanglement~\cite{tarabunga2025quantifying}.
Another open problem is to extend the notion of non-Gaussianity to operators~\cite{debertolis2025naturalsuperorbitalsrepresentationmanybody}. In the Schr\"odinger picture, entanglement, nonstabilizerness, and fermionic non-Gaussianity are three distinct facets of complexity and may even exhibit contrasting behavior. For example, FGSs can exhibit volume-law entanglement and nonstabilizerness while having zero non-Gaussianity. Surprisingly, recent results suggest a more unified picture in the Heisenberg representation. In particular, the operator stabilizer R\'enyi entropy~\cite{dowling2025magicheisenbergpicture}, which quantifies the nonstabilizerness of an operator, has been shown to upper-bound, and in the average case coincide with, operator entanglement~\cite{dowling2025bridging}. Moreover, fermionic circuit simulation with Majorana propagation~\cite{miller2025simulationfermioniccircuitsusing} is essentially equivalent to Pauli propagation. These observations suggest that a more unified description of non-classical resources may arise in the Heisenberg picture. Clarifying this structure remains an important open question.

Rigorously treating fermionic non-Gaussianity within a resource-theory framework is a relatively new development, and thus invites analogues of concepts from the significantly more developed entanglement theory. For instance, in analogy with short- and long-range entanglement, and following the corresponding notions for stabilizer states~\cite{wei2025long, korbany2025longrange, tarabunga2023many,korbany2026longrangenonstabilizernesstopologicallyencoded}, one may ask whether there exist states whose non-Gaussianity cannot be removed by shallow-depth circuits. Other many-body questions also warrant study: how fermionic non-Gaussianity influences thermalization~\cite{projansky2024entanglement,froland2025entanglement,turkeshi2024magic,PaLu26,falcao2026fermionic,aditya2025growth,bera2015mbl}, which could be probed via quantum quenches with weak integrability breaking; and what role it plays in variational ground-state approximations, including whether a scaling law constrains the attainable ground-state-energy accuracy. Addressing these questions in concrete condensed-matter settings would extend our understanding of both the underlying physics and the operational meaning of these quantities.

Finally, our results raise several questions for classical simulation. The quantities introduced---particularly $\nGNO^{[\alpha<1]}$---upper-bound the cost of fermionic simulation methods based on orthonormal Gaussian-basis expansions~\cite{dias2024classical, ReardonSmith2024improved}. While such sparse (low-rank) orthonormal-basis methods remain little explored in condensed-matter and quantum-information settings~\cite{jaques2022leveraging,steiger2024sparse,mullinax2025large}, closely related approaches are well established in quantum chemistry, such as selected configuration interaction~\cite{Holmes2016,Garniron2018} and coupled-cluster methods~\cite{Taube2005,Taube2008,Grneis2011}; cross-fertilization between the two could prove fruitful. Relatedly, $\nGNO^{[\alpha<1]}$ bounds the entanglement entropy needed to represent a state as an MPS in the natural-orbital basis, although for volume-law-non-Gaussian ground states this rotation may instead increase the entanglement. Since variationally optimized single-particle rotations can reduce entanglement and improve ground-state-energy estimates~\cite{krumnow2016fermionic, li2025entanglement, huang2025augmenting}, systematically understanding the interplay between basis optimization and non-Gaussianity is an important direction for future work.

\emph{Note added.}---While finalizing this manuscript, we became aware of a related work~\cite{haug2026practical}, which also develops bounds on the Gaussian fidelity and single-copy testing of Gaussianity.
We have also coordinated the submission of this manuscript with another independent related work~\cite{swietek2026}, which studies the occupation number entropies (therein referred to as non-Gaussianity and one-body purity) in random superpositions of orthogonal Gaussian states with fixed number of particles and in highly excited eigenstates of many-body lattice Hamiltonians.

\begin{acknowledgments}
    We thank X. Turkeshi, P. Sierant, and B. Dias for useful discussions. This work was supported by the Deutsche Forschungsgemeinschaft (DFG, German Research Foundation) under Germany's Excellence Strategy EXC-2111-390814868, TRR 360 (project ID 492547816), FOR 5522 (project ID 499180199), and the Munich Quantum Valley, which is supported by the Bavarian state government with funds from the Hightech Agenda Bayern Plus. P.S.T. acknowledges funding from the European Research Council (ERC) under the European Union (ERC, DynaQuant, No. 101169765). M.L. and B.K. acknowledge funding from the BMW endowment fund and the Horizon Europe programmes HORIZON-CL4-2022-QUANTUM-02-SGA via the project 101113690 (PASQuanS2.1) and HORIZON-CL4-2021-DIGITAL-EMERGING-02-10 under grant agreement No. 101080085 (QCFD). We acknowledge the use of Claude Fable 5 in the development of the proof of the strong monotonicity of the occupation number entropy for $\alpha=2$. The final proof presented in the manuscript was carefully verified and refined by the authors.

\end{acknowledgments}

\textbf{Data and materials availability:}
Raw data and
simulation codes are available on Zenodo~\cite{zenodo}.

\appendix

\section{Properties of the occupation number entropy} 

In this appendix, we collect several facts and properties of the $\alpha$-occupation number entropies. We start by investigating the resource-theoretic properties. Then, we comment on the complexity of computing these quantities for both state vector, as well as MPS simulations. We then focus especially on the case $\alpha = 2$, for which we show that $\nGoccu^{[2]}(\ket\psi)$ can be written as an expectation value of an operator on two copies of the state, which also enables us to prove a concentration inequality for typical states. We furthermore present bounds on the Gaussian fidelity in terms of $\nGoccu^{[2]}(\ket\psi)$ and an efficient measurement protocol for the latter, and show that those can be used to test whether a given pure state is Gaussian or not. We also show that the occupation number entropies bound the deviation of Majorana correlators from the values predicted by Wick's theorem. Finally, we address a connection between $\nGoccu^{[\alpha]}(\ket{\psi})$ and a family of global entanglement monotone.

Recall that the $\alpha$-occupation number entropy is defined as %
\begin{equation}
    \nGoccu^{[\alpha]}(\ket{\psi}) = \sum_{j=1}^N T_{\alpha}\left(\left\lbrace \lambda_j, 1-\lambda_j\right\rbrace\right),
\end{equation}
where $\lambda_j=(1+r_j)/2$ are the occupation numbers obtained from the Williamson eigenvalues $r_j$ of the covariance matrix, and $T_\alpha$ is the (normalized) $\alpha$-Tsallis entropy
\begin{equation} 
T_\alpha(\left\lbrace p_i\right\rbrace) = \frac{1}{1-2^{1-\alpha}} \left(1 - \sum_i p_i^\alpha\right) ,
\end{equation}
for a discrete probability distribution $p_i$.

\subsection{Resource theory properties}
\label{sec:occu_properties}

In this section, we prove the properties \ref{prop:faithful}--\ref{prop:composition_fgs} for $\nGoccu^{[\alpha]}$.
In particular, we prove the stronger additivity (\ref{prop:add}) property.
Note that all properties hold for $\alpha\geq0$, with the exception of property~(\ref{prop:strong}), which we establish only for $\nGoccu^{[1]}$ and $\nGoccu^{[2]}$.

\begin{enumerate} [label={P.\arabic*}]
\item[\ref{prop:faithful}] \textit{Faithfulness}: A pure FGS has occupation numbers $\lambda_j \in \{0,1\}$, so every $T_{\alpha}(\{\lambda_j, 1-\lambda_j\})=0$ and hence $\nGoccu^{[\alpha]}(\ket{\psi})=0$. Conversely, any non-Gaussian pure state has at least one $0<\lambda_j<1$~\cite{surace2022fgs}, giving $\nGoccu^{[\alpha]}(\ket{\psi})>0$.

\item[\ref{prop:stability}] \textit{Invariance under FGUs}: An FGU acts on the covariance matrix by the orthogonal similarity transformation $\Gamma\mapsto Q \Gamma Q^T$, which leaves its eigenvalues $\pm i r_j$---and hence the occupation numbers $\lambda_j=(1+r_j)/2$ and $\nGoccu^{[\alpha]}(\ket{\psi})$---invariant.

\item[\ref{prop:add}] \textit{Additivity}: The covariance matrix of a tensor product factorizes as $\Gamma(\ket{\psi}_A \otimes \ket{\psi}_B) = \Gamma(\ket{\psi}_A)\oplus \Gamma(\ket{\psi}_B)$, so its Williamson eigenvalues are the union of those of the two factors. Since $\nGoccu^{[\alpha]}$ is a sum of Tsallis entropies over these eigenvalues, it splits accordingly, $\nGoccu^{[\alpha]}(\ket{\psi}_A \otimes \ket{\psi}_B) = \nGoccu^{[\alpha]}(\ket{\psi}_A) + \nGoccu^{[\alpha]}(\ket{\psi}_B)$. 

\item[\ref{prop:strong}] \textit{Non-increasing, on average, under computational basis measurements}: For $\nGoccu^{[1]}$, since it coincides with $\nGR$ restricted to pure states, the result immediately follows from the strong monotonicity of $\nGR$, which will be shown in Appendix~\ref{sec:Mr_properties}. For $\nGoccu^{[2]}$, this is established directly from the covariance matrix in Appendix~\ref{sec:strong_mono_occu2}.

\item[\ref{prop:composition_fgs}] \textit{Invariance under composition with FGS}: It immediately follows from additivity under tensor product.
\end{enumerate}

\subsection{Strong monotonicity of \texorpdfstring{$\bm{\nGoccu^{[2]}}$}{NG occu [2]}}
\label{sec:strong_mono_occu2}

Here we prove that $\nGoccu^{[2]}$ satisfies property~(\ref{prop:strong}): for a pure state $\ket\psi$ it is non-increasing on average under computational-basis measurements. Combined with the other properties shown in Appendix~\ref{sec:occu_properties}, this shows that $\nGoccu^{[2]}$ is a strong monotone of fermionic non-Gaussianity for pure states. Throughout, $\rho=\ketbra{\psi}{\psi}$, $\langle X\rangle=\Tr[\rho X]$, and $\Sigma_{jk}:=-i\gamma_j\gamma_k$ denotes the Majorana quadratics, so that $\Gamma_{jk}=\langle\Sigma_{jk}\rangle$ for $j \neq k$, while $\Gamma_{jj}=0$. We use the covariance-matrix form $\nGoccu^{[2]}(\ket\psi)=N-\tfrac12\Tr(\Gamma^T\Gamma)=N-\tfrac12\lVert\Gamma\rVert_F^2$, with $\lVert\Gamma\rVert_F^2=\sum_{j,k}\Gamma_{jk}^2$.

Without loss of generality, consider a computational-basis measurement of the first mode, with measured operator $\Sigma_{12}$. Its outcomes $\lambda\in\{\pm1\}$ occur with probabilities $p_\lambda=\Tr(V_\lambda\rho)=\tfrac12(1+\lambda m)$, where
$V_\lambda=\tfrac12(\Id+\lambda\Sigma_{12})$ and $m:=\langle\Sigma_{12}\rangle=\Gamma_{12}$, leaving the post-measurement states $\rho_\lambda=V_\lambda\rho V_\lambda/p_\lambda$, which are again pure, with covariance matrices $\Gamma_\lambda$ and expectation values $\langle X\rangle_\lambda=\Tr[\rho_\lambda X]$. Our goal is to show that $\sum_\lambda p_\lambda\,\nGoccu^{[2]}(\ket{\psi_\lambda})\le\nGoccu^{[2]}(\ket\psi)$, which, since $\nGoccu^{[2]}=N-\tfrac12\lVert\Gamma\rVert_F^2$, is equivalent to
\begin{equation}\label{eq:strong2_goal}
    \sum_{\lambda=\pm1}p_\lambda\,\lVert\Gamma_\lambda\rVert_F^2 \;\ge\; \lVert\Gamma\rVert_F^2 .
\end{equation}
If $m=\pm1$ the measurement leaves the state unchanged and monotonicity is trivial; we therefore assume $|m|<1$ and denote $s:=1-m^2>0$.

We split the Majorana indices into the measured pair $A=\{1,2\}$ and the remainder $B=\{3,\dots,2N\}$. Since $\lVert\Gamma_{AA}\rVert_F^2=2m^2$, the Frobenius norm decomposes as
\begin{equation}\label{eq:strong2_split}
    \lVert\Gamma\rVert_F^2=2m^2+2\lVert\Gamma_{AB}\rVert_F^2+\lVert\Gamma_{BB}\rVert_F^2 .
\end{equation}
For the post-measurement state, the outcome fixes $(\Gamma_\lambda)_{12}=\lambda$ and all remaining entries in rows $1$ and $2$ vanish, $(\Gamma_\lambda)_{1k}=(\Gamma_\lambda)_{2k}=0$ for $k\ge3$; thus $\Gamma_\lambda$ is block-diagonal across $A$ and $B$, and
\begin{equation}\label{eq:strong2_blockdiag}
    \lVert\Gamma_\lambda\rVert_F^2 = 2 + \lVert(\Gamma_\lambda)_{BB}\rVert_F^2 .
\end{equation}
Moreover, since $\gamma_j,\gamma_k$ with $j,k\in B$ commute with $\Sigma_{12}$, the two-point functions on $B$ are left invariant on average,
\begin{equation}\label{eq:strong2_avg}
    \sum_\lambda p_\lambda(\Gamma_\lambda)_{BB}=\Gamma_{BB} .
\end{equation}
Introducing the outcome variance $\mathcal{V}:=\sum_\lambda p_\lambda\lVert(\Gamma_\lambda)_{BB}-\Gamma_{BB}\rVert_F^2\ge0$ and expanding the square,
\begin{equation}\label{eq:strong2_var}
    \begin{aligned}
        \mathcal{V} &=\sum_\lambda p_\lambda\Big(\lVert(\Gamma_\lambda)_{BB}\rVert_F^2+\lVert\Gamma_{BB}\rVert_F^2-2\Tr\big((\Gamma_\lambda)_{BB}^T\Gamma_{BB}\big)\Big)\\
        &= \sum_\lambda p_\lambda\lVert(\Gamma_\lambda)_{BB}\rVert_F^2-\lVert\Gamma_{BB}\rVert_F^2\\
        &= \sum_\lambda p_\lambda\lVert\Gamma_\lambda\rVert_F^2-\lVert\Gamma\rVert_F^2-2+2m^2+2\lVert\Gamma_{AB}\rVert_F^2 ,
    \end{aligned}
\end{equation}
where the second equality uses Eq.~\eqref{eq:strong2_avg} and the third uses Eqs.~\eqref{eq:strong2_blockdiag} and~\eqref{eq:strong2_split}. Eq.~\eqref{eq:strong2_goal} is thus equivalent to
\begin{equation}\label{eq:strong2_reduced}
    \mathcal{V} \;\ge\; 2\lVert\Gamma_{AB}\rVert_F^2 - 2s .
\end{equation}

The task is now to lower-bound the variance. To this end, we apply a further FGU supported on $B$ (which leaves $\nGoccu^{[2]}$ unchanged) to align the couplings between $A$ and $B$ so that the only nonzero entries of $\Gamma_{AB}$ are $\Gamma_{13}$, $\Gamma_{23}$, and $\Gamma_{24}$; equivalently, $\Gamma_{14}=0$ and $\Gamma_{1k}=\Gamma_{2k}=0$ for $k\ge5$, so that $\lVert\Gamma_{AB}\rVert_F^2=\Gamma_{13}^2+\Gamma_{23}^2+\Gamma_{24}^2$. This is always possible: an FGU acts by the orthogonal similarity $\Gamma\mapsto R\Gamma R^{T}$ with $R\in O(2N)$, and one supported on $B$ takes the form $R=\Id_A\oplus R_B$, sending $\Gamma_{AB}\mapsto\Gamma_{AB}R_B^{T}$. The two rows of $\Gamma_{AB}$, $x=(\Gamma_{1k})_{k\in B}$ and $y=(\Gamma_{2k})_{k\in B}$, are thereby rotated to $R_B x$ and $R_B y$; since they span at most a two-dimensional subspace, choosing $R_B$ so that $\gamma_3$ points along $x$ and $\gamma_4$ along the component of $y$ orthogonal to $x$ (a Gram--Schmidt step) puts $\Gamma_{AB}$ in the stated form.

We can thus lower-bound $\mathcal{V}$ by retaining only the $(3,4)$ entry of the $B$-block, giving $\mathcal{V}\ge 2\sum_\lambda p_\lambda(\langle\Sigma_{34}\rangle_\lambda-\langle\Sigma_{34}\rangle)^2$. Since $\Sigma_{34}$ commutes with $\Sigma_{12}$, one has $\sum_\lambda p_\lambda\langle\Sigma_{34}\rangle_\lambda=\langle\Sigma_{34}\rangle$ and $\sum_\lambda \lambda\,p_\lambda\langle\Sigma_{34}\rangle_\lambda=\langle\Sigma_{34}\Sigma_{12}\rangle$; writing $d_\lambda:=\lambda-m$, with $\sum_\lambda p_\lambda d_\lambda=0$ and $\sum_\lambda p_\lambda d_\lambda^2=s$, the Cauchy--Schwarz inequality gives
\begin{equation}
    \sum_\lambda p_\lambda\big(\langle\Sigma_{34}\rangle_\lambda-\langle\Sigma_{34}\rangle\big)^2
    \;\ge\; \frac{\big(\sum_\lambda p_\lambda d_\lambda\langle\Sigma_{34}\rangle_\lambda\big)^2}{\sum_\lambda p_\lambda d_\lambda^2}
    = \frac{W^2}{s},
\end{equation}
where
\begin{equation}
    W:=\langle\Sigma_{34}\Sigma_{12}\rangle-\langle\Sigma_{34}\rangle\langle\Sigma_{12}\rangle .
\end{equation}
Hence $\mathcal{V}\ge 2W^2/s$, so it suffices to establish
\begin{equation}\label{eq:strong2_W}
    W^2 \;\ge\; s(\Gamma_{13}^2+\Gamma_{23}^2+\Gamma_{24}^2) - s^2 ,
\end{equation}
from which Eq.~\eqref{eq:strong2_reduced}---and with it the claim---follows.

This inequality involves only $\gamma_1,\gamma_2,\gamma_3,\gamma_4$, i.e., the reduced state $\rho_{1234}$ of the two modes they span. For a global pure state of definite parity, $\rho_{1234}$ is an even state and reads
\begin{equation}
    \rho_{1234}=\tfrac14\Big(\Id+\!\!\sum_{1\le j<k\le4}\!\!\Gamma_{jk}\,\Sigma_{jk}+q\,Q\Big),
\end{equation}
where
\begin{equation}
     Q:=-\gamma_1\gamma_2\gamma_3\gamma_4,\quad q:=\langle Q\rangle .
\end{equation}
Using $\Sigma_{34}\Sigma_{12}=Q$, we have $W=q-\Gamma_{12}\Gamma_{34}=q-m\,\Gamma_{34}$. To lower-bound $W$, we use that $\rho_{1234}$, being a physical reduced state, is positive semidefinite. It is block-diagonal in the two eigenspaces of $Q$---the parity-even ($Q=+1$) and parity-odd ($Q=-1$) sectors, with projectors $P_\pm=\tfrac12(\Id\pm Q)$---each two-dimensional and spanned by four operators: the projected identity together with a Pauli-like triple of $\mathfrak{su}(2)$ generators,
\begin{equation}
\begin{aligned}
    \vec L&=\tfrac12\big(\Sigma_{23}+\Sigma_{14},\,-\Sigma_{13}+\Sigma_{24},\,\Sigma_{12}+\Sigma_{34}\big),\\
    \vec R&=\tfrac12\big(\Sigma_{23}-\Sigma_{14},\,-\Sigma_{13}-\Sigma_{24},\,\Sigma_{12}-\Sigma_{34}\big),
\end{aligned}
\end{equation}
for the parity-even and parity-odd sector, respectively. These obey $[L_i,L_j]=2i\,\epsilon_{ijk}L_k$, $[R_i,R_j]=2i\,\epsilon_{ijk}R_k$, and $L_i R_j=0$, with $\vec L$ ($\vec R$) acting as the Pauli vector $\vec\sigma$ on the parity-even (parity-odd) sector and vanishing on the other. Writing the quadratic term as $\sum_{j<k}\Gamma_{jk}\Sigma_{jk}=2(\vec a\cdot\vec L+\vec b\cdot\vec R)$, with
\begin{equation}
\begin{aligned}
    \vec a&=\tfrac12\big(\Gamma_{23}+\Gamma_{14},\,-\Gamma_{13}+\Gamma_{24},\,\Gamma_{12}+\Gamma_{34}\big),\\
    \vec b&=\tfrac12\big(\Gamma_{23}-\Gamma_{14},\,-\Gamma_{13}-\Gamma_{24},\,\Gamma_{12}-\Gamma_{34}\big),
\end{aligned}
\end{equation}
the two blocks of $\rho_{1234}$ reduce to
\begin{equation}
\begin{aligned}
P_+\rho_{1234}P_+&=\tfrac14\big((1+q)\,P_+ +2\,\vec a\cdot\vec L\big),\\
    P_-\rho_{1234}P_-&=\tfrac14\big((1-q)\,P_- +2\,\vec b\cdot\vec R\big),
\end{aligned}
\end{equation}
which can be interpreted as unnormalized single-spin density matrices of the form $c\,\Id+\vec v\cdot\vec\sigma$. Since the latter has eigenvalues $c\pm|\vec v|$, positivity of $\rho_{1234}$ is equivalent to $2\lvert\vec a\rvert\le 1+q$ and $2\lvert\vec b\rvert\le1-q$. Squaring and adding these, and using $\Gamma_{14}=0$, gives $\Gamma_{13}^2+\Gamma_{23}^2+\Gamma_{24}^2+m^2+\Gamma_{34}^2\le 1+q^2$; multiplying by $s=1-m^2$ then yields $s(\Gamma_{13}^2+\Gamma_{23}^2+\Gamma_{24}^2)-s^2\le s\,(q^2-\Gamma_{34}^2)$. Combined with the algebraic identity $(q-m\,\Gamma_{34})^2-(1-m^2)(q^2-\Gamma_{34}^2)=(\Gamma_{34}-m\,q)^2\ge0$, this gives
\begin{equation}
\begin{aligned}
    W^2&=(q-m\,\Gamma_{34})^2\\
    &\ge s(q^2-\Gamma_{34}^2)\\
    &\ge s(\Gamma_{13}^2+\Gamma_{23}^2+\Gamma_{24}^2)-s^2 ,
\end{aligned}
\end{equation}
which is Eq.~\eqref{eq:strong2_W}. This establishes property~(\ref{prop:strong}) for $\nGoccu^{[2]}$, and hence its strong monotonicity.

\subsection{Properties of \texorpdfstring{$\bm{\nGoccu^{[2]}(|\psi\rangle)}$}{NG occu [2] ( psi)}}
\label{sec:property_mlin}

In this section, we show that
\begin{equation}
\label{eq:occu_Lambda_2}
    \nGoccu^{[2]}(\ket{\psi}) = \frac{1}{2}
    \lVert \Lambda |\psi \rangle \otimes |\psi \rangle\rVert^2_2,
\end{equation}
where $\Lambda$ is defined in Eq.~\eqref{eq:Lambda_def}. Indeed, we have
\begin{equation}
\begin{split}
    \frac{1}{2} \lVert \Lambda |\psi \rangle &\otimes |\psi \rangle\rVert^2_2
    =  \frac{1}{2} \bra{\psi} \otimes \bra{\psi} \Lambda^\dagger \Lambda \ket{\psi} \otimes \ket{\psi} \\
    &= \frac{1}{2} \sum_{j,k}  \bra{\psi} \otimes \bra{\psi} (\gamma_j \otimes \gamma_j) (\gamma_k \otimes \gamma_k)\ket{\psi} \otimes \ket{\psi}\\
    &= N - \frac{1}{2}\sum_{j\neq k} \bra{\psi} i \gamma_j \gamma_k \ket{\psi}^2 \\
    &= N - \frac{1}{2} \Tr(\Gamma(\ket{\psi})^\dagger\Gamma(\ket{\psi})) \\
    &= N - \sum_{j=1}^N r_j^2.
\end{split}
\end{equation}
The latter coincides with $\nGoccu^{[2]}(\ket{\psi})$ (see Eq.~\eqref{eq:ngoccu2_main}). The faithfulness of $\nGoccu^{[2]}(\ket{\psi})$ is thus equivalent to the Gaussianity condition in Eq.~\eqref{eq:Lambda_condition}.

\subsection{Construction of covariance matrix} \label{sec:cov_mat_construction}

A direct numerical evaluation of the covariance matrix requires computing expectation values of $N(2N-1)$ distinct Pauli operators. For quantum states represented as a state vector, a straightforward approach would evaluate each expectation value independently: for each Pauli operator, one computes $\ket{\phi}=i\gamma_j \gamma_k \ket{\psi}$ and then evaluates the overlap $\braket{\psi}{\phi}$. Computing $\ket{\phi}$ in this way involves applying up to $N$ single-qubit Pauli operators, represented as sparse matrices. To improve efficiency, we instead construct the covariance matrix one row at a time: fixing the first Majorana index $j$ and sweeping over the second index $k$, the successive Pauli strings $i\gamma_j\gamma_k$ differ on only two qubits, so each state $i\gamma_j\gamma_k\ket{\psi}$ is obtained from the previous one, $i\gamma_j\gamma_{k-1}\ket{\psi}$, by applying a constant-support operator rather than rebuilding the full Pauli string. As a result, the full covariance matrix can be constructed with an overall computational cost of $\mathcal{O}(N^2 2^N)$ when using an exact state-vector representation.

With a similar strategy, this idea can be implemented efficiently in the MPS formalism by reusing partial contractions when computing the environments required for expectation values. In this way, all elements in a given row are obtained in a single sweep along the chain. This yields a total cost of $\mathcal{O}(N^2\chi^3)$ to construct the full covariance matrix.

With similar considerations, the covariance matrix for stabilizer states can be constructed in $\mathcal{O}(N^3)$ time (see Appendix~\ref{sec:ng_stabilizer_appendix} for the explicit algorithm).

\subsection{Simulation with MPSs}
\label{sec:mps_algos}
Here, we introduce algorithms to compute $\nGoccu^{[\alpha]}$ for matrix product states (MPSs). We consider a system of $N$ qubits in a pure state $|\psi \rangle$ given by an MPS of bond dimension $\chi$:
\begin{equation} \label{eq:mps}
|\psi \rangle=\sum_{s_1,s_2,\cdots,s_N} A^{s_1}_1 A^{s_2}_2 \cdots A^{s_N}_N |s_1,s_2,\cdots s_N \rangle
\end{equation}
with $A_i^{s_i}$ being $\chi \times \chi$ matrices, except at the left (right)
boundary where $A_1^{s_1}$ ($A_n^{s_n}$) is a $1 \times \chi$ ($\chi \times 1$) row (column) vector. Here $s_i \in \left \lbrace 0, 1 \right \rbrace$  is a local computational basis. 

\subsubsection{From covariance matrix}
The computation of $\nGoccu^{[\alpha]}$ can be done by diagonalizing the covariance matrix $\Gamma(\ket{\psi})$. As discussed in Appendix~\ref{sec:cov_mat_construction}, the covariance matrix can be constructed with cost $\mathcal{O}(N^2\chi^3)$ for an MPS. Then, it can be diagonalized with an additional cost of $O(N^3)$. Therefore, the overall computational cost for evaluating $\nGoccu^{[\alpha]}$ is $\mathcal{O}(N^2\chi^3+N^3)$. Note that the case $\alpha=2$ can be obtained directly from the covariance matrix without diagonalization, reducing the cost to $\mathcal{O}(N^2\chi^3)$.

Additionally, the computation can be further simplified by exploiting symmetry. In particular, with translational symmetry, the covariance matrix is completely fixed by the first two rows, which can be obtained at the computational cost of $\mathcal{O}(N\chi^3)$ following the algorithm above. Furthermore, by exploiting the special structure of the covariance matrix, the diagonalization can be performed with cost $\mathcal{O}(N \log N)$ by means of fast Fourier transform, as detailed in the main text.

\subsubsection{Matrix product operator technique}
For the case $\alpha=2$, we can write (see Eq.~\eqref{eq:occu_Lambda_2})
\begin{equation}
    \nGoccu^{[2]} = \frac{1}{2}\bra{\psi} \otimes \bra{\psi} \Lambda^2 \ket{\psi} \otimes \ket{\psi}.
\end{equation}
The operator $\Lambda^2/2$ can be represented as an MPO with bond dimension 3:
\begin{equation}
    \frac{1}{2} \Lambda^2 = \widehat{\ell}\, \underbrace{\widehat{W}\widehat{W}\cdots\widehat{W}}_{N}\, \widehat{r},
\end{equation}
where the local MPO tensor is given by
\begin{equation}
    \widehat{W} = \begin{pmatrix}
        I & X \otimes X + Y \otimes Y & I - Z \otimes Z \\
        0  & Z \otimes Z & -X \otimes X - Y \otimes Y \\
        0 & 0 & I
    \end{pmatrix},
\end{equation}
and the boundary tensors are
\begin{equation}
\begin{split}
    \widehat{\ell} &= (I \quad  0  \quad 0),  
    \\ 
    \quad \widehat{r} &= (0\quad 0 \quad I)^T.
\end{split}
\end{equation}
This is obtained following the well-known construction of an MPO for a Hamiltonian using a finite state machine~\cite{schollwock2011,parker2020localmpo}, where each Pauli string contribution to the operator can be associated with a path from state 0 to state 2 as follows:
\begin{equation*}
\begin{tikzpicture}[
    >=stealth,
    node distance=3.2cm,
    state/.style={
        circle,
        draw,
        minimum size=1.0cm,
        font=\small
    },
    every loop/.style={looseness=8},
    edge label/.style={font=\small, align=center}
    ]
    
    \node[state] (q1) {$0$};
    \node[state, right of=q1] (q2) {$1$};
    \node[state, right of=q2] (q3) {$2$};
    
    \path[->]
    (q1) edge[loop above] node[edge label] {$I$} (q1)
    (q2) edge[loop above] node[edge label] {$Z\otimes Z$} (q2)
    (q3) edge[loop above] node[edge label] {$I$} (q3);
    
    \path[->]
    (q1) edge[bend left=18]
        node[edge label, above]
        {$X\otimes X + Y\otimes Y$}
        (q2)
    
    (q1) edge[bend left=65]
        node[edge label, above]
        {$I - Z\otimes Z$}
        (q3)
    
    (q2) edge[bend left=18]
        node[edge label, above]
        {$-X\otimes X - Y\otimes Y$}
        (q3);

\end{tikzpicture}%
\hspace{1mm}\raisebox{2.5mm}{.}
\end{equation*}
This allows the calculation of $\nGoccu^{[2]}$ using standard MPO-MPS contraction algorithm with a computational cost scaling as $\mathcal{O}(N\chi^5)$.

For the evaluation of $\nGoccu^{[2]}$, the MPO formulation is particularly advantageous in the regime of large system sizes and small bond dimension, i.e., when $N \gtrsim \chi^2$. Conversely, the covariance matrix approach is more favorable when $\chi^2 \gtrsim N$. %

\subsection{Lipschitz-continuity of \texorpdfstring{$\bm{\nGoccu^{[2]}(|\psi\rangle)}$}{NG occu [2] (psi)} and concentration inequality}
\label{app:concent_ineq}

Here, we show that $\nGoccu^{[2]}$ is Lipschitz-continuous and use this to derive a concentration inequality for $\nGoccu^{[2]}$. In particular, the Lipschitz constant $K$ is bounded by $K\leq 2N(2N-1)$, i.e., it holds that
\begin{equation}
    |\nGoccu^{[2]}(\rho) - \nGoccu^{[2]}(\ket{\psi}) | \leq 2N(2N-1) \|\rho-\ketbra{\psi}{\psi}  \|_1.
\end{equation}
To show this, we have 
\begin{equation}
    \begin{split}
        &|\nGoccu^{[2]}(\rho) - \nGoccu^{[2]}(\ket{\psi}) |= \\
        &= \left|\sum_{1\leq j<k\leq 2N} \lvert \Tr(i\gamma_j \gamma_k \rho) \rvert^2  -  \sum_{1\leq j<k\leq 2N} \lvert \Tr(i\gamma_j \gamma_k \ketbra{\psi}{\psi}) \rvert^2 \right| \\
        &\leq  \sum_{1\leq j<k\leq 2N} \left| \lvert\Tr(i\gamma_j \gamma_k \rho) \rvert^2 - \lvert\Tr(i\gamma_j \gamma_k \ketbra{\psi}{\psi}) \rvert^2\right| \\
        &=  \sum_{1\leq j<k\leq 2N} \left(\left| \Tr(i\gamma_j \gamma_k \rho)  + \Tr(i\gamma_j \gamma_k \ketbra{\psi}{\psi})\right| \right. \\
        &\left.\hspace{6em}\left| \Tr(i\gamma_j \gamma_k \rho)  - \Tr(i\gamma_j \gamma_k \ketbra{\psi}{\psi})\right|\right) \\
        &\leq  2\sum_{1\leq j<k\leq 2N} \left| \Tr(i\gamma_j \gamma_k \rho)  - \Tr(i\gamma_j \gamma_k \ketbra{\psi}{\psi})\right| \\
        &\leq  2\sum_{1\leq j<k\leq 2N} \| i\gamma_j \gamma_k \|_\infty \| \rho - \ketbra{\psi}{\psi}\|_1 \\
        &= 2N(2N-1)\| \rho - \ketbra{\psi}{\psi}\|_1,
    \end{split}
\end{equation}
as claimed. Here, the first inequality uses triangle inequality, while the last inequality follows from H\"older's inequality, $|\Tr(AB)|\leq\|A\|_\infty\|B\|_1$, together with $\|i\gamma_j\gamma_k\|_\infty=1$. 

As an immediate consequence of this property, we can easily show that for a Haar random state $\nGoccu^{[2]}$ exhibits strong typicality: any deviation $\epsilon \geq \Omega(1/\text{poly}(N))$ of $\nGoccu^{[2]}$ from its expected value has an exponentially small probability. This can be seen by applying Levy's lemma~\cite{Ledoux2005}, such that
\begin{equation} \label{eq:concent_ineq}
\begin{split}
    \text{Pr}_{\mathbb{H}}&\left[ \left|\nGoccu^{[2]}(\ket{\psi}) -\mathbb{E}_{\mathbb{H}}[\nGoccu^{[2]}(\ket{\psi})] \right| 
    \geq \epsilon  \right]\\ &\leq \exp\left(-\frac{C2^N\epsilon^2}{4N^2(2N-1)^2}\right),
\end{split}
\end{equation}
where $\mathbb{E}_{\mathbb{H}}$ denotes the Haar average and $C$ is a constant. It has been shown that $\mathbb{E}_{\mathbb{H}}[\nGoccu^{[2]}(\ket{\psi})]\simeq N$ up to an exponentially small correction~\cite{sierant2025fermionicmagicresourcesquantum,tarabunga2026fermionic}.

\subsection{Bounds on the Gaussian fidelity}
\label{sec:bounds_Fg}

In this section, we prove the inequality
\begin{equation} \label{eq:occu_bound_Fg_2}
    1-\frac{\nGoccu^{[2]}(\ket{\psi})}{2}
    \leq F_\mathcal{G}(\ket{\psi}) \leq
    1 - \frac{1}{4}\!\left(\! 1 - \sqrt{1-\frac{\nGoccu^{[2]}}{N}}\right)^{\!\!2}\!,
\end{equation}
which bounds the Gaussian fidelity $F_\mathcal{G}(\ket{\psi}) = \max_{\ket{\phi} \in \G_\text{p}} |\langle\phi|\psi\rangle|^2$ in terms of $\nGoccu^{[2]}$. 

To show the lower bound, we use that $r_j^2\leq r_j$ so that $\nGoccu^{[2]}(\ket{\psi})\geq \sum_j (1-r_j)$. The fidelity has been shown to satisfy $1-F_\mathcal{G}(\ket{\psi})\leq (\sum_j (1-r_j))/2$ in Ref.~\cite{bittel2025optimal}, from which Eq.~\eqref{eq:occu_bound_Fg_2} immediately follows. Note however that the inequality is non-trivial only when $\nGoccu^{[2]}(\ket{\psi})\leq2$. 

For the upper bound, we have
\begin{equation}
    r_{\text{min}} \leq \frac{\sum_{j=1}^N r_j}{N} \leq \sqrt{\frac{\sum_{j=1}^N r_j^2}{N}} = \sqrt{1-\frac{\nGoccu^{[2]}}{N}},
\end{equation}
where $r_{\text{min}}$ is the smallest Williamson eigenvalue of the covariance matrix and the second inequality is obtained by Jensen's inequality. Combining with the inequality $F_\mathcal{G} \leq 1- (1-r_{\text{min}})^2/4$ shown in Ref.~\cite{bittel2025optimal}, we obtain the upper bound of Eq.~\eqref{eq:occu_bound_Fg_2}.

\subsection{Experimental measurement of \texorpdfstring{$\bm{\nGoccu^{[2]}(|\psi\rangle)}$}{NG occu [2] (psi)}}
\label{sec:exp_meas}

In this section, we show that for any $N$-qubit state $\ket{\psi}$, $\nGoccu^{[2]}(\ket{\psi})$ can be measured within $\epsilon$ accuracy via single-copy measurements with $\mathcal{O}(N^4\epsilon^{-2})$ samples. To this end, $\lvert \Tr(\gamma_j \gamma_k \rho) \rvert^2$ can be estimated for $1\leq j<k\leq 2N$ by measuring $\ket{\psi}$ in the eigenbasis of the Pauli operator $i\gamma_j \gamma_k$. We measure $2C$ copies of $\ket{\psi}$ and obtain the outcomes $s_m\in \{+1,-1\}$, where $C$ will be chosen later. We define $\hat{X}_m = s_{2m-1}s_{2m}$ for $m=1,\dots,C$ such that $\hat{\Gamma}_{j,k}=\frac{1}{C}\sum_{m=1}^C \hat{X}_m$ is an unbiased estimator of $\lvert \Tr(\gamma_j \gamma_k \rho) \rvert^2$. Then, $\hat{M} = N - \sum_{j>k} \hat{\Gamma}_{j,k}$ is an unbiased estimator of $\nGoccu^{[2]}$. We apply Hoeffding's inequality~\cite{Hoeffding01031963} to obtain
\begin{equation}
    \text{Pr}\left( \lvert\hat{M} - \nGoccu^{[2]} \rvert\geq\epsilon \right) \leq 2\exp\left( -\frac{2\epsilon^2 C}{M(b-a)^2} \right),
\end{equation}
where  $M=N(2N-1)$, $a=-1$ and $b=1$. Thus, $\nGoccu^{[2]}$ can be estimated within $\epsilon$ accuracy and $\delta$ failure probability using 
\begin{equation}
    C \geq \frac{2N(2N-1)}{\epsilon^2} \ln \left(\frac{2}{\delta}\right),
\end{equation}
where the total number of copies is $2CM$. The sample complexity is thus $\mathcal{O}(N^4\epsilon^{-2})$.

\subsection{Algorithm for testing Gaussian states}
\label{sec:testing}

Here, we present the algorithm for Gaussian testing, which is the task of deciding, from copies of an unknown pure state, whether it is close to or far from the set of fermionic Gaussian states. We will use the bounds in Eq.~\eqref{eq:occu_bound_Fg_2} proven in Appendix~\ref{sec:bounds_Fg}.

Let $\ket{\psi}$ be an $N$-qubit state where it is promised that 
\begin{align*}
\mathrm{either}\quad (a)& \,\,F_\G(\ket{\psi})\geq 1 -\epsilon_1 \,,\\
\mathrm{or}\quad (b)& \,\, F_\G(\ket{\psi})\leq 1- \epsilon_2\,.
\end{align*} 
We assume that $\epsilon_s=\epsilon_2-\frac{N}{2}(1-(1-2\sqrt{\epsilon_1})^2) \geq 1/\text{poly}(N)$. With this condition, we provide an efficient algorithm to distinguish case ($a$) and ($b$) using $\mathcal{O}(N^4 \epsilon_s^{-2})$ single-copy measurements of $\ket{\psi}$ with high probability. 

We set 
\begin{equation}
    \epsilon<\epsilon_2-\frac{N}{2}(1-(1-2\sqrt{\epsilon_1})^2)
\end{equation}
 and 
 \begin{equation}
     \epsilon_t=2\epsilon_2-\epsilon.
 \end{equation}
  We can measure $\hat{M}$ as an estimate for $\nGoccu^{[2]}$ using $\mathcal{O}(N^4\epsilon^{-2})$ copies such that $|\nGoccu^{[2]}-\hat{M}|<\epsilon$ with failure probability $\delta$ (see Appendix~\ref{sec:exp_meas}). We then make the following decision: if $\hat{M}\leq\epsilon_t$ then we output (a), otherwise (b). 

Let us now verify the correctness of the algorithm. If we are in case (a), then the upper bound in Eq.~\eqref{eq:occu_bound_Fg_2} implies 
\begin{equation}
    \nGoccu^{[2]}/N \leq 1-(1-2\sqrt{\epsilon_1})^2.
\end{equation}
If $\hat{M}>\epsilon_t$ then
\begin{equation}
    \hat{M} - \nGoccu^{[2]} >  \epsilon_t  - N(1-(1-2\sqrt{\epsilon_1})^2) > \epsilon,
\end{equation}
which occurs with probability at most $\delta$. 
 
If we are in case (b), then the lower bound in Eq.~\eqref{eq:occu_bound_Fg_2} implies 
 \begin{equation}
     2\epsilon_2 \leq \nGoccu^{[2]}.
 \end{equation}
If $\hat{M}\leq\epsilon_t$ then 
\begin{equation}
    \nGoccu^{[2]} - \hat{M} \geq 2\epsilon_2 -\epsilon_t = \epsilon,
\end{equation} 
which occurs with probability at most $\delta$. Therefore, the algorithm efficiently distinguishes case (a) and (b) with probability at least $1-\delta$ using $\mathcal{O}(N^4\epsilon^{-2})$ single-copy measurements.

\subsection{Connection to the deviation from Wick's theorem}
\label{app:wick}

Here we derive the bounds relating $\nGoccu^{[2]}(\ket{\psi})$ to the deviation of correlators from the values predicted by Wick's theorem. Throughout, $\langle X\rangle=\Tr[\rho X]$, and $G_{jk}=\langle\gamma_j\gamma_k\rangle=\delta_{jk}+i\Gamma_{jk}$ is the two-point function, so that $G_{jk}+G_{kj}=2\delta_{jk}$ and $|G_{jk}|=|\Gamma_{jk}|$ for $j\neq k$. The strategy is to relate the Gaussianity condition $[\Lambda,\rho^{\otimes2}]=0$ of Eq.~\eqref{eq:Lambda_condition_mixed} to a recursive form of Wick's theorem.

Let \(S\subset[2N]\) be an ordered set with \(|S|=2m\), and we write
\[
S=(s_1,\ldots,s_{2m}).
\]
Singling out the last element \(c=s_{2m}\), we write $S^- = S \backslash c$, so that $S=S^-\sqcup\{c\}$. Accordingly,
\[
\gamma_{S^-}:=\gamma_{s_1}\cdots\gamma_{s_{2m-1}},
\qquad
\gamma_{S^-\setminus s_p}
:=
\gamma_{s_1}\cdots\widehat{\gamma_{s_p}}\cdots\gamma_{s_{2m-1}} ,
\]
and
\( \gamma_S = \gamma_{S^-}  \gamma_c\), where $\gamma_{S^-\setminus s_p}$ is the Majorana string with $\gamma_{s_p}$ removed from $\gamma_{S^-}$ (with the remaining ordering preserved). %
We define the deviation of a single Wick contraction by
\begin{equation}
    \label{eq:wick_violation}
\delta(S^-, c)
:=
\langle \gamma_{S}\rangle
-
\sum_{p=1}^{2m-1}
(-1)^{p-1}G_{s_p c}
\langle \gamma_{S^-\setminus s_p}\rangle .    
\end{equation}
When the condition $\delta(S^-, c)=0$ holds, we can express the $2m$-point correlator $\langle \gamma_{S}\rangle$ through the lower, $(2m-2)$-point correlators $\langle \gamma_{S^-\setminus s_p}\rangle$;
applied recursively, it expresses every higher-order correlator through two-point functions alone---the Pfaffian of Eq.~\eqref{eq:wick_pfaffian}---thereby recovering Wick's theorem. Indeed, this recursion form is precisely the cofactor expansion of the Pfaffian, and hence $\delta(S^-,c)=0$ for all $S^-,c$ if and only if $\rho$ is an FGS. By expressing each $\delta(S^-,c)$ in terms of the commutator $[\Lambda,\rho^{\otimes2}]$, its vanishing
for all $S^-, c$ is equivalent to the Gaussianity condition and bounding the
deviation $\delta(S^-,c)$ can be achieved by bounding this commutator. We then bound $\delta(S^-, c)$ for pure
states by $\nGoccu^{[2]}$, and finally propagate
this bound recursively through the Wick contractions to bound the full deviation $W(\gamma_S)$ of
an arbitrary $2m$-point correlator from the value predicted by Wick's theorem.

We first establish 
\begin{equation} \label{eq:wick_violation_app}
\delta(S^-,c)=\tfrac12\Tr[[\Lambda,\rho^{\otimes2}](\gamma_{S^-}\otimes\gamma_c)],
\end{equation}
with $\delta(S^-, c)$ given by Eq.~\eqref{eq:wick_violation}. We note that here $\gamma_{S^-}$ and $\gamma_c$ act on different copies of the Hilbert space, so $\gamma_{S^-}\otimes\gamma_c$ should not be confused with $\gamma_{S^-}\gamma_c=\gamma_S$, which acts on a single copy. To this end, expanding $\Lambda=\sum_j\gamma_j\otimes\gamma_j$ and taking $\Tr[\rho^{\otimes2}\,\cdot\,]$ yields the following relation,
\begin{align}
   &\Tr[[\Lambda,\rho^{\otimes2}](O_1 \otimes O_2)] \nonumber
   \\
   &\qquad =\sum_{j=1}^{2N}\big(\langle O_1\gamma_j\rangle\langle O_2\gamma_j\rangle-\langle\gamma_j O_1\rangle\langle\gamma_j O_2\rangle\big),
\label{eq:app_master}
\end{align}
valid for any operators $O_1, O_2$. Setting $O_1=\gamma_{S^-},O_2=\gamma_c$ and using $\langle\gamma_j\gamma_c\rangle=G_{jc}$ and $\langle\gamma_c\gamma_j\rangle=2\delta_{jc}-G_{jc}$,
\begin{equation}
\label{eq:app_close}
   \Tr[[\Lambda,\rho^{\otimes2}](\gamma_{S^-}\otimes\gamma_c)]
   =2\langle \gamma_{S}\rangle-\sum_{j}G_{jc}\,\langle\{\gamma_j, \gamma_{S^-}\}\rangle.
\end{equation}
For a Majorana string of odd length $\gamma_{S^-}=\gamma_{a_1}\cdots\gamma_{a_{2m-1}}$, moving $\gamma_j$ through $\gamma_{S^-}$ with the anticommutation relations $\gamma_j\gamma_a=2\delta_{ja}-\gamma_a\gamma_j$ gives 
\begin{equation}
    \{\gamma_j, \gamma_{S^-}\}=2\sum_{p}(-1)^{p-1}\delta_{j s_p}\gamma_{S^-\setminus s_p}.
\end{equation}
Substituting to Eq.~\eqref{eq:app_close} collapses the sum to
\begin{equation}
\label{eq:app_collapse}
\begin{aligned}
   &\Tr[[\Lambda,\rho^{\otimes2}](\gamma_{S-}\otimes\gamma_c)]\\
   &=2\langle \gamma_{S}\rangle-2\sum_{p=1}^{2m-1}(-1)^{p-1}G_{s_p c}\langle \gamma_{S^-\setminus s_p}\rangle\\
   &=2\,\delta({S^-},c),
\end{aligned}
\end{equation}
yielding the claim.

Next, we bound $\delta({S^-},c)$ in terms of $\nGoccu^{[2]}$ for pure states. For pure $\rho=\ketbra{\psi}{\psi}$ write $\ket{\Phi}=\ket{\psi}^{\otimes2}$ and $X=\gamma_{S^-}\otimes\gamma_c$, so that $\rho^{\otimes2}=\ketbra{\Phi}{\Phi}$ and Eq.~\eqref{eq:wick_violation_app} becomes
\begin{equation}
\label{eq:app_pure_comm}
\begin{aligned}
   \delta({S^-},c)&=-\tfrac12\bra{\Phi}
   [\Lambda,X]\ket{\Phi}\\
   &=\tfrac12\big(\bra{\Phi}X\Lambda\ket{\Phi}-\bra{\Phi}\Lambda X\ket{\Phi}\big)
   .
\end{aligned}
\end{equation}
Using that $\Lambda$ is Hermitian, $\bra{\Phi}\Lambda X\ket{\Phi}=\langle\Lambda\Phi|X\Phi\rangle$ and $\bra{\Phi}X\Lambda\ket{\Phi}=\langle X^{\dagger}\Phi|\Lambda\Phi\rangle$, so by the Cauchy--Schwarz inequality,
\begin{equation}
\label{eq:app_pure_cs}
\begin{aligned}
   |\bra{\Phi}\Lambda X\ket{\Phi}|\leq\big\|\Lambda\ket{\Phi}\big\|_2\,\big\|X\ket{\Phi}\big\|_2, \\
   \qquad
   |\bra{\Phi}X\Lambda\ket{\Phi}|\leq\big\|X^\dagger\ket{\Phi}\big\|_2\,\big\|\Lambda\ket{\Phi}\big\|_2.
\end{aligned}
\end{equation}
Since $X$ is unitary, $\|X\ket{\Phi}\|_2=\|X^{\dagger}\ket{\Phi}\|_2=1$. Combining with Eq.~\eqref{eq:app_pure_comm} and applying the triangle inequality gives
\begin{equation}
\label{eq:app_pure}
   |\delta(S^-,c)|\leq\big\|\Lambda\ket{\Phi}\big\|_2
   =\sqrt{2\,\nGoccu^{[2]}(\ket{\psi})},
\end{equation}
where $\|\Lambda\ket{\Phi}\|_2^2=\bra{\Phi}\Lambda^2\ket{\Phi}$ and $\nGoccu^{[2]}=\tfrac12\|\Lambda\ket{\psi}^{\otimes2}\|_2^2$ from Eq.~\eqref{eq:occu_Lambda}. This bounds  $\delta(S^-,c)$ in terms of $\nGoccu^{[2]}$.

We now proceed to bound $W(\gamma_S)=\langle\gamma_S\rangle-\Tr[\G(\rho)\gamma_S]$ in terms of $\nGoccu^{[2]}$. Recall that $W(\gamma_S)$ is the full deviation of the $2m$-point correlator $\gamma_S$ ($|S|=2m$)  from the value predicted by Wick's theorem, given by the Pfaffian of the covariance matrix. The actual correlator obeys $\langle \gamma_{S}\rangle=\delta(S^-,c)+\sum_{p}(-1)^{p-1}G_{s_p c}\langle \gamma_{S^- \setminus s_p}\rangle$ by Eq.~\eqref{eq:wick_violation}. Subtracting by $\Tr[\G(\rho)\,\gamma_{S}]$,
\begin{equation}
\label{eq:app_W_recursion}
\begin{aligned}
   W(&\gamma_{S})
   =\langle \gamma_{S}\rangle-\Tr[\G(\rho)\,\gamma_{S}]\\
   &=\delta(S^-,c)\\
   &\qquad+\sum_{p=1}^{2m-1}(-1)^{p-1}G_{s_p c}\big(\langle \gamma_{S^- \setminus s_p}\rangle-\Tr[\G(\rho)\,\gamma_{S^- \setminus s_p}]\big)\\
   &=\delta(S^-,c)+\sum_{p=1}^{2m-1}(-1)^{p-1}G_{s_p c}\,W(\gamma_{S^- \setminus s_p}),
\end{aligned}
\end{equation}
where we used that $\G(\rho)$ shares the two-point function $G$ with $\rho$ and applied the Wick's theorem to $\G(\rho)$ in the second equality.
This relates the deviation $W(\gamma_{S})$  of a
$2m$-point Majorana string $\gamma_{S}$ to Wick deviations of $(2m-2)$-point
strings. %
The base case is $W(\gamma_j\gamma_c)=0$ since  $\rho$ and $\G(\rho)$ have the same
two-point function $\langle\gamma_j\gamma_c\rangle=G_{jc}$.

Let $B_{2m}=\max_{|S|=2m}|W(\gamma_S)|$. The rows of $\Gamma$ are sub-normalized, $\sum_j\Gamma_{jc}^2=(\Gamma\Gamma^{T})_{cc}\leq1$ (since the eigenvalues of $\Gamma$ are at most of modulus one). Taking the absolute value of Eq.~\eqref{eq:app_W_recursion},
\begin{equation}
\label{eq:app_B_chain}
\begin{aligned}
   |W(&\gamma_S)|
   \leq |\delta(S^-,c)| + \sum_{p=1}^{2m-1}|G_{s_p c}|\,|W(S^- \setminus s_p)|\\
   &\leq \sqrt{2\,\nGoccu^{[2]}(\ket{\psi})} + B_{2m-2}\sum_{p=1}^{2m-1}|G_{s_p c}|\\
   &\leq \sqrt{2\,\nGoccu^{[2]}(\ket{\psi})} + B_{2m-2}\sqrt{(2m-1)\sum_{p=1}^{2m-1}|G_{s_p c}|^2}\\
   &\leq \sqrt{2\,\nGoccu^{[2]}(\ket{\psi})} + \sqrt{2m-1}\,B_{2m-2},
\end{aligned}
\end{equation}
where we have used the triangle inequality in the first inequality, the bound of Eq.~\eqref{eq:app_pure} together with $|W(S^- \setminus s_p)|\leq B_{2m-2}$ in the second inequality, the Cauchy--Schwarz inequality in the third inequality, and $\sum_{p}|G_{s_p c}|^2\leq\sum_j\Gamma_{jc}^2\leq1$ in the last inequality. Maximizing the left-hand side over all $2m$-strings yields the recursion
\begin{equation}
\label{eq:app_B_recursion}
   B_{2m}\leq \sqrt{2\,\nGoccu^{[2]}}+\sqrt{2m-1}\,B_{2m-2},\qquad B_2=0.
\end{equation}
Dividing by $K=\sqrt{2(2m-1)!!\nGoccu^{[2]}}$ and setting $e_m=B_{2m}/K$ turns this into $e_m\leq e_{m-1}+1/\sqrt{(2m-1)!!}$ with $e_1=0$, hence
\begin{equation}
\label{eq:app_em}
   e_m\;\leq\;\sum_{k=2}^{m}\frac{1}{\sqrt{(2k-1)!!}}\;\leq\;\sum_{k=2}^{\infty}\frac{1}{\sqrt{(2k-1)!!}} .
\end{equation}
We now show that the infinite sum is bounded by one. Consecutive terms have ratio
\begin{equation}
\label{eq:app_ratio}
   \frac{1}{\sqrt{(2k+1)!!}}\Big/\frac{1}{\sqrt{(2k-1)!!}}=\frac{1}{\sqrt{2k+1}}\leq\frac{1}{\sqrt7}\qquad(k\geq3),
\end{equation}
so the terms from $k=3$ onward are dominated by an infinite geometric series with ratio $1/\sqrt7$. Summing it explicitly,
\begin{equation}
\label{eq:app_sum_bound}
\begin{aligned}
   \sum_{k=2}^{\infty}\frac{1}{\sqrt{(2k-1)!!}}
   \;&\leq\;\frac{1}{\sqrt3}+\frac{1}{\sqrt{15}}\sum_{j=0}^{\infty}7^{-j/2} \\
   &=\frac{1}{\sqrt3}+\frac{1}{\sqrt{15}}\,\frac{\sqrt7}{\sqrt7-1}\\
   &<1 .
\end{aligned}
\end{equation}
Hence $e_m<1$ for every $m$, and therefore 
\begin{equation}
    |W(\gamma_S)|\leq \sqrt{(2m-1)!!}\,\sqrt{2\,\nGoccu^{[2]}(\ket{\psi})},
\end{equation}
which is Eq.~\eqref{eq:wick_higher}.

\subsection{Connection to global entanglement}
\label{sec:connection_multipartite}

In this section, we show that $\nGoccu^{[\alpha]}$ has a direct connection to entanglement measures computed in the natural-orbital basis.

We first define a family of global entanglement monotones as
\begin{equation} \label{eq:ME_def}
    \ME^{[\alpha]}(\ket{\psi}) = \sum_{j=1}^N T_{\alpha}\left(\rho_j\right),
\end{equation}
where $\rho_j$ is the single-qubit reduced density matrix at site $j$ and $T_\alpha(\rho)$ is defined as
\begin{equation}
T_\alpha(\rho) = \frac{1}{1-\frac{1}{2^{\alpha-1}}} \left(1 - \Tr( \rho^\alpha)\right) ,
\end{equation}
analogously to $T_\alpha(\{p_i\})$ in Eq.~\eqref{eq:tsallis_prob} for a discrete probability distribution $p_i$. %

For even states, it is straightforward to verify that the single-qubit reduced density matrix is diagonal:
\begin{equation} \label{eq:single_qubit_diag}
    \rho_j = \frac{1}{2} (I + \Gamma_{2j-1,2j} Z).
\end{equation}
Therefore, we have that 
\begin{equation}
    T_\alpha(\rho_j) = T_\alpha(\{p_j, 1-p_j\}),
\end{equation}
where $p_j = (1+\Gamma_{2j-1,2j})/2$. In the natural-orbital basis, the set $\{p_j\}$ corresponds exactly to the set of occupation numbers. This immediately establishes that the occupation number entropies $\nGoccu^{[\alpha]}$ are precisely the global entanglement monotones $\ME^{[\alpha]}$ of the state when expressed in the natural-orbital basis. 

 More strongly, we can show that the natural-orbital basis is the Gaussian basis that minimizes the global entanglement monotones in Eq.~\eqref{eq:ME_def}. To see this, recall that the occupation numbers are the eigenvalues of the matrix $H = \frac{1}{2} \big(\Id + i\Gamma\big)$. Without loss of generality, we assume that the state $\ket{\psi}$ is in the natural-orbital basis. Let $D$ be the diagonal correlation matrix of $\ket\psi$ with entries $D=\begin{pmatrix}
     \boldsymbol{\lambda} & 0\\ 0 & 1-\boldsymbol{\lambda}
 \end{pmatrix}$. For a given FGU $U_Q$,  there exists a unitary $V$ such that the diagonal elements of $VDV^\dagger$ in the rotated Gaussian basis are given by $p_j = (1+\Gamma_{2j-1,2j})/2$~\cite{gigena2015entanglement}. We denote by $q_j$ and $\nu_j$ the full set of diagonal elements of $VDV^\dagger$ and the diagonal entries of $D$, respectively, such that $\{q_j\}$ contains $\{p_j\}$ and $\{1-p_j\}$, and similarly for $\{\nu_j\}$. We have that $q_j=\sum_{k}|V_{jk}|^2\nu_k$. Combined with the fact that $x^\alpha$ is convex for $\alpha>1$ and concave for $0<\alpha<1$, it follows that
  \begin{equation}
  \begin{aligned}
\ME^{[\alpha]}(U_Q \ket{\psi}) &=  \frac{1}{1-2^{1-\alpha}}\left(N-\sum_{j=1}^{2N}q_j^\alpha \right) \\
&\geq \frac{1}{1-2^{1-\alpha}}\left(N-\sum_{j=1}^{2N}\nu_j^\alpha \right)\\
&= \nGoccu^{[\alpha]}(\ket{\psi}), \\
\end{aligned}
\end{equation}
It similarly holds for $\alpha=1$ due to $-x\log x$ being concave, and for $\alpha=1$ by continuity.
This shows the claim.

\section{Properties of the orthogonal Gaussian participation entropy}  
\label{sec:NGOG_app}
The orthogonal Gaussian R\'enyi entropy is defined as
\begin{equation} 
\label{eq:M_og_app}
    \nGOG^{[\alpha]}(\ket{\psi} ) = \min_{U \in \G_\text{U}} S^{\text{part}}_\alpha(U^\dagger\ket{\psi}),
\end{equation}
where $S^{\text{part}}_\alpha (\ket{\psi})$ is the participation entropy, defined as the $\alpha$-R\'enyi entropy of the probability distribution $\{p_s\}$ arising from computational basis measurements, where $p_s = |\langle s | \phi \rangle|^2$ for $\ket{\phi}=U^\dagger\ket{\psi}$:
\begin{equation}
    H_\alpha(\{p\}) = \frac{1}{1-\alpha} \log_2 \Big(\sum_{s} p_s^\alpha\Big).
\end{equation}
In this appendix, we show the detailed proofs of the different resource-theoretic properties of this quantity and prove bounds in terms of the Gaussian fidelity.

\subsection{Resource theory properties}
\label{sec:og_properties}

In this section, we prove properties \ref{prop:faithful}--\ref{prop:sub-add} for $\nGOG^{[\alpha]}$.
\begin{enumerate} [label={P.\arabic*}]
\item[\ref{prop:faithful}] \textit{Faithfulness}: A pure FGS satisfies $\ket{\psi}=U_G \ket{0}^{\otimes N}$ for some FGU $U_G$; choosing $U=U_G$ gives $S^{\text{part}}_\alpha(U_G^\dagger\ket{\psi})=0$, hence $\nGOG^{[\alpha]}(\ket{\psi})=0$. Conversely, $\nGOG^{[\alpha]}(\ket{\psi})=0$ implies $U^\dagger\ket{\psi}=\ket{0}^{\otimes N}$ for some FGU $U$, so $\ket{\psi}$ is a pure FGS.

\item[\ref{prop:stability}] \textit{Invariance under FGUs}: As $\nGOG^{[\alpha]}$ minimizes the participation entropy over all FGUs, which form a group, the minimization is unaffected by first applying any FGU $U'$; hence $\nGOG^{[\alpha]}(U'\ket{\psi})=\nGOG^{[\alpha]}(\ket{\psi})$.

\item[\ref{prop:sub-add}] \textit{Sub-additivity}: Let $U_A$ and $U_B$ minimize the participation entropy of $\ket{\psi}_A$ and $\ket{\psi}_B$. Using $U_A\otimes U_B$ as a candidate in the minimization and the additivity of the participation entropy under tensor products,
\begin{equation}
\begin{aligned}
    \nGOG^{[\alpha]}&(\ket{\psi}_A \otimes \ket{\psi}_B) = \min_{U \in \G_\text{U}} S^{\text{part}}_\alpha(U^\dagger\ket{\psi}_A \otimes \ket{\psi}_B) \\
    &\leq S^{\text{part}}_\alpha(U_A^\dagger \otimes U_B^\dagger\ket{\psi}_A \otimes \ket{\psi}_B) \\
    &= S^{\text{part}}_\alpha(U_A^\dagger \ket{\psi}_A) + S^{\text{part}}_\alpha(U_B^\dagger \ket{\psi}_B) \\
    &= \nGOG^{[\alpha]}(\ket{\psi}_A) + \nGOG^{[\alpha]}(\ket{\psi}_B). 
\end{aligned}
\end{equation}
\end{enumerate}

\subsection{Bounds on the Gaussian fidelity}
\label{sec:bounds_Fg_ogre}

In this section, we show that $\nGOG^{[\alpha]}(\ket{\psi} )$ provide both an upper and lower bound on the logarithm of the Gaussian fidelity $F_{\G}(\ket{\psi})$. The lower bound is also inherited by $\nGNO^{[\alpha]}(\ket{\psi})$ (defined in the main text, see also Appendix~\ref{sec:ngno_app}).
Specifically, we will show
\begin{equation} \label{eq:upper_fid_2}
     \nGOG^{[\alpha]}(\ket{\psi}) \leq  - \frac{\alpha}{\alpha-1}\log F_{\G}(\ket{\psi} ), \quad (\alpha > 1),
\end{equation}
and
\begin{equation} \label{eq:lower_fid_2}
    -\log F_{\G}(\ket{\psi} )  \leq \nGOG^{[\alpha]}(\ket{\psi}),  \quad (\alpha \geq 0).
\end{equation}
To prove these bounds, we will use the following inequality of the R\'enyi entropies:
\begin{equation}
    \frac{\alpha}{\alpha-1}H_{\infty}(\{p\}) \geq H_{\alpha}(\{p\})  \geq H_{\infty}(\{p\}), \quad (\alpha > 1).
\end{equation}
Note that the second inequality also holds for $\alpha > 0$.

To show Eq.~\eqref{eq:upper_fid_2}, let $\ket{\phi}$ be the Gaussian state with the largest overlap with $\ket{\psi}$, i.e., such that $F_\G(\ket{\psi}) =  |\braket{\phi}{\psi}|^2$.
Let $\{p_j\} = \{ |\langle j |\psi \rangle |^2\}$ be the probability distribution induced by the coefficients of $\ket{\psi}$ in an orthonormal basis $\ket{j}$ with $\ket{j=0}=\ket{\phi}$. We have $H_\infty(\{  p_j \})=-\log F_\G(\ket{\psi})$ and
\begin{equation}
    \frac{\alpha}{\alpha-1} H_\infty(\{  p_j \}) \geq H_\alpha(\{  p_j \}) \geq  \nGOG^{[\alpha]}(\ket{\psi}), \quad (\alpha>1) .
\end{equation}

For the bound in Eq.~\eqref{eq:lower_fid_2}, consider the discrete distribution $\{ |d_j|^2\}$, where $d_j$ are the wave function coefficients in the orthonormal basis that minimizes $\nGOG^{[\alpha]}(\ket{\psi})$, such that $\nGOG^{[\alpha]}(\ket{\psi}) = H_\alpha(\{  |d_j|^2 \})$. We have
\begin{equation}
    H_\alpha(\{  |d_j|^2\}) \geq H_\infty(\{  |d_j|^2\}) \geq  -\log F_\G(\ket{\psi}).
\end{equation}

Finally, the natural-orbital participation entropy obeys the same lower bound,
\begin{equation} \label{eq:lower_fid_NO}
    -\log F_{\G}(\ket{\psi} )  \leq \nGNO^{[\alpha]}(\ket{\psi}),  \quad (\alpha \geq 0),
\end{equation}
since the natural-orbital basis is a particular orthonormal Gaussian basis, so that $\nGNO^{[\alpha]}(\ket{\psi}) \geq \nGOG^{[\alpha]}(\ket{\psi})$. Combined with Eq.~\eqref{eq:lower_fid_2}, this gives Eq.~\eqref{eq:lower_fid_NO}.

\section{Properties of the natural-orbital participation entropy}
\label{sec:ngno_app}

The natural-orbital participation entropy is defined as
\begin{equation}
    \label{eq:M_no_app}
    \nGNO^{[\alpha]} (\ket{\psi})= \min_{U_Q} S^{\text{part}}_\alpha(U_Q^\dagger | \psi \rangle) 
\end{equation}
such that
$\Gamma(\rho) = Q \bigoplus_{j=1}^N
\begin{pmatrix}
    0 & r_j \\
    -r_j & 0 \\
\end{pmatrix}
Q^T$,
and $U_Q$ is an FGU associated to the orthogonal matrix $Q$. The minimization is required only when the Williamson eigenvalues have degeneracy. Here, we again prove the resource-theoretic properties and investigate its relation with the occupation number entropy. %

\subsection{Resource theory properties}
\label{sec:no_properties}
In this section, we prove the properties \ref{prop:faithful}--\ref{prop:sub-add} and \ref{prop:composition_fgs} for $\nGNO^{[\alpha]}$. Faithfulness~\ref{prop:faithful} and invariance under FGUs~\ref{prop:stability} hold irrespective of any degeneracy, whereas for sub-additivity~\ref{prop:sub-add} and composition with an FGS~\ref{prop:composition_fgs} the subtlety arises in the presence of degeneracies in the Williamson eigenvalues, which we treat explicitly below.

\begin{enumerate} [label={P.\arabic*}]
\item[\ref{prop:faithful}] \textit{Faithfulness}: A pure FGS satisfies $\ket{\psi}=U_G \ket{0}^{\otimes N}$ for some FGU $U_G$, and $\ket{0}^{\otimes N}$ has a covariance matrix in normal form; choosing $U=U_G$ gives $S^{\text{part}}_\alpha(U_G^\dagger\ket{\psi})=0$, hence $\nGNO^{[\alpha]}(\ket{\psi})=0$. Conversely, $\nGNO^{[\alpha]}(\ket{\psi})=0$ requires $U^\dagger\ket{\psi}=\ket{0}^{\otimes N}$ for some FGU $U$, so $\ket{\psi}$ is a pure FGS.

\item[\ref{prop:stability}] \textit{Invariance under FGUs}: Let $\ket{\phi}=(U^*)^\dagger\ket{\psi}$ be the normal-form state achieving the minimum, $\nGNO^{[\alpha]}(\ket{\psi})=S^{\text{part}}_\alpha(\ket{\phi})$. For $\ket{\psi'}=U'\ket{\psi}$ with $U'$ an FGU, the same $\ket{\phi}$ is recovered by $U=U'U^*$, so the minimization is unchanged and $\nGNO^{[\alpha]}(U'\ket{\psi})=\nGNO^{[\alpha]}(\ket{\psi})$.

\item[\ref{prop:sub-add}] \textit{Sub-additivity}: Let $U_A$ and $U_B$ achieve the minima for $\ket{\psi}_A$ and $\ket{\psi}_B$. Since $\Gamma(\ket{\phi}_A \otimes \ket{\phi}_B) = \Gamma(\ket{\phi}_A)\oplus \Gamma(\ket{\phi}_B)$ with $\ket{\phi}_{A(B)}=U_{A(B)}^\dagger\ket{\psi}_{A(B)}$, the product $U_A\otimes U_B$ is a valid candidate in the minimization, so
\begin{equation}
    \begin{split}
        \nGNO^{[\alpha]}&(\ket{\psi}_A \otimes \ket{\psi}_B) = \min_{U} S^{\text{part}}_\alpha(U^\dagger\ket{\psi}_A \otimes \ket{\psi}_B) \\
        &\leq S^{\text{part}}_\alpha(U_A^\dagger \ket{\psi}_A) + S^{\text{part}}_\alpha(U_B^\dagger \ket{\psi}_B) \\
        &= \nGNO^{[\alpha]}(\ket{\psi}_A) + \nGNO^{[\alpha]}(\ket{\psi}_B),
    \end{split}
\end{equation}
where the second line uses the additivity of the participation entropy. Whether the inequality can be strict when the two covariance matrices share an eigenvalue remains an open question.
\end{enumerate}
\begin{enumerate}
    \item[\ref{prop:composition_fgs}] \textit{Invariance under composition with an FGS}:
     Suppose $\ket \psi$ has $k$ noninteracting modes, i.e., its covariance matrix $\Gamma$ has $+1$ as a $k$-fold degenerate Williamson eigenvalue. Let $V$ be an FGU bringing $\Gamma$ to its normal form, so that $V\ket \psi= \ket\phi \otimes \ket 0^{\otimes k}$, where the $iY=\begin{pmatrix} 0 & 1 \\ -1 & 0 \end{pmatrix}$ blocks of the normal form correspond to disentangled qubits in the state $\ket 0$. Any FGU in the minimization likewise brings $\Gamma$ to normal form, and hence---up to a reordering of modes---takes the form $U_Q = (U_{R} \otimes U_{S})V$, with $U_R$ and $U_S$ acting on the interacting and noninteracting modes. By additivity of $S^{\text{part}}$,
    \begin{equation}
    \begin{split}
    \nGNO^{[\alpha]} (\ket{\psi})&= \min_{U_Q} S^{\text{part}}_\alpha(U_Q^\dagger | \psi \rangle) \\ &= \min_{U_R \otimes U_S} S^{\text{part}}(U_R \ket \phi \otimes U_S \ket 0^{\otimes k}) \\
    &= \min_{U_R} S^{\text{part}}(U_R \ket \phi),
    \end{split}
    \end{equation}
    since $\ket 0^{\otimes k}$ already minimizes the participation entropy on the noninteracting modes. An arbitrary FGS $\ket \chi = W \ket{0}^{\otimes N}$ contributes only further $\ket 0$ blocks, so the same argument gives $\nGNO^{[\alpha]} (\ket{\psi} \otimes \ket\chi) = \nGNO^{[\alpha]} (\ket{\psi})$, as claimed.
\end{enumerate}

\subsection{Relation with \texorpdfstring{$\bm{\nGoccu^{[1]}}$}{NG occu [1]}}
\label{sec:relation_occu}

In this section, we show that the two families introduced in this work are closely related---specifically, we show that the occupation number entropy upper-bounds the natural-orbital participation entropy for ${\alpha=1}$:
    \begin{equation}
        \nGoccu^{[1]} \geq \nGNO^{[1]}.
    \end{equation}
This inequality shows that the easily computable $\nGoccu^{[1]}$ is an upper bound on $\nGNO^{[1]}$, and hence on the cost of classically representing the state in an orthonormal Gaussian basis, although the bound may generally be loose.
    
To this end, consider the full dephasing channel $\Delta_{\{s\}}(\rho)=\sum \bra{s} \rho \ket{s} \ketbra{s}{s}$ in the natural-orbital basis $\{\ket{s}\}$. The participation entropy in the natural-orbital basis can be expressed as $\nGNO^{[1]} = S(\rho \Vert \Delta(\rho))$~\cite{baumgratz2014quantifying}. Without loss of generality, we assume that the natural-orbital basis is the computational basis, which implies that the covariance matrix $\Gamma(\rho)$ is in a normal form, i.e., $\Gamma=\bigoplus_{j=1}^N  \begin{pmatrix}0 & r_j\\ -r_j & 0\end{pmatrix}$. One can see that $\Delta_{\{s\}}(\rho)$ has the same covariance matrix as $\rho$, i.e., $\Gamma(\Delta_{\{s\}}(\rho))=\Gamma(\rho)$. Recalling that $\G(\sigma)$ denotes the fermionic Gaussian state with the same covariance matrix as $\sigma$, this implies $\G(\Delta_{\{s\}}(\rho))=\G(\rho)$. Thus, we have
\begin{equation}
\begin{split}
    \nGoccu^{[1]} - \nGNO^{[1]} &= S(\G(\rho)) - S(\Delta_{\{s\}}(\rho)) \\
    &= S(\G(\Delta_{\{s\}}(\rho))) - S(\Delta_{\{s\}}(\rho)) \\
    &= S(\Delta_{\{s\}}(\rho) \Vert \G(\Delta_{\{s\}}(\rho))) \\
    & \geq 0,
\end{split}
\end{equation}
where the third line is shown in Ref.~\cite{lyu2024fermionicgaussiantesting}.
This proves the claim.

\section{Resource theory properties of the relative entropy of fermionic non-Gaussianity}
\label{sec:Mr_properties}

The relative entropy of fermionic non-Gaussianity for mixed states is defined as~\cite{lyu2024fermionicgaussiantesting}
\begin{equation}
    \nGR(\rho) = \min_{\sigma \in \G} S(\rho \Vert \sigma) = S(\rho \Vert \G(\rho)) = S(\G(\rho)) - S(\rho).
\end{equation}
This quantity is known to satisfy the properties~\ref{prop:faithful}--\ref{prop:sub-add}~\cite{lyu2024fermionicgaussiantesting}.

Here, we consider the resource theory whose free operations are Gaussian completely positive trace-preserving (CPTP) maps (without feedforward), which differs from the convex resource theory built on Gaussian protocols with feedforward.
By the Gaussian Stinespring dilation~\cite{bravyi2005lagrangian,caruso2008multimode}, any Gaussian CPTP map can be realized by appending a Gaussian ancilla, applying an FGU to the enlarged system, and tracing out a subset of modes. Monotonicity under all such maps therefore follows from invariance under FGUs~(\ref{prop:stability}), invariance under composition with an FGS~(\ref{prop:composition_fgs}), and monotonicity under partial trace (we denote as P.6 below). In this section, we establish these properties---together with the strong monotonicity~(\ref{prop:strong}) under computational-basis measurements---thereby completing the requirements for a valid non-Gaussianity monotone in the mixed-state setting.

We note that any Gaussian measurement can be obtained by composing the Gaussian primitives: appending an FGS, applying an FGU, performing a computational-basis measurement, and discarding modes. Since strong monotonicity is established below for computational-basis measurements~(\ref{prop:strong}), and the remaining primitives are deterministic free operations under which $\nGR$ does not increase, it follows that $\nGR$ is non-increasing on average under any general Gaussian measurement, i.e., it is a strong monotone in this resource theory.%

\begin{enumerate} [label={P.\arabic*}]
\setcounter{enumi}{3}
\item[\ref{prop:strong}] \textit{Non-increasing, on average, under computational basis measurements}: Without loss of generality, consider a computational-basis measurement of the first mode, $V_{\lambda} = (I + \lambda\, i \gamma_1 \gamma_2)/2$, which projects $\rho$ onto $\rho_{\lambda} = V_{\lambda} \rho V_{\lambda} / p_{\lambda}$ with $p_{\lambda} = \Tr \rho V_{\lambda}$. Applying the same measurement to $\G(\rho)$ yields $\sigma_{\lambda} = V_{\lambda} \G(\rho) V_{\lambda} / q_{\lambda}$, which is again an FGS since computational-basis measurements preserve Gaussianity. Therefore,
\begin{equation}
\begin{split}
     \nGR(\rho) &= S(\rho \Vert \G(\rho)) \\
     &\geq \sum_{\lambda} p_\lambda S(\rho_\lambda \Vert \sigma_\lambda) \\
     &\geq \sum_{\lambda} p_\lambda S(\rho_\lambda \Vert \G(\rho_\lambda)) \\
     &= \sum_{\lambda} p_\lambda \nGR(\rho_\lambda),
\end{split}
\end{equation}
where the second line is the monotonicity of the relative entropy on average under measurements~\cite{vedral1998entanglement}, and the third uses that $\G(\rho_\lambda)$ is the nearest FGS to $\rho_\lambda$~\cite{lyu2024fermionicgaussiantesting}. 

\item[\ref{prop:composition_fgs}] \textit{Invariance under composition with FGS}: It immediately follows from additivity under tensor product.
\end{enumerate}

We now prove an additional property of monotonicity under partial tracing of the first $m$ qubits:
\begin{enumerate} [label={P.\arabic*}]
\setcounter{enumi}{5}
\item \textit{Non-increasing under partial trace}: Without loss of generality, we consider tracing out the first mode. By monotonicity of the relative entropy under partial trace, $S(\Tr_1(\rho) \Vert \Tr_1(\G(\rho)))\leq S(\rho \Vert \G(\rho))$, and since $\Tr_1(\G(\rho)) = \G(\Tr_1(\rho))$, this gives $\nGR(\Tr_1(\rho))\leq\nGR(\rho)$. The identity $\Tr_1(\G(\rho)) = \G(\Tr_1(\rho))$ holds only when tracing the first or last consecutive modes~\cite{Friis2015modeentanglement}; an arbitrary mode is handled by first fermionically swapping it to that position, so monotonicity holds for tracing of any mode.

\end{enumerate}

\section{Properties of \texorpdfstring{$\bm{\mathcal{M}_\Lambda(\rho)}$}{M Lambda (rho)}}
\label{sec:M_Lambda_properties}

In this appendix, we investigate another quantity which quantifies fermionic non-Gaussianity for mixed states. It is based on the Gaussianity condition stated in Eq.~\eqref{eq:Lambda_condition_mixed}, namely that an even state $\rho$ is Gaussian if and only if $[\Lambda, \rho \otimes \rho] = 0$. We investigate again resource-theoretic properties, and then discuss how this quantity can be evaluated on MPS. Before proceeding, we derive a convenient expression for the quantity.

We recall the definition:
\begin{equation}
    \mathcal{M}_\Lambda(\rho) = \frac{1}{4 \Tr(\rho^2)^2} \lVert [\Lambda, \rho \otimes \rho] \rVert_2^2.
\end{equation}
We can simplify the expression as follows,
\begin{equation}
\begin{split} \label{eq:a}
    \lVert [\Lambda, \rho \otimes \rho] \rVert_2^2
    &= \Tr\left[(\Lambda \rho\otimes\rho - \rho\otimes\rho \Lambda)^\dagger(\Lambda \rho\otimes\rho - \rho\otimes\rho \Lambda ) \right] \\
    &= \Tr(\rho\otimes\rho \Lambda^2 \rho\otimes\rho) - \Tr(\rho\otimes\rho \Lambda \rho\otimes\rho \Lambda) \\
    &\quad - \Tr(\Lambda \rho\otimes\rho \Lambda \rho\otimes\rho ) + \Tr(\Lambda \rho^2\otimes\rho^2 \Lambda) \\
    &= 2(\Tr(\Lambda^2 \rho^2\otimes\rho^2) - \Tr(\rho\otimes\rho \Lambda \rho\otimes\rho \Lambda)).
\end{split}
\end{equation}
Next, we substitute the definition of $\Lambda$ in Eq. \eqref{eq:Lambda_def} to obtain
\begin{equation}
\begin{split} \label{eq:b}
    \Tr(\Lambda^2 \rho^2\otimes\rho^2) &= \sum_{j,k} \Tr(\gamma_j \otimes \gamma_j \gamma_k \otimes \gamma_k \rho^2\otimes\rho^2) \\
    &= 2N \Tr(\rho^2)^2 - \sum_{j \neq k} \Tr(i \gamma_j \gamma_k \rho^2)^2,
\end{split}
\end{equation}
and 
\begin{equation}
\begin{split} \label{eq:c}
    \Tr(\rho\otimes\rho \Lambda \rho\otimes\rho \Lambda) &= \sum_{j,k} \Tr(\rho\otimes\rho\gamma_j \otimes \gamma_j \rho\otimes\rho  \gamma_k \otimes \gamma_k ) \\
    &= \sum_{j,k} \Tr(\rho\gamma_j  \rho \gamma_k )^2 .
\end{split}      
\end{equation}
Finally, combining Eqs. \eqref{eq:a}, \eqref{eq:b}, and \eqref{eq:c}, we arrive at 
\begin{equation}
\begin{split} \label{eq:d}
    \mathcal{M}_\Lambda(\rho) = &N - \frac{1}{\Tr(\rho^2)^2} \sum_{j>k}  |\Tr(\gamma_j \gamma_k \rho^2)|^2  \\
    &-\frac{1}{2\Tr(\rho^2)^2} \sum_{j,k} \Tr(\rho \gamma_j \rho \gamma_k)^2.
\end{split}
\end{equation}

For a pure even state $\ket{\psi}$ and density matrix $\rho=\ketbra{\psi}{\psi}$, we have $\rho^2=\rho$ and $\Tr(\rho^2)=1$. Furthermore, $\Tr(\rho \gamma_j \rho \gamma_k)= \bra{\psi} \gamma_j \ket{\psi} \bra{\psi} \gamma_k \ket{\psi}=0$, since $\ket{\psi}$ is an even state. Therefore, the right hand side of Eq.~\eqref{eq:d} reduces to the expression of $\nGoccu^{[2]}(\ket{\psi})$ (Eq.~\eqref{eq:occu_exp}), thus yielding 
\begin{equation}
    \mathcal{M}_\Lambda(\ketbra{\psi}{\psi})=\nGoccu^{[2]}(\ket{\psi}).
\end{equation}

We define the $2N \times 2N$ matrices $\tilde{\Gamma}(\rho)$ and $\Sigma(\rho)$ as  
\begin{equation}
    \tilde{\Gamma}(\rho)\vert_{j,k} = -\frac{i}{2\Tr(\rho^2)} \Tr([\gamma_j,\gamma_k] \rho^2),
\end{equation}
and 
\begin{equation}
    \Sigma(\rho)\vert_{j,k} = \frac{1}{\Tr(\rho^2)} \Tr(\rho\gamma_j\rho\gamma_k),
\end{equation}
where $j,k=1,\dots,2N$. We can write $\mathcal{M}_\Lambda(\rho)$ as
\begin{equation} \label{eq:e}
    \mathcal{M}_\Lambda(\rho) = N - \frac{1}{2}\Tr(\tilde{\Gamma}(\rho)^\dagger \tilde{\Gamma}(\rho)) - \frac{1}{2}\Tr(\Sigma(\rho)^\dagger \Sigma(\rho)).
\end{equation}

\subsection{Resource theory properties}

Now, we prove the properties~\ref{prop:faithful}--\ref{prop:sub-add}.

\begin{enumerate} [label={P.\arabic*}]
\item \textit{Faithfulness}: Faithfulness of $\mathcal{M}_\Lambda(\rho)$ follows immediately from the Gaussianity condition in Eq.~\eqref{eq:Lambda_condition_mixed}.
\item \textit{Invariance under FGUs}: Under an FGU $U$, the state transforms as $\rho\mapsto U\rho U^\dagger$, and since $\Lambda$ commutes with $U\otimes U$, the commutator transforms covariantly, $[\Lambda,\rho\otimes\rho]\mapsto (U\otimes U)[\Lambda,\rho\otimes\rho](U\otimes U)^\dagger$. Its Hilbert-Schmidt norm, and hence $\mathcal{M}_\Lambda$, is therefore unchanged.
\item \textit{Additivity}: For a tensor product $\rho_A \otimes \rho_B$, one can verify that $\tilde{\Gamma}(\rho_A \otimes \rho_B) = \tilde{\Gamma}(\rho_A)\oplus \tilde{\Gamma}(\rho_B)$ and $\Sigma(\rho_A \otimes \rho_B) = \Sigma(\rho_A)\oplus \Sigma(\rho_B)$. Plugging these to Eq.~\eqref{eq:e}, we immediately get additivity $\mathcal{M}_\Lambda(\rho_A \otimes \rho_B) = \mathcal{M}_\Lambda(\rho_A) + \mathcal{M}_\Lambda(\rho_B)$.
\end{enumerate}

\subsection{Computation with MPSs}

Here, we discuss the computation of $\mathcal{M}_\Lambda$ when the state $\rho$ represents the leftmost $N_A$ sites of an MPS. According to Eq.~\eqref{eq:e},
$\mathcal{M}_\Lambda$ can be evaluated by constructing the matrices $\tilde{\Gamma}$ and $\Sigma$. The construction of $\tilde{\Gamma}$ can be performed similarly as in Appendix~\ref{sec:mps_algos}, with complexity $\mathcal{O}(N_A^2 \chi^3)$.
This section focuses on the method to construct $\Sigma$.

\begin{figure}
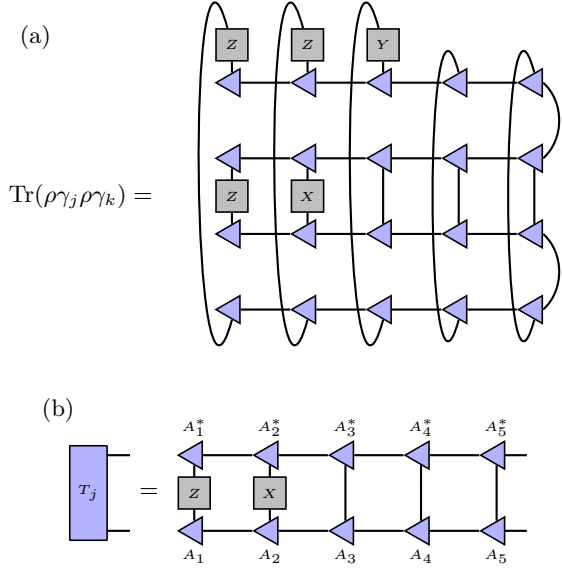

    \centering
    \begin{subfigure}
        \centering
        \mixedMPS
    \end{subfigure}
    \vspace{-10pt}
    \begin{subfigure}
        \centering
        \Tj
    \end{subfigure}
    \caption{(a) MPS contraction for $\Tr(\rho \gamma_j \rho \gamma_k)$, where $\rho$ describes the leftmost $N_A$ qubits. The figure shows an example with $j=3$ and $k=6$. (b) Definition of tensor $T_j$. The figure shows an example with $j=3$.}
    \label{fig:sigma_mps}
\end{figure}

\begin{figure}
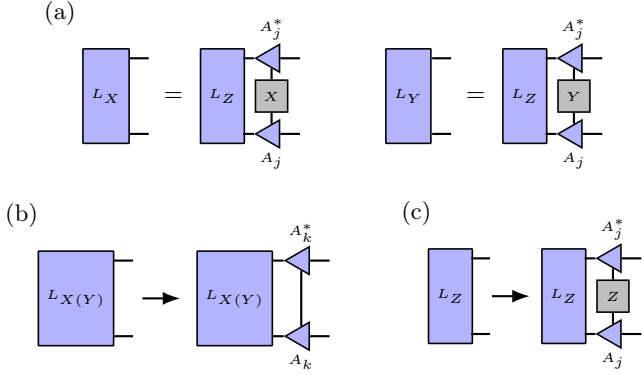

    \centering
    \begin{subfigure}
        \centering
        \initLSigma
    \end{subfigure}
    \begin{subfigure}
        \centering
        \updateLSigma
    \end{subfigure}
    \hfill
    \begin{subfigure}
        \centering
        \updateLZSigma
    \end{subfigure}
    \caption{MPS algorithm to construct the matrix $\Sigma$. (a) Initialization of the environment matrices $L_X$ and $L_Y$. (b) Update of the environment matrices $L_{X(Y)}$. (c) Update of the environment matrix $L_{Z}$.}
    \label{fig:sigma_algo}
\end{figure}

To this end, we need to evaluate the matrix elements in the form of 
\begin{equation}
    \widetilde{\Sigma}_{j,k}= \text{Tr}(\rho \gamma_j \rho \gamma_k),
\end{equation}
 for $j,k=1,\dots,2N$, so that $\Sigma_{j,k}=\widetilde{\Sigma}_{j,k}/\Tr(\rho^2)$. Note that $\Tr(\rho^2)$ is easily accessible in MPS, so this immediately gives $\Sigma_{j,k}$.
We assume that the MPS is in the right canonical form. Then, $\text{Tr}(\rho \gamma_j \rho \gamma_k)$ is graphically represented as in Fig.~\ref{fig:sigma_mps}, which can be computed with cost $O(N_A\chi^3)$. Since there are $N_A(2N_A+1)$ matrix elements, a naive approach results in a total cost of $O(N_A^3 \chi^3)$. This complexity can again be reduced to $O(N_A^2\chi^3)$ using the algorithm detailed below.

The key idea of the algorithm is to first construct and store the tensors $T_j$ illustrated in Fig.~\ref{fig:sigma_mps}(b) for $j=1,\dots,2N_A$. Then, the elements $\Sigma_{j,k}$ can be computed as
\begin{equation} \label{eq:Sigma_jk_mps}
 \widetilde{\Sigma}_{j,k} = \text{Tr}\left[ T_j T_k\right].
\end{equation}
We will now construct all $T_{j}$ by a sweeping procedure.

We first initialize the environment tensor $L_Z=1$. Starting from the first site, $j=1$, we initialize the environment tensors
\begin{equation} \label{eq:init_L_mps_2}
\begin{split}
    L_{X}&=\sum_{s_j, s_j'} X_{s_j, s_j'}(A^{s'_j}_j)^\dagger L_Z  A^{s_j}_j,   \\
    L_{Y}&=\sum_{s_j, s_j'} Y_{s_j, s_j'} (A^{s'_j}_j)^\dagger L_Z A^{s_j}_j,
\end{split}
\end{equation}
which encode the insertion of an $X$ or $Y$ operator at site $j$. 

Next, we sweep forward along the chain, where the environment tensors are updated recursively as 
\begin{equation} \label{eq:update_L_mps_2}
    L_{X(Y)} \to \sum_{s_k} 
     (A_k^{s_k})^\dagger L_{X(Y)} A_k^{s_k},    
\end{equation} 
After reaching the final site of the subsystem, we store the tensors $T_{2j-1}=L_X$ and $T_{2j}=L_Y$.  Then, $L_Z$ is updated as
 \begin{equation} \label{eq:update_Lz_mps}
    L_{Z} \to \sum_{s_j, s_j'} Z_{s_j, s_j'} 
        (A_j^{s_j})^\dagger L_{Z} A_j^{s_j},    
\end{equation}
and the procedure is repeated for the subsequent sites $j$. Once all tensors $T_j$ for $j=1,\dots,2N_A$ have been constructed, we can compute all the elements of $\Sigma$ using Eq.~\eqref{eq:Sigma_jk_mps}.

This procedure, summarized in Algorithm~\ref{alg:sigma_mps}, constructs the full matrix $\Sigma$ with the computational cost of $\mathcal{O}(N_A^2\chi^3)$, the same complexity as the construction of $\tilde{\Gamma}$. Consequently, the evaluation of $\mathcal{M}_\Lambda$ scales as $\mathcal{O}(N_A^2\chi^3)$.

\begin{algorithm}[H]
\caption{Construction of $\widetilde{\Sigma}$ for left subsystems of an MPS}
\label{alg:sigma_mps}
\begin{flushleft}
\textbf{Input}: an MPS $\ket{\psi}$ with local tensors $\{A_i^{s_i}\}_{i=1}^N$, the number of sites $N_A$ \\
\textbf{Output}: The matrix $\widetilde{\Sigma}_{j,k}= \text{Tr}(\rho \gamma_j \rho \gamma_k)$ corresponding to the leftmost $N_A$ sites of the MPS
\end{flushleft}
\begin{algorithmic}[1]
\State Bring MPS into right canonical form 
\State Initialize $L_Z = 1$
\For{$j = 1$ to $N_A$}
    
    \State Compute environments $L_X$ and $L_Y$ as in Eq.~\eqref{eq:init_L_mps_2}
    \For{$k = j+1$ to $N_A$}
    \State Update environments $L_X$ and $L_Y$ as in Eq.~\eqref{eq:update_L_mps_2}

    \EndFor
    \State $T_{2j-1} \gets L_X$  
    \State $T_{2j} \gets L_Y$  
    \State Update environment $L_Z$ as in Eq.~\eqref{eq:update_Lz_mps}
\EndFor

\For{$j = 1$ to $2N_A$}
    \For{$k = 1$ to $2N_A$}
        \State Compute $\widetilde{\Sigma}_{j,k}$ as in Eq.~\eqref{eq:Sigma_jk_mps}
    \EndFor
\EndFor
\end{algorithmic}
\end{algorithm}

\section{Occupation number entropy in translation-invariant systems}
\label{sec:ti_appendix}

In this section, we provide more technical details on the convergence of the occupation number entropy density $\nGoccu^{[\alpha]}/N$ to its value in the thermodynamic limit in translation-invariant systems, as originally discussed in Sec.~\ref{sec:TI} in the main text. In particular, we show that in an infinite system, it is enough to consider two-point correlators up to a cutoff-distance $\cutoff$ to accurately recover the occupation number entropy density if the correlations decay sufficiently fast algebraically or exponentially. Further, we also establish that finite-size corrections to the thermodynamic-limit occupation number density decay exponentially with system size in a gapped phase.
Both results are obtained in two steps: We first view the relevant approximation (either distance cutoff or the finite system size) as an approximation of the momentum-space covariance matrix, and show that the error of this approximation is small. Next, we apply the entropy functional to the approximated momentum-space blocks of the covariance matrix to compute the occupation number entropies, and control how the approximation error propagates through this entropy functional.

Even though the statements in the main text were presented for one-dimensional systems, since the arguments straightforwardly extend to higher-dimensional systems, we will directly adapt our notation to the more general case. This means we write the covariance matrix of the $D$-dimensional infinite system as
\begin{equation}
    \Gamma_{\bm{j},\alpha;\bm{k},\beta} = -\frac{i}{2} \langle [\gamma_{\bm{j},\alpha}, \gamma_{\bm{k},\beta}] \rangle = t_{\alpha\beta}(\bm{k} - \bm{j}),
\end{equation}
where $\bm{j}$ and $\bm{k}$ denote the $D$-dimensional vectors of the lattice sites, and due to translation invariance the elements only depend on the distance vector $\bm{\ell} = \bm{k} - \bm{j}$ between lattice sites. The indices $\alpha$ and $\beta$ correspond to additional degrees of freedom within a unit cell, and label the elements within the $2B\times2B$ block $t(\bm{k} - \bm{j})$. The covariance matrix is block-diagonalized by a multi-dimensional Fourier transform, analogous to Eq.~\eqref{eq:ti_momentum_inf} in the main text, defining the $2B\times2B$ momentum blocks
\begin{equation}
    \label{eq:ti_multidim_momentum_inf}
    \tilde{t}(\bm{p}) = \sum_{\bm{\ell} \in \mathbb{Z}^D} e^{-i2\pi \bm{p} \cdot \bm{\ell}} \, t(\bm{\ell}),
\end{equation}
where the momentum $\bm{p}$ takes continuous values in the Brillouin zone $2\pi\bm{p}\in [-\pi, \pi)^D$.

The occupation number entropy density in the thermodynamic limit $m_{(\infty)}^{[\alpha]} = \lim_{N\to\infty} \nGoccu^{[\alpha]}(N)/N$ is analogous to Eqs.~\eqref{eq:ti_ngoccu_inf_main} and~\eqref{eq:ti_ngoccu_vne_inf_main} in the main text. We define the Hermitian matrix ${H(\bm{p}) = \frac{1}{2}\big(\Id_{2B} + i\tilde{t}(\bm{p})\big)}$ and the integration volume $\mathcal{P} = [-1/2, \, 1/2)^D$, to write the occupation number entropy density for $\alpha\neq1$ as
\begin{equation}
    \label{eq:ti_multidim_ngoccu_inf}
    m_{(\infty)}^{[\alpha]}
    = \frac{1}{1 - 2^{1-\alpha}}
    \left(
    1 - \frac{1}{B} \int_{\mathcal{P}} \d\bm{p} \, \Tr[H(\bm{p})^{\alpha}]
    \right)\!,
\end{equation}
and in the limit $\alpha\to1$ as
\begin{equation}
    \label{eq:ti_multidim_ngoccu_vne_inf}
    m_{(\infty)}^{[1]}
    = - \frac{1}{B} \int_{\mathcal{P}} \d\bm{p} \, \Tr[H(\bm{p})\log_2 H(\bm{p})].
\end{equation}

In general, our strategy will be to approximate the Fourier transform in Eq.~\eqref{eq:ti_multidim_momentum_inf} by either truncating the summation range to some cutoff, or by instead considering the two-point correlators of the finite system, i.e., the finite-size correlators. We then derive a bound
\begin{equation}
    \label{eq:ti_multidim_block_approx}
    \| \tilde{t}(\bm{p}) - \tilde{t}_{\text{appr.}}(\bm{p}) \|_F \leq \epsilon,
\end{equation}
where the Frobenius norm of a matrix $M$ is defined through $\| M \|_F^2 = \sum_{jk} |M_{jk}|^2$, to show that the approximated blocks $\tilde{t}_{\text{appr.}}(\bm{p})$ are close to the exact blocks in the thermodynamic limit. This also translates to a small error for the Hermitian matrix $H(\bm{p})$ that appears in the integrand of Eqs.~\eqref{eq:ti_multidim_ngoccu_inf} and~\eqref{eq:ti_multidim_ngoccu_vne_inf}. To ensure that this small error also leads to a small error after the integral, we need to propagate it through the Tsallis entropies. For notational simplicity, we define the family of functions $f_{\alpha}(x) = x^{\alpha}$ if $\alpha\neq1$ and $f_1(x) = x \log_2 x$ if $\alpha=1$. Then, we need to consider the matrix function $\Tr[f_{\alpha}(H(\bm{p}))]$. We define the action of $f_{\alpha}$ on the Hermitian matrix $H(\bm{p})$ through its spectral decomposition: writing $A = U \, D \, U^{\dagger}$ with $D$ diagonal, we set $f_{\alpha}(A) = U \, f_{\alpha}(D) \, U^{\dagger}$, applying $f_{\alpha}$ to the eigenvalues~\cite{Bhatia1996, Hiai2014}. This spectral definition applies to any continuous $f_{\alpha}$, in particular to the cases $x^{\alpha}$ with $0<\alpha<1$ and $x\log_2 x$, which is non-analytic at $x=0$~\cite{Hiai2014}.
Note that $f_{\alpha}(x)$ is Lipschitz continuous with some Lipschitz constant $K_{\alpha}$ for $\alpha>1$, and also for $\alpha\leq1$ if its domain is a compact interval excluding zero~\footnote{%
    For a differentiable function, the Lipschitz constant is given by the maximal absolute value of the derivative. Here, we are interested in the interval $[\mu,1]$ for some $\mu>0$. For $f_{\alpha}(x)$, this yields the Lipschitz constants $K_{\alpha} = \alpha$ independent of $\mu$ for $\alpha>1$, $K_1=\max \, \{|\log_2(e\mu)|, \, \log_2(e)\}$ with Euler's number $e$ for $\alpha=1$, and $K_{\alpha} = \alpha\mu^{\alpha-1}$ for $0<\alpha<1$.
}. Lipschitz continuity also translates to the corresponding matrix function with respect to the Frobenius norm when applied to normal matrices, if the eigenvalues lie within the domain~\cite[Lemma VII.5.5]{Bhatia1996}. In our case, this means that the normal matrix $H(\bm{p})$ cannot have any eigenvalues that are zero, i.e., any remaining Gaussian modes~\footnote{%
     Since $H(\bm{p})$ is Hermitian and a periodic and continuous function of the momentum $\bm{p}$, its eigenvalues are continuous on the compact set $[-1/2, \,1/2]^D$~\cite{Kato1966, Bhatia1996}. By the extreme value theorem, there exists some momentum for which the smallest eigenvalue takes on its minimal value, which must still be strictly larger than zero. Thus, requiring that the smallest eigenvalue of $H(\bm{p})$ be strictly larger than zero is equivalent to there being some number $\mu>0$ such that all eigenvalues are larger than or equal to $\mu$ for all momenta.
}. Then, we can bound the approximation error introduced in the integrands in Eqs.~\eqref{eq:ti_multidim_ngoccu_inf} and~\eqref{eq:ti_multidim_ngoccu_vne_inf} as
\begin{equation}
\begin{aligned}
    &\big|\Tr[f_{\alpha}(H(\bm{p})) - f_{\alpha}(H_{\text{appr.}}(\bm{p}))] \, \big|\\
    &\qquad\leq \sqrt{2B} \| f_{\alpha}(H(\bm{p})) - f_{\alpha}(H_{\text{appr.}}(\bm{p})) \|_F\\
    &\qquad\leq \sqrt{2B} K_{\alpha} \| H(\bm{p}) - H_{\text{appr.}}(\bm{p}) \|_F
    = \sqrt{2B} K_{\alpha} \epsilon.\hspace{-1em}
\end{aligned}
\end{equation}
In the first step we made use of an inequality relating the trace of a general $2B\times2B$ matrix $M$ to its Frobenius norm: $|\hspace{-0.15em}\Tr[M]|\leq \sqrt{2B} \, \|M\|_F$~\footnote{%
    This bound is obtained by applying the Cauchy-Schwarz inequality to the Hilbert-Schmidt inner product of the identity matrix $\Id$ and the $n \times n$ matrix $M$:
    $|\Tr[M]\,| = |\Tr[\Id^{\!\dagger} \, M]| \leq \sqrt{|\Tr[A^{\dagger} A]|} \, \sqrt{|\Tr[\Id \Id^{\!\dagger}]|} \leq \|\Id\|_F \, \|A\|_F = \sqrt{n} \|A\|_F$.
}. In the second step we applied the Lipschitz continuity of $f_{\alpha}(x)$ with Lipschitz constant $K_{\alpha}$.

If we are interested in obtaining the occupation number entropy density from finite-range data, we can simply evaluate Eqs.~\eqref{eq:ti_multidim_ngoccu_inf} and~\eqref{eq:ti_multidim_ngoccu_vne_inf} with respect to the approximated truncated covariance matrix instead of the exact thermodynamic-limit one; the error introduced by this approximation can be bounded by
\begin{equation}
\begin{aligned}
    \label{eq:ti_multidim_ngoccu_approx}
    &\Big|
    \int_{\mathcal{P}} \! \d\bm{p} \, \Tr[f_{\alpha}( H(\bm{p}) )]
    - \int_{\mathcal{P}} \! \d\bm{p} \, \Tr[f_{\alpha}( H_{\text{appr.}}(\bm{p}) )]
    \Big|\\
    &\qquad\leq \sqrt{2B} K_{\alpha} \epsilon = \mathcal{O}(\epsilon)
\end{aligned}
\end{equation}
using that $\int_\mathcal{P}\d\bm{p} = 1$. Thus, up to a constant, the approximation error of the momentum blocks in Eq.~\eqref{eq:ti_multidim_block_approx} directly propagates through to the approximation error of the occupation number entropy density.

In the finite-size approximation, we run into the additional problem that the correct formula for the occupation number entropy density does not make use of an integral, but rather a sum over the finite-size momenta. Still, for analytic functions this sum converges exponentially fast to the value of the integral~\cite{trefethen2014exponentiallyconvergent}, and for strictly positive inputs $x>0$ the function $f_{\alpha}(x)$ is also analytic.

In the following sections, we give the remaining calculations to derive the bound on the momentum block approximations in Eq.~\eqref{eq:ti_multidim_block_approx} for the different cases, and elaborate further on the convergence of the sum to an integral in the case of finite systems.

\subsection{Finite-range approximation for algebraic decay}
\label{sec:ngoccu_critical}

First, we consider an algebraic decay of the correlation functions in the thermodynamic limit. We require that the two-point correlators that appear in the covariance matrix decay asymptotically as $\mathcal{O}(1/|\bm{\ell}|^{D+\eta})$ with the separation distance $|\bm{\ell}|$ for some $\eta>0$. For large enough distances we can then write $|t_{\alpha\beta}(\bm{\ell})| \leq C/|\bm{\ell}|^{D+\eta}$ for some constant $C>0$. Introducing a cutoff $\cutoff$, we approximate the Fourier transform in Eq.~\eqref{eq:ti_multidim_momentum_inf} by setting all correlators with distance $|\bm{\ell}|>\cutoff$ to zero, which leads to a truncation of the series. The Frobenius norm of the difference between the actual momentum block $\tilde{t}(\bm{p})$ and the finite-range truncated one $\tilde{t}^{(\cutoff)}\!(\bm{p})$ is then given by the weight of the discarded entries. We thus find
\begin{equation}
\begin{aligned}
    \|\tilde{t}(\bm{p}) - \tilde{t}^{(\cutoff)}\!(\bm{p})\|_F^2
    &= \!\sum_{\alpha,\beta=1}^{2B} \Big| \!\sum_{|\bm{\ell}|>\cutoff}\! e^{-i2\pi \bm{p} \cdot \bm{\ell}} \, t_{\alpha\beta}(\bm{\ell})\Big|^2\\
    &\leq (2B)^2 \Big(C\! \sum_{|\bm{\ell}|>\cutoff} 1/|\bm{\ell}|^{D+\eta}\Big)^2.
\end{aligned}
\end{equation}
To more easily bound the sum over $\bm{\ell}$, we want to turn it into a sum over the individual components $\ell_j$.
To this end, we use the inequality of arithmetic and geometric means~\footnote{%
    The inequality of arithmetic and geometric means is a consequence of Jensen's inequality. The logarithm function is concave, and thus for $D$ positive numbers $x_j$ we have $\log(\frac{1}{D}\sum_{j=1}^D x_j) \geq \frac{1}{D}\sum_{j=1}^D \log(x_j) = \frac{1}{D}\log(\prod_{j=1}^D x_j) = \log(\prod_{j=1}^D x_j^{1/D})$. Reexponentiating both sides with the base of the logarithm gives the result. The result also holds for nonnegative numbers, as the inequality trivially holds if any number is zero.
}: we find $|\bm{\ell}|^2 = \sum_{j=1}^D \ell_j^2 \geq D \prod_{j=1}^D \ell_j^{2/D}$, and therefore have $1/|\bm{\ell}|^{D+\eta} \leq D^{-(D+\eta)/2} \prod_{j=1}^D \big(1/\ell_j\big)^{1+\eta/D}$. Further, instead of summing over all $\bm{\ell}$ outside the ball $|\bm{\ell}|\leq\cutoff$, we sum over all $\bm{\ell}$ outside the cube $|\ell_j|\leq\cutoff/\sqrt{D}$. This cube is fully contained inside the ball and all summands are positive, so this constitutes a bound on the sum. Further using symmetry to sum over positive $\ell_j$ only, we have
\begin{equation}
    \label{eq:ti_multidim_alg_intermediate_step}
    \|\tilde{t}(\bm{p}) - \tilde{t}^{(\cutoff)}\!(\bm{p})\|_F
    \hspace{-0.1em}\leq\hspace{-0.16em}
    \frac{2BC}{\sqrt{D}^{D+\eta}}
    \Bigg(
    2 \!\!\sum_{\ell_j>\frac{\cutoff}{\sqrt{D}}} \!\!\!
    \ell_j^{-(1+\frac{\eta}{D})}\!
    \Bigg)^{\!\!D}\!.\!\!\!
\end{equation}
Note that we must sum over integer-valued $\ell_j$, so that the actual summation range starts at $\lceil \cutoff/\sqrt{D}\rceil$ and goes to infinity. To bound the remaining sum, we follow the same basic idea as in Ref.~\cite{Jobst2024}, where similar sums are considered. First, we shift the summation index to start the sum at zero, which also allows to get rid of the integer rounding:
\begin{equation}
    \sum_{\ell_j=\left\lceil\frac{\cutoff}{\sqrt{D}}\right\rceil}^{\infty}
    \ell_j^{-(1+\frac{\eta}{D})}
    \leq \sum_{\ell_j=0}^{\infty}
    \left(
    \ell_j + \tfrac{\cutoff}{\sqrt{D}}
    \right)^{\!-(1+\frac{\eta}{D})}.
\end{equation}
This is then precisely the form of the series representation of the Hurwitz zeta function $\zeta(s,a) = \sum_{j=0}^{\infty} (j+a)^{-s}$ with $s=1+\frac{\eta}{D}>1$ and $a=\cutoff/\sqrt{D}>0$~\cite{Hurwitz_zeta}. For $s>1$ and large positive $a$ the Hurwitz zeta function asymptotically decays as $\mathcal{O}(a^{1-s})$~\footnote{%
    This can be seen by bounding the series by an integral as $\zeta(s,a) = \sum_{j=0}^{\infty} (j+a)^{-s} \leq \int_0^{\infty} \!\d x \, (x-1+a)^{-s} = \frac{1}{1-s} (x-1+a)^{1-s}\big|_0^{\infty} = \frac{1}{s-1} (a-1)^{1-s} = \mathcal{O}(a^{1-s})$.
}. In our case, $1-s = -\eta/D$ leads to an asymptotic decay of the sum with the cutoff $\cutoff$ as $\mathcal{O}\big(\cutoff^{-\eta/D}\big)$.

Plugging this result into Eq.~\eqref{eq:ti_multidim_alg_intermediate_step}, we find that the finite-range approximation leads to momentum blocks that deviate as
\begin{equation}
    \hspace{-1em}\|\tilde{t}(\bm{p}) - \tilde{t}^{(\cutoff)}\!(\bm{p})\|_F
    \hspace{-0.1em}\leq\hspace{-0.1em}
    \frac{2BC\,2^D}{\sqrt{D}^{D+\eta}}
    \mathcal{O}\big(\cutoff^{\!-\frac{\eta}{D}}\big)^{\!D} \!\! = \mathcal{O}(\hspace{-0.08em}\cutoff^{\hspace{-0.04em}-\eta}\hspace{-0.08em})\hspace{-1em}
\end{equation}
from the exact result. This bound is what we required in Eq.~\eqref{eq:ti_multidim_block_approx} in the previous section. From the results there, particularly Eq.~\eqref{eq:ti_multidim_ngoccu_approx}, we have seen that the asymptotic behavior of the error of the occupation number entropy density is given by this error. We thus have that the finite-range approximation error of the occupation number entropy density decays as $\mathcal{O}(\cutoff^{-\eta})$ with the cutoff range $\cutoff$, as claimed in Sec.~\ref{sec:ngoccu_thermo} in the main text.

\subsection{Finite-range approximation for exponential decay}
\label{sec:ngoccu_approx_infinite}

Next, we consider an exponential decay of the correlation functions in the thermodynamic limit, as is the case, for example, in gapped phases of matter. In this case, we can refine the error bound derived above to also decay exponentially with the cutoff range $\cutoff$. We introduce a correlation length $\xi>0$, such that the two-point correlators appearing in the covariance matrix decay asymptotically as $\mathcal{O}(e^{-|\bm{\ell}|/\xi})$ with the separation distance $|\bm{\ell}|$. For large enough distances we can then write $|t_{\alpha\beta}(\bm{\ell})| \leq C e^{-|\bm{\ell}|/\xi}$ for some constant $C>0$. As before, we approximate the Fourier transform in Eq.~\eqref{eq:ti_multidim_momentum_inf} by retaining only terms within the cutoff $|\bm{\ell}|\leq\cutoff$. The Frobenius norm of the difference between the actual momentum block $\tilde{t}(\bm{p})$ in Eq.~\eqref{eq:ti_multidim_momentum_inf} and the finite-range truncated one, $\tilde{t}^{(\cutoff)}\!(\bm{p})$, is then given by
\begin{equation}
\begin{aligned}
    \|\tilde{t}(\bm{p}) - \tilde{t}^{(\cutoff)}\!(\bm{p})\|_F^2
    &= \!\!\sum_{\alpha,\beta=1}^{2B} \Big| \!\sum_{|\bm{\ell}|>\cutoff}\!\! e^{-i2\pi \bm{p} \cdot \bm{\ell}} \, t_{\alpha\beta}(\bm{\ell})\Big|^2\\
    &\leq (2B)^2 \Big(C\! \sum_{|\bm{\ell}|>\cutoff}\! e^{-|\bm{\ell}|/\xi} \Big)^2.
\end{aligned}
\end{equation}
As before, we want to simplify the sum over $\bm{\ell}$ by turning it into a component-wise sum over its individual entries~$\ell_j$. We lower-bound the separation distance as ${|\bm{\ell}| \geq \sum_{j=1}^D |\ell_j| / \sqrt{D}}$~\footnote{%
    This is a consequence of the Cauchy-Schwarz inequality. For a general $D$-dimensional vector $\bm{x}$, consider the $D$-dimensional vector of phases $\bm{\phi}$ with entries $\phi_j = x_j/|x_j|$ such that $\|x\|_1 = \bm{\phi}^{\dagger} \cdot \bm{x}$. Then, we have $\sqrt{D} \, \|\vec{x}\|_2 = \sqrt{\bm{\phi}^{\dagger} \cdot \bm{\phi}} \, \sqrt{\bm{x}^{\dagger} \cdot \bm{x}} \geq |\bm{\phi}^{\dagger} \cdot \bm{x}| = \|x\|_1$.
}, and therefore have that $e^{-|\bm{\ell}|/\xi} \leq \prod_{j=1}^D e^{-|\ell_j|/(\xi\sqrt{D})}$. We also again replace the sum over all $\bm{\ell}$ outside the ball $|\bm{\ell}|\leq\cutoff$ by the sum over all $\bm{\ell}$ outside the cube $|\ell_j|\leq\cutoff/\sqrt{D}$. This gives the bound
\begin{equation}
    \label{eq:ti_multidim_exp_intermediate_step}
    \|\tilde{t}(\bm{p}) - \tilde{t}^{(\cutoff)}\!(\bm{p})\|_F
    \hspace{-0.1em}\leq\hspace{-0.1em}
    2BC \Bigg(2 \!\! \sum_{\ell_j>\frac{\cutoff}{\sqrt{D}}} \!\! e^{-\ell_j/(\xi\sqrt{D})}\Bigg)^{\!\!D}\!.
\end{equation}
Note that, as before, the sum must still run over integer-valued $\ell_j$, so that it actually ranges from $\lceil \cutoff/\sqrt{D} \rceil$ to infinity. We can shift the summation index to start at zero by factoring out the term corresponding to $\lceil \cutoff/\sqrt{D} \rceil$, and also get rid of the integer rounding:
\begin{equation}
    \sum_{\ell_j=\left\lceil\frac{\cutoff}{\sqrt{D}}\right\rceil}^{\infty} e^{-\ell_j/(\xi\sqrt{D})}
    \leq e^{-\cutoff/(\xi D)} \sum_{\ell_j=0}^{\infty} e^{-\ell_j/(\xi\sqrt{D})}.
\end{equation}
The remaining series is then simply a geometric series with solution $\sum_{j=0}^{\infty} r^j = (1-r)^{-1}$ if $|r|<1$. In our case, $r=e^{-1/(\xi\sqrt{D})} < 1$ since both $\xi>0$ and $D>0$, and the result $(1-r)^{-1}$ is a constant as a function of the cutoff $\cutoff$. The result of the series thus decays asymptotically as $\mathcal{O}\big(e^{-\cutoff/(\xi D)}\big)$.

Plugging this result into Eq.~\eqref{eq:ti_multidim_exp_intermediate_step}, we find that the finite-range approximation leads to momentum blocks that deviate as
\begin{equation}
    \hspace{-1em}\|\tilde{t}(\bm{p}) - \tilde{t}^{(\cutoff)}\!(\bm{p})\|_F
    \hspace{-0.1em}\leq\hspace{-0.1em}
    2BC\,2^D \mathcal{O}\Big(\! e^{-\frac{\cutoff}{\xi D}} \!\Big)^{\!D} \!\!
    = \mathcal{O}\Big(\! e^{-\frac{\cutoff}{\xi}} \!\Big)\hspace{-1em}
\end{equation}
from the exact result. This bound corresponds to Eq.~\eqref{eq:ti_multidim_block_approx} in the previous subsection, and is the error that sets the asymptotic behavior of the error of the occupation number entropy density as in Eq.~\eqref{eq:ti_multidim_ngoccu_approx}. We thus have that the finite-range approximation error of the occupation number entropy density decays as $\mathcal{O}\big(e^{-\cutoff/\xi}\big)$ with the cutoff range $\cutoff$, as claimed in Sec.~\ref{sec:ngoccu_gapped_main} in the main text.

\subsection{Finite-size approximation in gapped phases}
\label{sec:ngoccu_gapped}

Finally, we consider a gapped phase and study the convergence of the occupation number entropy density in a finite system to its value in the thermodynamic limit. As in the main text, we denote the size of the finite system as $N=L^DB$, where $L$ is the number of unit cells in each of the $D$ dimensions and $B$ is the number of modes per unit cell. We want to show that the occupation number entropy density in the infinite system can be approximated by its value in the finite system with an exponentially small error in system size, under the assumptions that the two-point correlators in the finite system converge exponentially fast to their value in the infinite system~\cite{hastings2006spectralgap, nachtergaele2006liebrobinson} and that no eigenvalues $\pm i$ exist in the covariance matrix (i.e., there are no Gaussian modes). We first use the triangle inequality to split the approximation error into the two contributions arising from discretizing the momentum integrals by finite sums and from approximating the thermodynamic limit by the finite-size system, and thus bound
\begin{equation}
\begin{aligned}
    &\bigg| \int_{\mathcal{P}} \! \d\bm{p} \, \Tr[f_{\alpha}(H(\bm{p}))]
    - \frac{1}{L^D}\sum_{\bm{q}} \Tr\!\big[
    f_{\alpha}\big( H^{(L)}\!\big(\tfrac{\bm{q}-\bm{\theta}}{L}\big) \big) \big]
    \bigg|  \\
    &\quad\leq\bigg| \int_{\mathcal{P}} \! \d\bm{p} \, \Tr[f_{\alpha}(H(\bm{p}))]
    - \frac{1}{L^D}\sum_{\bm{q}} \Tr\!\big[
    f_{\alpha}\big( H\big(\tfrac{\bm{q}-\bm{\theta}}{L}\big) \big) \big]
    \bigg|  \\
    &\qquad+\bigg|\frac{1}{L^D} \sum_{\bm{q}} \Tr\big[f_{\alpha}\big( H\big( \tfrac{\bm{q}-\bm{\theta}}{L} \big) \big) - f_{\alpha}\big( H^{(L)}\!\big( \tfrac{\bm{q}-\bm{\theta}}{L} \big) \big)\big]
    \Big|,
\end{aligned}
\label{eq:triangle_inequality_in_proof_for_finite_size}
\end{equation}
where $H^{(L)}\!(\tfrac{\bm{q}-\bm{\theta}}{L})$ denotes the finite-size analog of the matrix ${H(\bm{p}) = \frac{1}{2}\big(\Id_{2B} + i\tilde{t}(\bm{p})\big)}$ at momentum $\tfrac{\bm{q}-\bm{\theta}}{L}$ for system size $L$, to be defined more precisely below.
In the following, we first show that the error made by summing over $H(\tfrac{\bm{q}-\bm{\theta}}{L})$ instead of $H^{(L)}\!(\tfrac{\bm{q}-\bm{\theta}}{L})$ in a finite system is only exponentially small in $L$.
Then, we show that summation over $\Tr\!\big[f_{\alpha}\big(H\big(\tfrac{\bm{q}-\bm{\theta}}{L}\big)\big)\big]$ converges exponentially fast to the integral.

We make an assumption on how fast the two-point correlators in a finite system converge to their value in the thermodynamic limit. In gapped systems, there exists a correlation length $\xi$ beyond which the correlations decay exponentially fast as $e^{-|\bm{\ell}|/\xi}$, and we thus expect distant parts to not influence local quantities~\cite{hastings2006spectralgap, nachtergaele2006liebrobinson}. For systems much larger than the correlation length $L\gg2\xi$, we therefore expect local operators, such as two-point correlators with distance $\ell\leq\xi$, to be already converged up to exponentially small corrections. Conversely, one expects that the two-point correlators with distance $\ell>\xi$ are exponentially small in a finite system. Formally, we write this as the elements $t_{\alpha\beta}^{(L)}(\bm{\ell})$ of the finite-system covariance matrix being close to the elements $t_{\alpha\beta}(\bm{\ell})$ of the infinite-system covariance matrix at the same distance:
\begin{equation}
    | t_{\alpha\beta}(\bm{\ell}) - t_{\alpha\beta}^{(L)}(\bm{\ell}) | = \mathcal{O}\Big(e^{-\frac{L}{2\xi}}\Big).
\end{equation}
Note that this holds for two-point correlators with separation distance $\bm{\ell}$ that elementwise satisfies $|\ell_j|\leq\lfloor L/2 \rfloor$. For every lattice direction $\bm{\hat{e}_j}$ along which the distance is $|\ell_j|>\lfloor L/2 \rfloor$, one should instead use translation-invariance with corresponding (anti)periodic boundary conditions $e^{i2\pi\theta_j}$ through the relation $\gamma_{\bm{k},\alpha} \to e^{i2\pi\theta_j} \gamma_{\bm{k}+L\bm{\hat{e}_j},\alpha}$ to reduce the two-point correlator to one whose distance does fulfill $|\ell_j|\leq\lfloor L/2 \rfloor$.

Taking the Fourier transform of the finite system, we now have a discrete set of phases $\omega_{\bm{q}} = e^{-i \frac{2\pi}{L} (\bm{q}-\bm{\theta})}$, with $\bm{q}\in \{ -\lfloor L/2 \rfloor, - \lfloor L/2 \rfloor +1, \dots, \lceil L/2 \rceil -1 \}^D$ labeling the $D$-dimensional discrete momenta and the vector $\bm{\theta}$ keeping track of periodic ($\theta_j=0$) or antiperiodic ($\theta_j=\pi$) boundary conditions along the $j$th dimension. The finite Fourier transform, summing over $\bm{\ell}\in \{ -\lfloor L/2 \rfloor, - \lfloor L/2 \rfloor +1, \dots, \lceil L/2 \rceil -1 \}^D$, then reads
\begin{equation}
\begin{aligned}
    &\tilde{t}^{(L)}_{\alpha\beta}\big(\tfrac{\bm{q}-\bm{\theta}}{L}\big)
    = \sum_{\bm{\ell}} e^{-i\frac{2\pi}{L} (\bm{q}-\bm{\theta}) \cdot \bm{\ell}} \, t_{\alpha\beta}^{(L)}(\bm{\ell}) \\
    &\qquad= \sum_{\bm{\ell}} e^{-i2\pi \frac{\bm{q}-\bm{\theta}}{L} \cdot \bm{\ell}} \, t_{\alpha\beta}(\bm{\ell}) + \mathcal{O}\Big(L^D \, e^{-\frac{L}{2\xi}}\Big).
\end{aligned}
\end{equation}
Up to small corrections, the finite-size momentum blocks $\tilde t^{(L)}\!(\bm{p})$ with momentum $\bm{p} = \frac{\bm{q}-\bm{\theta}}{L}$ are thus equivalent to approximating the infinite-system momentum blocks as given in Eq.~\eqref{eq:ti_multidim_momentum_inf} at the same momentum $\bm{p}$ by a finite-range approximation with cutoff $\cutoff=\lfloor L/2 \rfloor$. In the previous section, we showed that the result of such a finite-range approximation for systems with exponentially decaying correlations is an error of order $\mathcal{O}\big(e^{-\cutoff/\xi}\big)$. To leading order, we thus obtain
\begin{equation}
\begin{aligned}
    \label{eq:ti_multidim_ngoccu_block_approx_finite}
    \|\tilde{t}(\bm{p}) - \tilde{t}^{(L)}\!(\bm{p})\|_F
    \leq \epsilon
\end{aligned}
\end{equation}
with $\epsilon = \mathcal{O}\big(L^D \, e^{-L/(2\xi)}\big)$. This is the finite-size version of Eq.~\eqref{eq:ti_multidim_block_approx}.
From the momentum blocks $\tilde{t}^{(L)}\!(\bm{p})$ of the covariance matrix in the finite system, we can then define the finite-size analog $H^{(L)}\!(\bm{p})$ of the Hermitian matrix $H(\bm{p})$ defined for the infinite system; for valid finite-size momenta $\bm{p} = \frac{\bm{q}-\bm{\theta}}{L}$ it is given by ${H^{(L)}\!(\bm{p}) = \frac{1}{2}\big(\Id_{2B} + i\tilde{t}^{(L)}\!(\bm{p})\big)}$.

In the finite-range approximation, we only needed Lipschitz continuity to directly plug in the above approximation result in Eq.~\eqref{eq:ti_multidim_ngoccu_approx_finite} into the integral formulas in Eqs.~\eqref{eq:ti_multidim_ngoccu_inf} and~\eqref{eq:ti_multidim_ngoccu_vne_inf} to establish the small approximation error of the occupation number entropy density in Eq.~\eqref{eq:ti_multidim_ngoccu_approx}. This time, however, we consider a finite system and so the integrals need to be replaced by sums. We can apply the same argument, though, and get the analog of Eq.~\eqref{eq:ti_multidim_ngoccu_approx} with the integrals replaced by sums:
\begin{equation}
\begin{aligned}
    \label{eq:ti_multidim_ngoccu_approx_finite}
    \bigg|
    &\frac{1}{L^D} \sum_{\bm{q}} \Tr\big[f_{\alpha}\big( H\big( \tfrac{\bm{q}-\bm{\theta}}{L} \big) \big) \big]\\[-0.28cm]
    &\qquad- \frac{1}{L^D} \sum_{\bm{q}} \Tr\big[f_{\alpha}\big( H^{(L)}\!\big( \tfrac{\bm{q}-\bm{\theta}}{L} \big) \big) \big]
    \Big| = \mathcal{O}(\epsilon).
\end{aligned}
\end{equation}
This establishes a bound on the second term in Eq.~\eqref{eq:triangle_inequality_in_proof_for_finite_size}.
We additionally need to investigate how fast the summation over $\Tr\!\big[f_{\alpha}\big(H\big(\tfrac{\bm{q}-\bm{\theta}}{L}\big)\big)\big]$ converges to the integral over $\Tr[f_{\alpha}(H(\bm{p}))]$. To this end, we proceed as follows. Note that in principle we need to integrate $\Tr[f_{\alpha}(H(\bm{p}))]$ over real momenta. We show that when inserting a complex argument $\bm{p} \in \mathbb{C}^D$, the function is holomorphic in a small strip around the real axes. This allows us to use a result presented in Ref.~\cite{trefethen2014exponentiallyconvergent} which guarantees exponentially fast convergence of the sum to the integral.

In the infinite system, the matrix $H(\bm{p})$ is obtained through the Fourier series in Eq.~\eqref{eq:ti_multidim_momentum_inf}. Since the coefficients of this Fourier series decay exponentially fast, this defines a function that is holomorphic in each momentum component separately on the open strip around the real line defined by $|\text{Im}(\bm{p})|<(2\pi\xi)^{-1}$~\footnote{%
    A Fourier series with exponentially decaying coefficients defines a holomorphic function~\cite[Exercise I.4.4]{Katznelson2004}. In our particular case, the Fourier series appearing in Eq.~\eqref{eq:ti_multidim_momentum_inf} has summands that we can bound as $\big| e^{-i2\pi \bm{p} \cdot \bm{\ell}} \, C \, e^{-|\bm{\ell}|/\xi} \big| \leq C \, e^{(2\pi|\text{Im}(\bm{p})| - 1/\xi) |\bm{\ell}|}$, and decay exponentially with $|\bm{\ell}|$ if $2\pi|\text{Im}(\bm{p})| \leq 1/\xi$. In this regime, the Fourier series thus converges uniformly (by the Weierstrass M-test) and the limit of a uniformly convergent sequence of holomorphic functions is again holomorphic.
}. Note that the functions $f_{\alpha}(x)$ are also holomorphic when expanded around $+1$ with a unit radius of convergence---for integer-valued $\alpha\geq2$ the radius of convergence is even infinite. This holomorphicity translates to the matrix function $f_{\alpha}(H(\bm{p}))$, as a function of one component $p_j$, in the regime where the operator norm $\|H(\bm{p})-\Id\|$ lies within the radius of convergence~\cite{Hiai2014}. We thus are interested in ensuring the singular values $\{\sigma_{\beta}(\bm{p})\}_{\beta=1}^{2B}$ of $H(\bm{p})-\Id$ remain smaller than $+1$ in some region around the real line. Note that the singular values are the square roots of the eigenvalues of the matrix function $(H(\bm{p})-\Id)^{\dagger}(H(\bm{p})-\Id)$; and since both the square root function and the individual eigenvalues of Hermitian matrices are continuous~\footnote{%
    Note that more generally, the eigenvalues of matrices are only continuous functions of the matrix entries as an unordered tuple, but not individually. However, for Hermitian matrices the eigenvalues are all real and can thus be uniquely ordered, so that also the individual eigenvalues are continuous functions of the matrix elements~\cite{Kato1966, Bhatia1996}.
}, the singular values $\sigma_{\beta}(\bm{p})$ are continuous functions of $\bm{p}$. For real-valued momenta in the compact set $\bm{p}\in[-1/2,\,1/2]^D$ the matrix $H(\bm{p})-\Id$ is Hermitian, so that the singular values are just the absolute values of the eigenvalues. As before, we assume that $H(\bm{p})$ has no remaining Gaussian modes, i.e., all eigenvalues lie strictly between zero and one. Then, by the extremal value theorem, the spectrum of $H(\bm{p})$ for real momenta is fully contained in the interval $[\mu,\,1-\mu]$ for some $\mu>0$, and also the singular values of $H(\bm{p})-\Id$ fulfill $\sigma_{\beta}(\bm{p})\in[\mu,\,1-\mu]$.
Away from the real line, we can consider any compact subset of $\text{Re}(\bm{p})\in[-1/2,\,1/2]^D$ and $|\text{Im}(\bm{p})|<(2\pi\xi)^{-1}$ that still contains the real momenta, e.g., by restricting to $|\text{Im}(\bm{p})|\leq \delta_0 < (2\pi\xi)^{-1}$, and have that the singular values are uniformly continuous functions of $\bm{p}$ due to the Heine-Cantor theorem. By uniform continuity, there exists a $\delta \leq \delta_0$ such that for any $\bm{p}\in\mathbb{C}^D$ and $\bm{p'} \in [-1/2,\,1/2]^D$ with $|\bm{p} - \bm{p'}| < \delta$ we have $|\sigma_{\beta}(\bm{p}) - \sigma_{\beta}(\bm{p'})| < \mu/2$. This immediately implies that all singular values remain separated from $+1$ in the strip defined by $|\text{Im}(\bm{p})| < \delta$. Note that, for integer-valued $\alpha \geq 2$, the function $f_{\alpha}(x)$ is analytic on $\mathbb{C}$, so that we can simply take $\delta=\delta_0<(2\pi\xi)^{-1}$ in this case.
We have thus shown that the matrix function $f_{\alpha}(H(\bm{p}))$ is holomorphic if the momentum vector lies in the strip defined by $|\text{Im}(\bm{p})| < \delta$, at least as a function of each momentum component separately. The function remains holomorphic under taking the trace, so that $\Tr[f_{\alpha}(H(\bm{p}))]$ is also holomorphic on the same domain. Finally, note that a function that is holomorphic in each variable separately, by Hartogs's theorem, is also holomorphic as a multivariable function.

We are now in position to establish that the sum in Eq.~\eqref{eq:ti_multidim_ngoccu_approx_finite} converges exponentially fast to its integral. For this, we want to iteratively use the following result derived for single-variable functions in Ref.~\cite{trefethen2014exponentiallyconvergent}: for an analytic and periodic function $f(x+1) = f(x)$ that is bounded $|f(x)|\leq M$ in the strip $|\text{Im}(x)| < a$ the sum $I_L = \sum_{j=1}^L f\big(\frac{j}{L}\big)$ converges to the integral $I = \int_0^1 \d x \, f(x)$ exponentially fast as $|I-I_L| \leq 2M /\big(e^{2\pi aL}-1\big)$.
Note that after shifting the summation and integration range sufficiently, which is possible since $\Tr[f_{\alpha}(H(\bm{p}))]$ is periodic in each component of the momentum, we are precisely considering the $D$-dimensional analogue of this problem. We consider the integration component-wise: treating $\Tr[f_{\alpha}(H(\bm{p}))]$ as an analytic function of the first component while keeping all others fixed, we can apply the above theorem with $a = \delta/\sqrt{D}$ to find
\begin{equation}
\begin{aligned}
    &\int_0^1 \! \d p_1 \, \Tr[f_{\alpha}(H(p_1,p_2,\ldots,p_D))] \\[-0.28cm]
    &\quad= \frac{1}{L}\sum_{q_1=1}^{L} \Tr\!\big[
    f_{\alpha}\big( H\big(\tfrac{q_1-\theta_1}{L}, p_2, \ldots, p_D\big) \big) \big] \\[-0.21cm]
    &\hspace{5em}+ \epsilon_1(p_2,\ldots,p_D)
\end{aligned}
\end{equation}
for some function $\epsilon_1(p_2,\ldots,p_D)$ of the remaining momenta that fulfills $|\epsilon_1(p_2,\ldots,p_D)| = \mathcal{O}\big( e^{-2\pi aL} \big)$ where $a = \delta/\sqrt{D}$.
Note that a finite sum of analytic functions remains analytic, so for the second variable we can replace the first integral by its summation while introducing only a small error, and then apply the same theorem for the integration of the second variable:
\begin{equation}
\begin{aligned}
    &\int_0^1 \!\d p_2 \int_0^1 \!\d p_1 \Tr[f_{\alpha}(H(p_1, p_2, p_3, \ldots, p_D))] \\[-0.14cm]
    &= \int_0^1 \!\d p_2\, \frac{1}{L}\!\sum_{q_1=1}^{L}\! \Tr\!\big[
    f_{\alpha}\big( H\big(\tfrac{q_1-\theta_1}{L}, p_2, p_3, \ldots, p_D\big) \big) \big] \hspace{-1em} \\[-0.28cm]
    &\hspace{8em}+ \mathcal{O}\big( e^{-2\pi\delta L/\sqrt{D}} \big) \\[-0.28cm]
    &= \frac{1}{L^2} \!\! \sum_{q_1,q_2=1}^{L} \!\! \Tr\!\big[
    f_{\alpha}\big( H\big(\tfrac{q_1-\theta_1}{L}, \tfrac{q_2-\theta_2}{L}, p_3, \ldots, p_D\big) \big) \big] \hspace{-1em} \\[-0.28cm]
    &\hspace{6em}+ \epsilon_2(p_3,\ldots,p_D)
    + \mathcal{O}\big( e^{-2\pi\delta L/\sqrt{D}} \big),\hspace{-1em}
\end{aligned}
\end{equation}
where again we have some function $\epsilon_2(p_3,\ldots,p_D)$ of the remaining momenta that pointwise decays to zero as $|\epsilon_2(p_3,\ldots,p_D)| = \mathcal{O}\big( e^{-2\pi aL} \big)$ with $a = \delta/\sqrt{D}$. Iterating this $D$ times, we find that
\begin{equation}
\begin{aligned}
    \label{eq:ti_multidim_ngoccu_approx_integral}
    \hspace{-1em}&\bigg|
    \int_{\mathcal{P}} \! \d\bm{p} \, \Tr[f_{\alpha}(H(\bm{p}))]
    - \frac{1}{L^D}\sum_{\bm{q}} \Tr\!\big[
    f_{\alpha}\big( H\big(\tfrac{\bm{q}-\bm{\theta}}{L}\big) \big) \big]
    \bigg|\hspace{-1em}\\[-0.14cm]
    \hspace{-1em}&\hspace{2em}= D \, \mathcal{O}\big( e^{-2\pi\delta L/\sqrt{D}} \big).
\end{aligned}
\end{equation}

In summary, we have shown two steps of approximating the occupation number entropy: first, instead of computing it from the momentum blocks of the finite-system covariance matrix, we can compute it from the blocks of the infinite system at the same momenta, leading to an approximation error in Eqs.~\eqref{eq:ti_multidim_ngoccu_block_approx_finite} and~\eqref{eq:ti_multidim_ngoccu_approx_finite} that decays exponentially with system size as $\mathcal{O}\big(L^D \, e^{-L/(2\xi)}\big)$ because we assumed the two-point correlators of the finite system were exponentially close to those of the infinite system. In a second step, we have shown that the sum over the momentum blocks, where the momentum values are set by the finite system size, also converges exponentially fast to the value of the integral in Eq.~\eqref{eq:ti_multidim_ngoccu_approx_integral}. Taken together, the deviation of the finite-size occupation number entropy density from the infinite system decays as $\mathcal{O}\big(L^D \, e^{-L/(2\xi)}\big) + \mathcal{O}\big( e^{-2\pi a L} \big)$ with $a = \delta / \sqrt{D}$. Since  $\delta < (2\pi\xi)^{-1}$ can be quite small, this decay is typically dominated by $\mathcal{O}\big( e^{-2\pi\delta L/\sqrt{D}} \big)$. This is the result presented in Sec.~\ref{sec:ngoccu_gapped_main} in the main text.

\subsection{Occupation number entropy in translation-invariant MPSs}
\label{sec:ti_mps}

As an explicit setting of gapped systems with translation invariance, we consider the TI MPS of Eq.~\eqref{eq:ti_mps}, with $\chi \times \chi$ tensors $A_s$ ($s \in \{0,1\}$), taken to be injective~\cite{cirac2017mpdo} and of definite parity.

We recall the elements of the covariance matrix in the thermodynamic limit (derived in Sec.~\ref{sec:ti_mps_main}) as
\begin{equation} \label{eq:t12_app}
    \langle Z_j \rangle  = \brap{L_1} E_Z \ketp{R_1} , \quad \langle O_j O'_{j+1} \rangle  = \brap{L_1} E_O E_{O'} \ketp{R_1},
\end{equation}
and
\begin{equation} \label{eq:f}
\begin{split}
    &\langle O_j Z_{j+1} \dots Z_{j+\ell-1} O'_{j+\ell} \rangle \\ & =  e^{i \phi_P(\ell-1)}\sum_j  \delta_{\nu_1,1-\nu_j} \lambda_j^{\ell-1} \brap{L_1} E_O \ketp{R_j^Z} \brap{L_j^Z} E_{O'} \ketp{R_1} .
\end{split}
\end{equation}
where $O,O' \in \{X,Y\}$. Here, we recall that $E_O$ for $O=\{I,X,Y,Z\}$ are the mixed transfer matrices of the MPS, $\lambda_j$ and $\nu_j$ the eigenvalues and parities of the transfer matrix $E$, $\brap{L_j^Z}$ and $\ketp{R_j^Z}$ the left and right eigenvectors of $E_Z$, and $\phi_P\in\{0,\pi\}$.

The momentum-space covariance matrix $\tilde{t}(p)$ is a $2\times2$ skew-Hermitian matrix with vanishing diagonal, so its only independent entry is $\tilde{t}_{12}(p)$. We obtain it by Fourier-transforming the correlators over the separation $\ell$. Using the correlators in Eq.~\eqref{eq:f}, each Fourier sum is an infinite geometric series. For example, for terms involving $\langle X_j Z_{j+1} \dots Z_{j+\ell-1} X_{j+\ell} \rangle$, we obtain
\begin{equation} \label{eq:t3}
\begin{split}
&\sum_{\ell=2}^{\infty} e^{i2\pi p\ell}
\langle
X_j Z_{j+1}\cdots Z_{j+\ell-1} X_{j+\ell}
\rangle \\
&=
e^{i4\pi p} e^{i \phi_P}
\sum_j \delta_{\nu_1,1-\nu_j}
\brap{L_1} E_X \ketp{R_j^Z}
\brap{L_j^Z} E_{X} \ketp{R_1}
\\[4pt]
&\times
\frac{\lambda_j}
{1-e^{i2\pi p} e^{i \phi_P} \lambda_j} ,
\end{split}
\end{equation}
which converges because the parity constraint $\nu_j\neq\nu_1$ excludes the dominant eigenvalue, leaving $|\lambda_j|<1$. Combining with the on-site and nearest-neighbor terms in Eq.~\eqref{eq:t12_app}, we arrive at
\begin{widetext}
\begin{equation}
\begin{split}
    \tilde{t}_{12}(p) =   &\brap{L_1} E_Z \ketp{R_1} - e^{i2\pi p} \brap{L_1} E_X E_X \ketp{R_1} - e^{-i2\pi p} \brap{L_1} E_Y E_Y  \ketp{R_1} \\
    &- e^{i4\pi p} e^{i \phi_P}\sum_j  \brap{L_1} E_X \ketp{R_j^Z} \brap{L_j^Z} E_X \ketp{R_1} \frac{\delta_{\nu_1,1-\nu_j}\lambda_j}{1-e^{i2\pi p} e^{i \phi_P}   \lambda_j} \\
    &- e^{-i4\pi p} e^{i \phi_P}\sum_j  \brap{L_1} E_Y \ketp{R_j^Z} \brap{L_j^Z} E_Y \ketp{R_1} \frac{\delta_{\nu_1,1-\nu_j}\lambda_j}{1-e^{-i2\pi p} e^{i \phi_P}   \lambda_j}.
\end{split}
\end{equation}
\end{widetext}
The eigenvalues of $\tilde{t}(p)$ are given by $\pm |\tilde{t}_{12}(p)|i$, from which the occupation number entropy density follows by numerical integration over $p$ using Eq.~\eqref{eq:ti_multidim_ngoccu_inf} or~\eqref{eq:ti_multidim_ngoccu_vne_inf}. The dominant cost in this computation comes from the diagonalization of the transfer matrix, which costs $\mathcal{O}(\chi^6)$.

For $\alpha=2$, we can use Eq.~\eqref{eq:ti_ngoccu2} to obtain
\begin{widetext}
\begin{equation}
\begin{split}
    \lim_{N\to\infty} \frac{\nGoccu^{[2]}}{N} =\, &1 - \brap{L_1} E_Z \ketp{R_1}^2 -  \frac{1}{2}\brap{L_1} E_X E_X \ketp{R_1}^2 -  \frac{1}{2}\brap{L_1} E_Y E_Y  \ketp{R_1}^2  \\
    &- \frac{1}{2}\sum_{j,j'} \delta_{\nu_1,1-\nu_j} \delta_{\nu_1,1-\nu_{j'}} \brap{L_1} E_Y \ketp{R_j^Z} \brap{L_j^Z} E_Y \ketp{R_1} \brap{L_1} E_Y \ketp{R_{j'}^Z} \brap{L_{j'}^Z} E_Y \ketp{R_1} \frac{   \lambda_j \lambda_{j'}}{1-   \lambda_j \lambda_{j'}} \\
    &- \frac{1}{2}\sum_{j,j'} \delta_{\nu_1,1-\nu_j} \delta_{\nu_1,1-\nu_{j'}} \brap{L_1} E_X \ketp{R_j^Z} \brap{L_j^Z} E_X \ketp{R_1} \brap{L_1} E_X \ketp{R_{j'}^Z} \brap{L_{j'}^Z} E_X \ketp{R_1} \frac{   \lambda_j \lambda_{j'}}{1-   \lambda_j \lambda_{j'}},
\end{split}
\end{equation}
\end{widetext}
for TI MPSs.

\subsubsection{Example} \label{sec:ti_mps_example}

As a concrete illustration, we consider the single-parameter family of MPSs with local tensors~\footnote{This MPS wavefunction is related to the one in Ref.~\cite{PhysRevLett.97.110403} by a local Hadamard transformation.}
\begin{align}
A_0=\frac{1}{\sqrt{2}}\begin{pmatrix}
1 &g\\
1 &1
\end{pmatrix},\;A_1=\frac{1}{\sqrt{2}}
\begin{pmatrix}
-1 &-g\\
1 &1
\end{pmatrix}.
\end{align}
The MPS has a $\mathbb{Z}_2$ symmetry with the relation $(-1)^s A_s =  U_P A_s U_P^\dagger$ with $U_P = \tfrac{1}{\sqrt{g}}\left(\begin{smallmatrix} 0 &g\\ 1 &0 \end{smallmatrix}\right)$.
Note that the states are FGS for any $g$~\cite{PhysRevLett.97.110403}.
These states are the ground-states of the cluster-Ising model on a solvable line~\cite{skeleton2021,PhysRevLett.97.110403,PhysRevResearch.4.L022020},
\begin{align}
\mathcal{H}_{\text{CI}} = 2(g^2-1)\sum_{\ell=1}^{L-1} X_\ell X_{\ell+1}
-(g+1)^2\sum_{\ell=1}^{L}Z_\ell \nonumber \\
+(g-1)^2\sum_{\ell=1}^{L-2} X_\ell Z_{\ell+1}X_{\ell+2}.
\label{TCIM_ham}
\end{align}
The model hosts a phase transition at $g=0$.
At special values of $g$, the ground state reduces to a cluster state for $g=-1$, to a paramagnetic state for $g=1$, and to a GHZ-type state (in the $X$ basis) at the critical point $g=0$.

In the following we focus on the gapped phase with $g<0$; the case $g>0$ can be
treated analogously. The nonzero correlators are
\begin{equation}
\begin{split}
    \langle X_j X_{j+1} \rangle &= \frac{1-g^2}{(1-g)^2}, \\
    \langle X_j Z_{j+1} \cdots Z_{j+\ell-1} X_{j+\ell} \rangle &= \frac{4g}{(1+g)^2} \left(\frac{1+g}{1-g}\right)^\ell.
\end{split}
\end{equation}
with all other correlators vanishing. This yields the momentum-space covariance matrix
\begin{equation}
\begin{split}
    \tilde{t}_{12}(p) =  - e^{i2\pi p}\frac{1-g^2}{(1-g)^2} - \frac{e^{i4\pi p} 4g}{(1-g)^2-e^{i2\pi p}(1-g^2)}.
\end{split}
\end{equation}
One can verify that $|\tilde{t}_{12}(p)|=1$ for all $2\pi p\in(-\pi,\pi]$.
Consequently, the eigenvalues of the matrix $t(p)$ are $\pm i$,
as expected for an FGS.

\section{Non-Gaussianity of pure stabilizer states} 
\label{sec:ng_stabilizer_appendix}

In this appendix we provide further details about the resource theory of stabilizer fermionic non-Gaussianity. In this framework, the free states are stabilizer fermionic Gaussian states (S-FGSs) and the free operations are finite sequences of stabilizer Gaussian protocols, i.e., those mapping pure S-FGSs to pure S-FGSs. We prove several resource-theoretic properties, in particular monotonicity, of the quantities studied in this paper within this restricted resource theory, and we give further details about the computation of the non-Gaussianity for stabilizer states, complementing Sec.~\ref{sec:pure_stab}. As in that section, we consider a stabilizer state $\ket{\psi}$ with $N$ qubits, described by a stabilizer group $\mathcal{S}$ generated by the Pauli strings $\{g_1,\dots, g_N\}$. Throughout this appendix, monotonicity always refers to monotonicity under stabilizer Gaussian protocols.

This appendix is organized as follows: We start by providing an algorithm to compute the covariance matrix for stabilizer states with a complexity scaling as $\mathcal{O}(N^3)$. Then, we discuss the monotonicity of the natural-orbital participation entropy $\nGNO^{\text{S-FGS}}$, which we show to be independent of the R\'enyi index $\alpha$. Subsequently, we discuss how to compute $\nGNO^{\text{S-FGS}}$ for pure stabilizer states, proving that minimizing the participation entropy is equivalent to finding the stabilizer FGS whose stabilizer group has the largest intersection with that of the original state. Then, we present a stochastic method to approximate the value of $\nGNO^{\text{S-FGS}}$, and show an example of its evolution for the doped SWAP model. Finally, we show how to construct a given stabilizer state as a superposition of stabilizer FGSs.

\subsection{Efficient construction of the covariance matrix for stabilizer states}
\label{subapp: construction covmat}

In this section we are interested in the complexity of computing the covariance matrix for a stabilizer state. A naive algorithm would consist on calculating the expectation value of each pair of Majoranas. Since there are $\mathcal{O}(N^2)$ such pairs, and computing each expectation value for a stabilizer state has complexity $\mathcal{O}(N^2)$~\cite{aaronson2004improved}, the total complexity of this method is $\mathcal{O}(N^4)$. However, we present an algorithm which serves to determine which elements in the covariance matrix are non-zero. At most $N$ of them will be nonzero, and for those the expectation value can be determined using the standard method with complexity $\mathcal{O}(N^2)$~\cite{aaronson2004improved}. Thus the overall complexity of the algorithm is $\mathcal{O}(N^3)$. We describe this algorithm in detail in the following.

Let $M$ denote the stabilizer table of the state, where rows are binary vectors representing the stabilizer generators. Each such row vector has the form $\bm{v} = (x_1, \dots, x_N, z_1, \dots, z_N)$. We then define $\bar{\bm{v}} = (z_1, \dots, z_N, x_1, \dots, x_N)$ to be the vector obtained by exchanging the $X$ and $Z$ components of $\bm{v}$. A Pauli string corresponding to $\bm{v}$ is a stabilizer of the state (up to an overall sign) if and only if $M \bar{\bm{v}} = \bm{0}_{2N}$, where $\bm{0}_{2N}$ denotes the zero vector of length $2N$. We define the vectors $\bm{e}_{X_i}$ and $\bm{e}_{Z_i}$ to be the basis vectors corresponding to the Pauli operators $X_i$ and $Z_i$ respectively. Then, the vectors $M \bar{\bm{e}}_{X_i}$ and $M \bar{\bm{e}}_{Z_i}$ correspond to the columns of the stabilizer table $M$.

The idea of the algorithm is to determine the quadratic operators that are stabilizers of the state, which is achieved by checking the expectation value of all operators of the form $\gamma_i\gamma_j$ for $j>i$ and fixed $i$. This can be done iteratively using the known values of $M \bar{\bm{e}}_{X_k}$ and $M \bar{\bm{e}}_{Z_k}$, since all the operators will have the form $O_kZ_{k+1}\dots Z_{\ell-1}O_\ell$, with $O=X,Y$ and $k<\ell$. We distinguish two cases. If $i$ is odd, the first case to consider is whether $\gamma_i\gamma_{i+1}=Z_{(i+1)/2}$ is a stabilizer, which can be done from the value of $M\bar{\bm{e}}_{Z_{(i+1)/2}}$. The other operators $\gamma_i\gamma_j$ for $j>i+1$ will act with the Pauli operator $Y_{(i+1)/2}$ on qubit $(i+1)/2$, thus we start the left environment as $\bm{L}_P= M\bar{\bm{e}}_{X_{(i+1)/2}}+M\bar{\bm{e}}_{Z_{(i+1)/2}}$. Instead, if $i$ is even each operator will act with the Pauli $X_{i/2}$ on qubit $i/2$, so we set the left environment to be $\bm{L}_P = M \bar{\bm{e}}_{X_{i/2}}$. Now, we loop over all qubits $k=(i+1)/2 +1, \dots, N$ for odd $i$ or over $k=i/2 +1, \dots, N$ for even $i$. At each step, the operator $\gamma_{i}\gamma_{2k-1}$ is a stabilizer if $\bm{L}_P + M \bar{\bm{e}}_{X_{k}} = \bm{0}_{2N}$ and $\gamma_{i}\gamma_{2k}$ is a stabilizer if $\bm{L}_P + M \bar{\bm{e}}_{X_{k}} + M \bar{\bm{e}}_{Z_{k}} = \bm{0}_{2N}$. Finally, $\bm{L}_P$ is updated to $\bm{L}_P\rightarrow \bm{L}_P + M \bar{\bm{e}}_{Z_k}$. Note that whenever we find $j$ such that $\expval{\gamma_i\gamma_j}=1$, we can stop the loop and continue to the next row, since there can be at most one quadratic stabilizer involving the Majorana operator $\gamma_i$. This procedure has a complexity $\mathcal{O}(N^2)$ for each $i$, therefore $\mathcal{O}(N^3)$ in total.

\subsection{Monotonicity of \texorpdfstring{$\bm{\nGNO^{\mathrm{S-FGS}}}$}{NG NO S-FGS}} \label{sec:monotone_OG_stab}

The stabilizer natural-orbital participation entropy $\nGNO^{\mathrm{S-FGS}}$ is defined to be the participation entropy in the natural-orbital basis, with the particularity that the minimization is performed over FGUs that are also Clifford unitaries.

In this section, we show that $\nGNO^{\mathrm{S-FGS}}$ satisfies monotonicity under stabilizer Gaussian protocols for stabilizer states. To this end, we will first define the quantity $\nGOG^{\mathrm{S-FGS}}$ as the stabilizer orthogonal Gaussian participation entropy, which is defined by Eq.~\eqref{eq:M_og} with the minimization performed over the set of braiding gates. Note that we do not specify the R\'enyi index $\alpha$, since for stabilizer states the participation entropy is independent of $\alpha$~\cite{SiTu22}. We start the discussion of the monotonicity of $\nGOG^{\mathrm{S-FGS}}$ by establishing two properties.

First, we prove that $\nGOG^{\mathrm{S-FGS}}$ is non-increasing, on average, under measurements of $i\gamma_j \gamma_k$. Without loss of generality we consider the measurement of $i\gamma_1 \gamma_2$. To show this, let $\{\ket s\}$ be the orthonormal Gaussian stabilizer basis which minimizes the Shannon relative entropy (see Appendix~\ref{sec:relation_occu}), such that $\nGOG^{\mathrm{S-FGS}}=S(\rho\Vert\Delta_{\{\ket s\}}(\rho))$, where $\Delta_{\{\ket s\}}(\rho)=\sum \bra{s} \rho \ket{s} \ket{s} \bra{s}$ is the full dephasing channel. Let $\sigma_{\lambda} = V_{\lambda} \Delta_{\{\ket s\}}(\rho) V_{\lambda} / q_{\lambda}$, where $q_{\lambda} = \Tr \Delta_{\{\ket s\}}(\rho) V_{\lambda}$. Since $\{\ket s\}$ is a stabilizer basis, the Pauli measurement maps the basis $\{\ket s\}$ to $\{\ket{s^{\prime}}\}$, which is again a stabilizer basis. This implies that $\sigma_{\lambda}$ is diagonal in the $\{\ket{s^{\prime}}\}$ basis~\cite{TaFr25}. Therefore,
\begin{equation} 
\begin{split} \label{eq:strong_og}
     \nGOG^{\mathrm{S-FGS}} &= S(\rho \Vert \Delta_{\{\ket s\}}(\rho)) \\ 
     &\geq \sum_{\lambda} q_\lambda S(\rho_\lambda \Vert \sigma_\lambda) \\
     &\geq \sum_{\lambda} q_\lambda S(\rho_\lambda \Vert \Delta_{\{\ket{s^{\prime}}\}}(\rho_\lambda)) \\
     &\geq \sum_{\lambda} q_\lambda \nGOG^{\mathrm{S-FGS}}(\ket{\psi_\lambda}) .
\end{split}
\end{equation}
The third line follows from the fact that $\Delta_{\{\ket{s^{\prime}}\}}(\rho)$
is the nearest diagonal state (in the $\{\ket{s^{\prime}}\}$ basis) to $\rho_\lambda$ with respect to quantum relative entropy~\cite{baumgratz2014quantifying}.

Next, we prove that $\nGOG^{\mathrm{S-FGS}}$ is invariant when discarding Gaussian stabilizer states, namely
\begin{equation} \label{eq:OG_inv_comp}
    \nGOG^{\mathrm{S-FGS}}(\ket{u}\otimes \ket{\psi})=\nGOG^{\mathrm{S-FGS}}(\ket{\psi})\,,
\end{equation}
for any Gaussian stabilizer state $\ket{u}$. By using sub-additivity (see Sec.~\ref{sec:og_properties}), it suffices to prove that
\begin{equation}\label{eq:ineq_1}
    \nGOG^{\mathrm{S-FGS}}(\ket{\psi}\bra{\psi})\leq  \nGOG^{\mathrm{S-FGS}}(\ket{u}\bra{u}\otimes \ket{\psi}\bra{\psi}).
\end{equation}
To see this, let $\{\ket s\}$ be the orthonormal Gaussian basis which minimizes the Shannon entropy, such that $\nGOG^{\mathrm{S-FGS}}=S(\rho\Vert\Delta_{\{\ket s\}}(\rho))$. As before, since $\{\ket s\}$ is also a stabilizer basis, the partial tracing  $\Tr_1(\Delta_{\{\ket s\}}(\rho)))$ is diagonal in a stabilizer basis, denoted by $\{\ket{s^{\prime}}\}$~\cite{TaFr25}. Thus, the proof of monotonicity proceeds similarly as in Eq.~\eqref{eq:strong_og}.
This shows Eq.~\eqref{eq:OG_inv_comp}.

As a consequence of the previous properties, it follows that $\nGOG^{\mathrm{S-FGS}}$ is a monotone with respect to Gaussian stabilizer protocols, including for arbitrary sequences of measurements followed by braiding operations.

Moreover, for stabilizer states, any Gaussian mode can be brought to the computational basis by braiding gates, which implies that the minimization of the participation entropy can be done on the remaining non-Gaussian modes. This precisely corresponds to the definition of $\nGNO^{\mathrm{S-FGS}}$, which implies $\nGNO^{\mathrm{S-FGS}}=\nGOG^{\mathrm{S-FGS}}$ for stabilizer states. It follows that $\nGNO^{\mathrm{S-FGS}}$ is a monotone with respect to stabilizer Gaussian protocols.

\subsection{Computation of \texorpdfstring{$\bm{\nGNO^{\mathrm{S-FGS}}}$}{NG NO S-FGS}}
\label{app: equivalence optimizations}

Here we will prove that computing $\nGNO^{\mathrm{S-FGS}}$ (i.e., minimizing the participation entropy of the stabilizer state over braiding gates) corresponds to finding the stabilizer FGS $\ket{\phi_\text{FGS}}$ whose stabilizer group $\mathcal{S}(\ket{\phi_\text{FGS}})$ has the maximal number of stabilizers in common with the stabilizer group of the original state, $\mathcal{S}(\ket{\psi})$. To prove this result, we will show that the participation entropy of the state $U\ket\psi$, where $U$ is a braiding gate, can be expressed in terms of the size of the intersection of two stabilizer groups: that of the original state and that of the stabilizer FGS obtained by applying $U$ to a computational-basis state. With this result, we show that minimizing the participation entropy over braiding gates is equivalent to maximizing the number of shared stabilizers over stabilizer FGSs.

Given a pure stabilizer state $\ket{\psi}$, we define the sign-free stabilizer group $\mathcal{S}(\ket{\psi})$ as the group of Pauli operators that stabilize $\ket{\psi}$ up to a sign. Namely,
\begin{equation} \label{eq:stab_group_pure}
\mathcal{S}(\ket{\psi}) = \left\lbrace P \in \mathcal{P}_N : P \ket{\psi}  = \pm\ket{\psi} \right\rbrace \, .
\end{equation}

We define a stabilizer Majorana basis to be a basis of the Hilbert space where each basis state is a stabilizer FGS, i.e., defined by $\{i\gamma_{\sigma(1)}\gamma_{\sigma(2)}, \dots, i\gamma_{\sigma(2N-1)}\gamma_{\sigma(2N)}\}$ for a given permutation $\sigma$. Each basis state is then determined by a choice of signs in the stabilizer operators. For $\sigma$ being the identity permutation, the stabilizer Majorana basis corresponds to the computational basis. 

Let now $U_\sigma$ be the braiding unitary that transforms the basis defined by $\sigma$ to the computational basis. This unitary performs the Majorana swaps $\sigma(i) \rightarrow i$. Now we will prove that the following identity holds:
\begin{equation}
    S^{\text{part}}(U_\sigma\ket{\psi}) = N-\log_2|\mathcal{S}(\ket{\psi})\cap\mathcal{S}_\sigma|,
\end{equation}
where $\mathcal{S}_\sigma$ is the stabilizer group that corresponds to the stabilizer FGS with permutation $\sigma$.  Note that the participation entropy for a pure stabilizer state is given by~\cite{SiTu22} 
\begin{equation}
    \label{eq:participation_stabilizer}
    S^{\text{part}}(\ket{\psi})= \text{rank}_{\mathbb{Z}_2} (M_X),
\end{equation}
where $M_X$ is the $X$ part of the stabilizer table, i.e., the binary matrix whose $(i,j)$ entry is 1 if the $i$th stabilizer generator contains an $X$ or $Y$ operator on qubit $j$, and 0 otherwise. 

Observe that the size of $\mathcal{S}(\ket{\psi})\cap\mathcal{S}_\sigma$ is invariant under unitary transformations in both states. Then, we use $U_\sigma$ to bring $\mathcal{S}_\sigma$ into the computational basis state, i.e., with stabilizers $\{Z_1,\dots,Z_N\}$. Thus intersection with $\mathcal{S}(U_\sigma\ket{\psi})$ corresponds to all the stabilizers that only contain $Z$ operators. Therefore, $N-\log_2|\mathcal{S}(\ket{\psi})\cap\mathcal{S}_\sigma|$ counts how many linearly independent stabilizers of $U_\sigma\ket\psi$ contain at least one $X$ operator. But this gives exactly the participation entropy of the stabilizer state $U_\sigma \ket{\psi}$, see Eq.~\eqref{eq:participation_stabilizer}. Thus, minimizing over the braiding circuits corresponds to finding the stabilizer FGS with maximum stabilizer group overlap.

\subsection{Heuristic algorithm for \texorpdfstring{$\bm{\nGNO^{\mathrm{S-FGS}}}$}{NG NO S-FGS}}
\label{app:nGNO approximate stabilizers}

Since the total number of stabilizer FGSs is $(2N-1)!!$, a brute-force algorithm to find $\nGNO^{\mathrm{S-FGS}}$ by trying all possibilities is exponentially hard in $N$. Here we provide an efficient algorithm to find an upper bound for this quantity by using a stochastic optimization, with a method similar to simulated annealing~\cite{KiGe83}. Let $\ket{\psi}$ be the stabilizer state. We want to find the braiding circuit $U_B$ that minimizes the participation entropy of the state $U_B\ket{\psi}$. 
\begin{enumerate}
    \item Select two random (and different) Majorana operators $\gamma_i$ and $\gamma_j$, and let $U_{ij}$ be the braiding gate that exchanges them.
    \item Compute the quantity
    \begin{equation}
        p = e^{-\left(S^{\text{part}}(U_{ij}U_B\ket{\psi})-S^{\text{part}}(U_B\ket{\psi})\right)/T},
    \end{equation}
    with a given temperature $T$.
    \item The braiding gate $U_{ij}$ is applied with probability $p$. 
\end{enumerate}
We define a sweep to be $N$ of such steps. After each sweep the temperature is reduced.

As an illustration of this method, we discuss the evolution of the non-Gaussianity in a random circuit. We consider the following set-up, as depicted in Figure~\ref{fig:circuit_evolution}a: At each time step we apply a random braiding gate acting on the whole system, which corresponds to randomly shuffling the columns in the stabilizer table $M_{\text{maj}}$. Then, we apply a SWAP gate in two consecutive qubits. We use the simulated annealing method with decreasing temperatures from 1 to $10^{-4}$.

\begin{figure}
    \centering
    \includegraphics{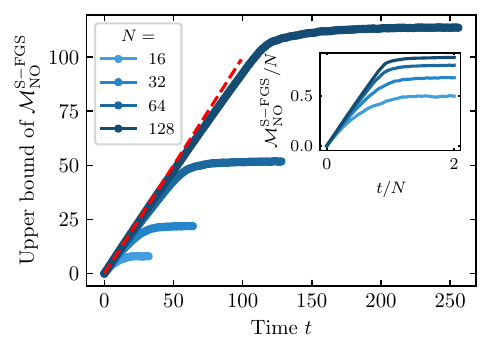}
    \caption{Evolution of the upper bound on $\nGNO^{\mathrm{S-FGS}}$ calculated via the simulated annealing method. The inset shows how this upper bound approaches $N$ as system size is increased. The red dashed line indicates linear corresponds to  ${\nGNO^{\mathrm{S-FGS}}(t)=t}$.}
    \label{fig:stabilizer_nOG}
\end{figure}

In Fig.~\ref{fig:stabilizer_nOG} we show the evolution of an upper bound for $\nGNO^{\mathrm{S-FGS}}$ obtained via the simulated annealing method. To calculate this quantity after each step we initialize the simulated annealing with the optimal braiding circuit found after the previous application of a SWAP gate. We observe that $\nGNO^{\mathrm{S-FGS}}$ can at most increase by 1 after each SWAP application, as also follows from nullity bound~\cite{MeHe25}. The initial growth follows this tendency.

\subsection{Construction of the FGS superposition}
\label{sec:ng_stabilizer_superposition}

In this appendix, we demonstrate how an arbitrary parity-preserving stabilizer state $\ket{\psi}$ with stabilizer group $\mathcal{S}$ can be constructed as a superposition of stabilizer FGSs.
This construction relies on choosing a basis of stabilizer FGSs. While the decomposition is valid for an arbitrary basis, the optimal choice is determined by the minimization required to compute $\nGNO^{\mathrm{S-FGS}}$, as it yields the minimum number of terms in the superposition.

Let $\ket{\phi}$ be a stabilizer FGS with stabilizer group  $\mathcal{S}_{\text{FGS}}$. Let $\{g_1,\dots,g_{N-m}\}$ be the generators of the intersection $\mathcal{S}\cap\mathcal{S}_{\text{FGS}}$. We complete the set of generators for the target state $\ket\psi$ by choosing $m$ additional stabilizers, denoted $g_{N-m+1},\dots,g_N$, such that $\mathcal{S}=\langle g_1,\dots,g_N\rangle$.

To construct the superposition, we must choose $m$ operators $h_1,\dots,h_m$ to complete the stabilizer set of the FGS, such that $\mathcal{S}_{\text{FGS}}=\langle g_1,\dots,g_{N-m}, h_1,\dots,h_m\rangle$. We furthermore choose these operators with the additional constraint that $\{h_j, g_{N-m+j}\}=0$ and $[h_j, g_i]=0$ for $i\neq N-m+j$. Such a choice of operators is guaranteed to exist~\cite{aaronson2004improved}.

Denote by $\ket{\phi_{n_1,\dots,n_m}}$ the stabilizer FGS with stabilizer group $\langle g_1,\dots,g_{N-m}, (-1)^{n_1}h_1,\dots,(-1)^{n_m}h_m\rangle$, with $n_k\in \{0,1\}$. Then, the state $\ket{\psi}$ can be written as:
\begin{equation}
    \ket\psi=\frac{1}{2^{m/2}}\sum_{n_1,\dots,n_m} \ket{\phi_{n_1,\dots,n_m}}.
\end{equation}
To prove this equation, we will show that the generating stabilizers of $\ket\psi$, $\mathcal{S}=\langle g_1,\dots,g_{N}\rangle$, are also stabilizers of the right-hand side (r.h.s.). First, note that the intersection generators $g_1,\dots,g_{N-m}$ are stabilizers of the r.h.s. by construction of the states.

Next, consider a generator $g_j$ with $j>N-m$. The key observation is that $$g_j\ket{\phi_{n_1,\dots,n_m}} = \ket{\phi_{n_1,\dots,n_j\oplus1,\dots,n_m}}.$$
This relation holds since $g_j$ commutes with all generators of $\ket{\phi_{n_1,\dots,n_m}}$ except for $(-1)^{n_j}h_j$, with which it anticommutes. The anticommutation introduces a minus sign, thus flipping the sign of the stabilizer. Therefore, applying $g_j$ to the superposition gives
$$g_j\sum_{n_1,\dots,n_m} \ket{\phi_{n_1,\dots,n_m}} = \sum_{n_1,\dots,n_m} \ket{\phi_{n_1,\dots,n_j\oplus1,\dots,n_m}}.$$
Since the summation covers all binary strings $(n_1,\dots,n_m)$, the operation simply permutes the terms in the sum, leaving the overall state invariant. Therefore, $g_j$ is a stabilizer of the r.h.s.

Since the superposition is stabilized by all generators of $\mathcal{S}$, it is equivalent to $\ket\psi$, up to a phase factor. Note that the constructed superposition consists of $2^m$ states. The optimal choice of stabilizer FGS minimizes $m$, yielding a complexity $2^{\nGNO^{\mathrm{S-FGS}}}$.

\section{Monotonicity of \texorpdfstring{$\bm{\M_\Lambda}$}{} under completely positive stabilizer Gaussian operations}
\label{sec:ng_stabilizer_mixed_state_monotone}

Here, we prove that $\M_\Lambda$ obeys monotonicity under general completely positive stabilizer Gaussian operations, where the free states are a subset of mixed stabilizer states~\cite{Audenaert2005} that arise under partial tracing of pure stabilizer states.
Note that any CP Gaussian operation $\mathcal{E}$ can be expressed as~\cite{SpSc18}
\begin{equation}
    \mathcal{E}(\rho) = \Tr_{23}\left[\rho_\mathcal{E}^{12} \rho^3 \ket{\Phi_{2N}^+}^{23}\bra{\Phi_{2N}^+} \right],
\end{equation}
where
\begin{equation}
    \rho_\mathcal{E} = \mathcal{E} \otimes I  \ketbra{\Phi_{2N}^+}{\Phi_{2N}^+}
\end{equation}
is a Gaussian state and $\ketbra{\Phi_{2N}^+}{\Phi_{2N}^+} \propto \prod_{a=1}^{2N} (I + i\gamma_a \gamma_{2N+a})$. For a stabilizer Gaussian operation, $\rho_\mathcal{E}$ is a stabilizer FGS, since $\ket{\Phi_{2N}^+}$ is; the operation $\mathcal{E}$ thus amounts to appending a stabilizer FGS, performing a post-selected Gaussian measurement, and partial tracing. As $\M_\Lambda$ is additive under tensor products, and hence invariant under appending an FGS, it suffices to establish monotonicity under (post-selected) measurement and partial tracing, which we do below.

To simplify notation, we will denote
\begin{equation}
    a_\Gamma = \frac{1}{\Tr(\rho^2)^2} \sum_{j>k}  |\Tr(\gamma_j \gamma_k \rho^2)|^2
\end{equation}
and
\begin{equation}
    a_\Sigma = \frac{1}{2\Tr(\rho^2)^2} \sum_{j,k} \Tr(\rho \gamma_j \rho \gamma_k)^2,
\end{equation}
such that $\M_\Lambda=N-a_\Gamma-a_\Sigma$ (see Eq.~\eqref{eq:d}).

Given a mixed stabilizer state $\rho_s$, we define the stabilizer group $\mathcal{S}(\rho_s)$ (without storing the signs) as
\begin{equation}
\mathcal{S}(\rho_s) = \left\lbrace P \in \mathcal{P}_N : P \rho_s  = \pm\rho_s \right\rbrace \, ,
\end{equation} 
similarly as in the pure state case (Eq.~\eqref{eq:stab_group_pure}). It follows from the standard theory of stabilizer states that $\mathcal{S}(\rho_s)$ contains $2^K$ elements, and is generated by $K$ mutually commuting Pauli operators~\cite{nielsen2011quantum}. In addition, we introduce the braid stabilizer set $\mathcal{S}_b(\rho_s)$ as the set of Pauli operators in the stabilizer group which are quadratic in the Majorana representation. It is easy to see that $a_\Gamma = |\mathcal{S}_b(\rho_{s})|$.

Let $M_{\text{maj}}^{ij}$ be the $K \times 2N$ stabilizer tableau in the majorana representation. One can show that $a_\Sigma$ counts the number of empty columns in $M_{\text{maj}}^{ij}$, divided by two. Notice that, while $M_{\text{maj}}^{ij}$ is not uniquely defined, the particular choice does not affect $a_\Sigma$ in this definition.

We will now show that $a_\Gamma+a_\Sigma$ is non-decreasing under the measurement of $i\gamma_j \gamma_k$ on a (possibly mixed) stabilizer state. Without loss of generality, we will consider the measurement of $Z_1$. Let $\sigma_s$ be the state after measurement.
There are several cases:
\begin{itemize}
    \item If $Z_1 \in \mathcal{S}(\rho_s)$, the measurement
outcome is fixed to $\pm 1$ and $\rho_s$ is unchanged.
\item If $Z_1 \notin \mathcal{S}(\rho_s)$, there are two possibilities. If $Z_1$ commutes with all elements of $\mathcal{S}(\rho_s)$, then $Z_1$ is added to the stabilizer group $\mathcal{S}(\rho_s)$ and hence to $\mathcal{S}_b(\rho_s)$, i.e., $\mathcal{S}_b(\sigma_s)= \mathcal{S}_b(\rho_s)\cup Z_1 $, which implies $a_\Gamma\to a_\Gamma+1$. Moreover, if any $P\in \mathcal{S}(\rho_s)$ contains $Z_1$, then $a_\Sigma$ is unchanged. Otherwise, $a_\Sigma \to a_\Sigma-1$. In both cases $a_\Gamma+a_\Sigma$ is non-decreasing. The second possibility is that $Z_1$ anticommutes with some element $P$ from $\mathcal{S}(\rho_s)$. In this case, $Z_1$ replaces $P$ in the stabilizer tableau. Since $Z_1$ consists of two Majoranas, and $P$ can have two or more Majoranas, $a_\Sigma$ is always non-decreasing. Moreover, we will see that $a_\Gamma$ is either unchanged or increased by $1$. We further divide into three cases:
\begin{itemize}
    \item If $Z_1$ commutes with all elements of $\mathcal{S}_b(\rho_s)$ then $Z_1$ is added to $\mathcal{S}_b(\rho_s)$, which implies $a_\Gamma\to a_\Gamma+1$.
    \item If $Z_1$ anticommutes with one element $P$ from $\mathcal{S}_b(\rho_s)$, then $Z_1$ replaces $P$ in $\mathcal{S}_b(\rho_s)$ and $a_\Gamma$ is unchanged.
    \item If $Z_1$ anticommutes with two elements $P$ and $Q$ from $\mathcal{S}_b(\rho_s)$, then one can check that $Z_1 P Q$ is a quadratic Majorana. Therefore, $Z_1$ and $Z_1 P Q$ replaces $P$ and $Q$ in $\mathcal{S}_b(\rho_s)$ and $a_\Gamma$ is unchanged.
\end{itemize}
Note that the three cases above exhaust all possibilities, since $Z_1$ can anticommute with no more than two elements in $\mathcal{S}_b(\rho_s)$. This is due to the fact that if $Z_1$ anticommutes with $P \in \mathcal{S}_b(\rho_s)$, then $Z_1$ and $P$ share precisely one Majorana operator. Since $Z_1$ consists of two Majorana operators, it can only anticommute with a maximum of two elements in $\mathcal{S}_b(\rho_s)$.
\end{itemize}

Next, we will show monotonicity of $\M_\Lambda$ under partial tracing. Without loss of generality, we consider the partial tracing of the first qubit. We have that $N\to N-1$, and it remains to show that $a_\Gamma+a_\Sigma$ cannot decrease by more than one. There are also several cases:
\begin{itemize}
    \item If $Z_1 \in \mathcal{S}(\rho_s)$, then the state is a product state between the first qubit and the rest of the system. It is thus easy to see that $a_\Gamma\to a_\Gamma-1$ while $a_\Sigma$ is unchanged.
    \item If $i\gamma_j \gamma_k \in \mathcal{S}_b(\rho_s)$ for $j\in\{1,2\}$ and $k>2$, then $i\gamma_j \gamma_k$ is removed from $\mathcal{S}_b(\rho_s)$, so that $a_\Gamma\to a_\Gamma-1$. $a_\Sigma$ is always non-decreasing since the column $k$ becomes empty, while the first and second columns can have at most one empty column.
    \item If two operators $i\gamma_{j_l} \gamma_{k_l} \in \mathcal{S}_b(\rho_s)$ for $l\in\{1,2\}$, $j_{1,2}\in\{1,2\}$ and $k_{1,2}>2$, then both operators are  removed from $\mathcal{S}_b(\rho_s)$, so that $a_\Gamma\to a_\Gamma-2$. We also have $a_\Sigma\to a_\Sigma+1$ since the two columns $k_{1,2}$ become empty.
\end{itemize}
Thus, in all three cases, $a_\Gamma+a_\Sigma$ cannot decrease by more than one, showing the monotonicity.

\section{Evolution of non-Gaussianity with doped braiding gates}
\label{app:evolution occu doped}

In this Appendix, we derive Eq.~\eqref{eq:evolution_swapdoped}, which describes the average evolution of the $\alpha=2$ occupation number entropy $\nGoccu$ for global random matchgate circuits doped with SWAP gates, as depicted in Fig.~\ref{fig:circuit_evolution}a. This quantity corresponds to the average evolution of the Gaussian nullity, $\nu_\mathcal{G}$, when the system dynamics are restricted to braiding gates interspersed with SWAP gates. To obtain this average behavior, we first characterize how braiding and SWAP gates transform the system's stabilizers. We then model the survival probability of Gaussian modes under the action of SWAP gates in order to compute the expected non-Gaussianity after $t$ time steps.

We begin by describing the stabilizer evolution under braiding and SWAP operations. A stabilizer state is fully characterized by its stabilizer group.  
A stabilizer corresponds to a Gaussian mode if it is quadratic in the Majorana operators, i.e., if it has the form $\pm i\gamma_i\gamma_j$, with $1\leq i,j \leq 2N$.  
The Gaussian nullity $\nu_\mathcal{G}$ is defined as the total number of qubits $N$ minus the number of independent Gaussian modes in the stabilizer group.

A braiding gate acting on a stabilizer state implements a permutation $\sigma$ of the Majorana operators. Therefore, if a stabilizer of the state is $s'\propto \gamma_{i_1}\ldots\gamma_{i_n}$, the transformed stabilizer becomes $s\propto \gamma_{\sigma(i_1)}\ldots\gamma_{\sigma(i_n)}$. Importantly, braiding operations preserve the number of Majorana operators in each stabilizer; in particular, quadratic stabilizers remain quadratic.

In contrast, a SWAP gate acting on qubits $k$ and $k+1$ involves the four Majorana operators $\{\gamma_{2k-1}, \gamma_{2k}, \gamma_{2k+1}, \gamma_{2k+2}\}$. The transformation rules are:
\begin{align*}
    \gamma_{2i-1} &\rightarrow \gamma_{2i-1}\gamma_{2i}\gamma_{2i+1}, \\
    \gamma_{2i}   &\rightarrow -\gamma_{2i-1}\gamma_{2i}\gamma_{2i+2}, \\
    \gamma_{2i+1} &\rightarrow -\gamma_{2i-1}\gamma_{2i+1}\gamma_{2i+2}, \\
    \gamma_{2i+2} &\rightarrow \gamma_{2i}\gamma_{2i+1}\gamma_{2i+2}.
\end{align*}
Unlike braiding operations, a SWAP gate maps single Majorana operators to products of three Majoranas. Consequently, if a quadratic stabilizer $\pm i \gamma_n\gamma_m$ involves one of the four Majoranas affected by the SWAP, it is mapped to an operator containing four Majoranas. This removes the corresponding Gaussian mode and increases the non-Gaussianity.

We model the system initially as a product state of $N$ disjoint pairs of Majoranas, representing $N$ quadratic stabilizers. At each time step, the global random braiding gate effectively randomizes the Majorana indices. The subsequent SWAP gate acts on two qubits, which corresponds to selecting four Majorana indices out of the total $2N$.

We adopt the following simplified destruction model:
\begin{enumerate}
    \item There are $2N$ particles partitioned into pairs (Gaussian modes).
    \item At each step, 4 distinct particles are chosen uniformly at random.
    \item If a chosen particle belongs to a surviving pair, that pair is ``broken'' (i.e., the mode becomes non-Gaussian).
\end{enumerate}
We neglect higher-order recovery processes (for example, a SWAP acting twice on the exact same set of Majoranas and accidentally restoring a quadratic mode), since such events are statistically suppressed in the large-$N$ limit.

Let $Y_t$ denote the expected number of Gaussian pairs remaining after $t$ steps. To determine the decay rate, we compute the probability that a specific pair is \textit{not} affected by the SWAP gate.  
The total number of ways to choose 4 Majoranas out of $2N$ is $\binom{2N}{4}$.  
The number of ways to choose 4 Majoranas such that neither element of a given pair is selected is $\binom{2N-2}{4}$, since the selection must be made from the remaining $2N-2$ Majoranas.  
Thus, the probability that a specific pair survives a single step is
\begin{equation*}
    P(\text{survival}) = \frac{\binom{2N-2}{4}}{\binom{2N}{4}} = \frac{(2N-4)(2N-5)}{(2N)(2N-1)}.
\end{equation*}
The probability $p$ that a pair is destroyed is the complement:
\begin{equation*}
    p = 1 - P(\text{survival}) = \frac{8N-10}{N(2N-1)} \approx \frac{4}{N},
\end{equation*}
where the approximation holds in the large-$N$ limit.

The expected number of surviving pairs evolves according to the recurrence relation $Y_{t+1} = Y_t - p Y_t = (1-p)Y_t$, with initial condition $Y_0 = N$. The solution to this recursion is an exponential decay,
\[
Y_t = N (1-p)^t \approx N e^{-pt}.
\]

Since the Gaussian nullity is the complement of the number of Gaussian modes ($\nu_\mathcal{G} = N - Y_t$), the resulting evolution of the non-Gaussianity is
\begin{equation}
    \overline{\nu}_\mathcal{G}(t) = N \left( 1 - \exp\left(-\frac{4t}{N}\right) \right).
\end{equation}

\section{Interacting quantum Ising model}
\label{app:Ising}

As a second example of non-Gaussianity in ground states of interacting many-body systems, we consider an interacting variant of the quantum Ising model. In particular, the Hamiltonian we investigate is given by
\begin{equation}
\begin{aligned}
    H = - \sum_{j=1}^{L} \Big[
    &(1-g) \, X_j X_{j+1}
    + g \, Z_j\\[-0.28cm]
    &+ {\textstyle\frac{g}{2}} \, Z_j Z_{j+1}
    + {\textstyle\frac{1-g}{2}} \, X_{j-1} X_{j+1}\Big]
    \label{eq:Ising_Hamiltonian}
\end{aligned}
\end{equation}
and we take periodic boundary conditions. At $g=0$, the model describes an interacting ferromagnet whose ground state would spontaneously break the $\mathbb{Z}_2$ symmetry in the thermodynamic limit; however, since we are interested in fermionic states with definite parity, we consider the symmetric states with definite parity. For $g=1$, the model describes an interacting paramagnet with a unique ground state. As we tune the parameter $g$ from $g=0$ to $g=1$, we cross a phase transition from the symmetry-broken to the symmetric phase.

After mapping this model to fermions through a Jordan--Wigner transform, the Hamiltonian becomes
\begin{equation}
\begin{aligned}
    H = \sum_{j=1}^{L} \Big[
    &(1-g) \, i \gamma_{2j} \gamma_{2j+1}
    + g \, i \gamma_{2j-1} \gamma_{2j}\\[-0.28cm]
    &+ {\textstyle\frac{g}{2}} \, \gamma_{2j-1} \gamma_{2j} \gamma_{2j+1} \gamma_{2j+2}\\
    &+ {\textstyle\frac{1-g}{2}} \, \gamma_{2j-2} \gamma_{2j-1} \gamma_{2j} \gamma_{2j+1}\Big]
\end{aligned}
\end{equation}
with periodic boundary conditions for odd-parity states and with antiperiodic boundary conditions for even-parity states. Note that with periodic boundary conditions this Hamiltonian has a Kramers--Wannier duality, that is, changing $g\to1-g$ and shifting all Majorana operators as $\gamma_j\to\gamma_{j+1}$ leaves the model invariant. Thus, the model's unique phase transition must lie at $g=0.5$ where it is self-dual, and the transition lies in the Ising universality class with $c=\frac{1}{2}$~\cite{Rahmani2015}.

\begin{figure}
    \centering
    \includegraphics[width=\linewidth]{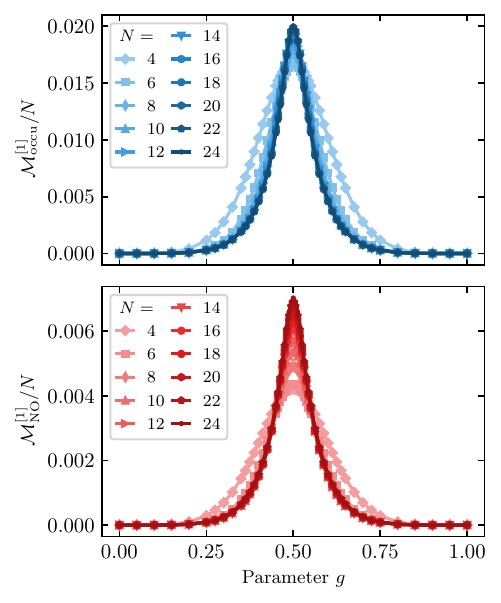}
    \caption{Occupation number entropy density $\nGoccu^{[1]}/N$ (top panel) and an upper bound on the natural-orbital participation entropy density $\nGNO^{[1]}/N$ (bottom panel) in the interacting quantum Ising model [see Eq.~\eqref{eq:Ising_Hamiltonian}] plotted against the parameter $g$. The two end-points $g=0$ and $g=1$ correspond to Gaussian states, and both quantities peak at the critical point $g=0.5$. States at $g$ and $1-g$ are related by a translation of Majorana operators, and thus both quantities are symmetric around $g=0.5$.
    Note that we compute the natural-orbital participation entropy by taking the minimium participation entropy in either the translation-invariant or inversion-symmetric natural-orbital basis (see text for additional details) instead of performing a costly direct minimization of the natural-orbital participation entropy within the degenerate subspaces; we found this to yield smaller values than a direct brute-force optimization.}
    \label{fig:Ising_nG}
\end{figure}

In contrast to the example in the main text, this model does not have a $U(1)$-symmetry of $z$-rotations and a $\mathbb{Z}_2$-symmetry of $\pi$-rotations around the $x$-axis, yet it still has both a translation and inversion symmetry which enforce degeneracies in the spectrum of the covariance matrix. As in the main text, we simultaneously bring the covariance matrix and one of the two symmetries into normal form, and select the smaller natural-orbital participation entropy of the two cases, as that yields better results than numerically optimizing the participation entropy within the degenerate subspaces.

Fig.~\ref{fig:Ising_nG} shows the results for the occupation number entropy density $\nGoccu^{[1]}/N$ (top panel) and the natural-orbital participation entropy density $\nGNO^{[1]}/N$ (bottom panel). Since $g$ and $1-g$ are related by shifting the Majorana operators by one site, both quantities are also the same for $g$ and $1-g$, i.e., symmetric around $g=0.5$. Thus, starting from Gaussian states at $g=0$ and $g=1$, the non-Gaussianity increases monotonically as one approaches the critical point at $g=0.5$. As the system size increases, the non-Gaussianity appears to converge rapidly to its density in the thermodynamic limit.

\begin{figure}[ht]
    \centering
    \includegraphics[width=\linewidth]{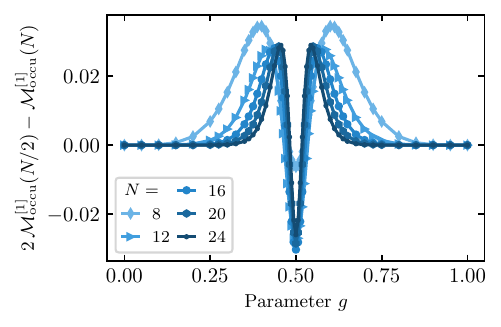}
    \caption{The subleading term of the occupation number entropy in the interacting quantum Ising model [see Eq.~\eqref{eq:Ising_Hamiltonian}] plotted against the parameter $g$, extracted by subtracting the occupation number entropy at a given system size $N$ from twice the value at half the system size. It goes to zero in the gapped phases where the correlation length is smaller than the system size and there are positive finite-size corrections around the critical point where the correlation length is larger than the system. At the critical point, the subleading contribution is negative and mostly independent of system size.}
    \label{fig:Ising_nG_subleading}
\end{figure}

We investigate this convergence by inspecting the subleading term of the occupation number entropy. For this, as in the main text, we consider the difference $2 \nGoccu^{[1]}(N/2) - \nGoccu^{[1]}(N)$, which should cancel out an extensive term and leave only a subleading constant term. The results are shown in Fig.~\ref{fig:Ising_nG_subleading}. There, we see that the subleading term quickly goes to zero with increasing system size in the gapped phases, and is only nonzero in a small window around the critical point, presumably where the correlation length is not smaller than the system size. As in the main text, we see that finite-size effects in the gapped phase lead to a positive subleading term, while the critical point has a negative subleading term that seems to not scale with system size.
\newpage

\begin{figure}[ht]
    \centering
    \includegraphics[width=\linewidth]{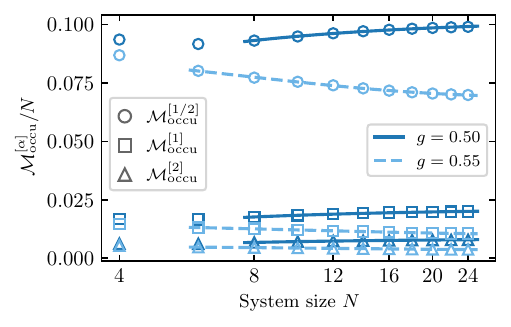}
    \caption{Convergence of the occupation number entropy density $\nGoccu^{[\alpha]}/N$ in the interacting quantum Ising model [see Eq.~\eqref{eq:Ising_Hamiltonian}] with system size $N$. The symbols show the occupation number entropy density for $\alpha=1/2$ (circles), for $\alpha=1$ (squares) and for $\alpha=2$ (triangles), both at the critical point $g=0.5$ (darker blue) and at one representative example in a gapped phase at $g=0.55$. Solid lines show fits of the form $f(N) = a - b/N$ to the data at the critical point, dashed lines show fits of the form $f(N) = a + b \, e^{-c \, N}$ to the data in the gapped phase. The exponential convergence to a constant with system size in the gapped phase is consistent with the result derived in Sec.~\ref{sec:TI}.}
    \label{fig:Ising_nG_convergence}
\end{figure}

We show the convergence with system size in Fig.~\ref{fig:Ising_nG_convergence}, where dark blue markers show data at the critical point $g=0.5$, and light blue markers show data at a representative point in the gapped phase at $g=0.55$. Different marker shapes show data for different values of $\alpha$, circles corresponding to $\alpha=1/2$, squares to $\alpha=1$ and triangles to $\alpha=2$. Solid lines show a fit of the form $\nGoccu^{[\alpha]}/N = m_{(\infty)}^{[\alpha]} + c/N$ to the data at the critical point. Dashed lines show a fit of the form $\nGoccu^{[\alpha]}/N = m_{(\infty)}^{[\alpha]} + c \, e^{-\delta N}$ to the data in the gapped phase. Data points not included in the range where the fitted function is plotted are excluded from the fit due to strong finite-size effects. In the gapped phase, the data is consistent with the exponential convergence predicted in Sec.~\ref{sec:TI}.

\bibliography{references}

\end{document}